\documentclass[fleqn,10pt]{olplainarticle}
\usepackage{algorithm}
\usepackage{algpseudocode}
\usepackage{amsthm}
\usepackage{mathrsfs}
\usepackage{amsmath,amssymb,mathtools,bm}
\usepackage[hidelinks]{hyperref}
\usepackage[nameinlink,noabbrev,capitalise]{cleveref}
\usepackage{microtype}
\usepackage{newtxtext}
\usepackage{newtxmath}

\newcommand{\Data}{y}

\newcommand{\Model}{\mathcal{M}}
\newcommand{\Z}{\mathcal{Z}}
\newcommand{\Zhat}{\widehat{\mathcal{Z}}}
\newcommand{\EE}{\mathbb{E}}
\newcommand{\PP}{\mathbb{P}}
\newcommand{\RR}{\mathbb{R}}

\providecommand{\Var}{\operatorname{Var}} 
\providecommand{\Cov}{\operatorname{Cov}}
\newcommand{\sone}{s_1}
\newcommand{\stwo}{s_2}
\newcommand{\Sone}{S_1}
\newcommand{\Stwo}{S_2}
\newcommand{\Xj}[1]{\theta^{\ast}_{#1}}  
\newcommand{\Yi}[1]{\tilde\theta_{#1}}   
\newcommand{\Hov}{H}                     
\newcommand{\Iov}{I}                     
\newcommand{\neff}{n_{\mathrm{eff}}}
\newcommand{\neffhat}{\widehat{n}_{\mathrm{eff}}}
\newcommand{\sigN}{\sigma_{N}}
\newcommand{\sighat}{\widehat{\sigma}_{\delta}}
\newcommand{\chat}{\widehat{c}}
\newcommand{\cstar}{c^{\star}}
\newcommand{\pistar}{\pi^{\star}}
\newcommand{\mix}{m}
\newcommand{\Hzero}{\widehat{H}_0}
\newcommand{\Hhat}{\widehat{H}}
\newcommand{\Vjack}{\mathcal{V}_{\mathrm{jack}}}
\newcommand{\Imax}{\mathcal{I}_{\max}}
\newcommand{\Gr}{\mathcal{G}_{\max}}
\newcommand{\JackVar}{\widehat{\Var}_{\mathrm{jack}}}
\newcommand{\Tp}{\mathcal{T}}
\newcommand{\PostBlock}[1]{\mathcal{B}^{\mathrm{post}}_{#1}}
\newcommand{\PropBlock}[1]{\mathcal{B}^{\mathrm{prop}}_{#1}}
\newcommand{\ind}{\mathbf{1}}
\newcommand{\dd}{\,\mathrm{d}}
\newcommand{\KL}{\mathrm{KL}}
\newcommand{\sigNP}{\sigma_{N,P}}
\newcommand{\neffP}{n_{\mathrm{eff}}^{P}}
\newcommand{\neffPhat}{\widehat{n}_{\mathrm{eff}}^{P}}
\newcommand{\sighatP}{\widehat{\sigma}_{\delta,P}}
\newcommand{\cplus}{c_{+}}
\newcommand{\cminus}{c_{-}}
\newcommand{\gs}{\gamma_{*}}

\newtheorem{theorem}{Theorem}      
\newtheorem{proposition}[theorem]{Proposition}
\newtheorem{corollary}[theorem]{Corollary}
\newtheorem{lemma}[theorem]{Lemma}

\newtheorem{assumption}{Assumption}

\theoremstyle{definition}
\newtheorem{definition}[theorem]{Definition}
\newtheorem{remark}[theorem]{Remark}
	
\title{Overlap, Not Moments: Sample Size and Standard Errors in Bridge Sampling}
\author[1]{Giorgio Micaletto}
\affil[1]{University of Chicago, Department of Statistics}

\keywords{Bayes factor; Berry--Esseen bound; Marginal likelihood; Markov chain Monte Carlo; Monte Carlo standard error.}

\begin{abstract}
Bridge sampling estimates of marginal likelihoods are routinely reported with a delta-method Monte Carlo standard error (MCSE). Existing theory is asymptotic and cannot say whether a given run is long enough to trust the estimate or its MCSE. We give finite-sample answers. The optimal bridge estimator fits a logistic regression that separates posterior from proposal draws, and its accuracy is governed by their overlap, within a factor of two of the classification's Bayes error. Because the estimating function is a sum of bounded terms, no moment conditions are needed, and all bounds hold uniformly over targets, proposals and dimensions. A number of draws inversely proportional to the overlap suffices and is necessary in the worst case, paralleling the exponential-in-Kullback--Leibler requirement of importance sampling. A Berry--Esseen bound with an explicit constant holds at the optimal rate, and the MCSE is accurate once the effective overlap is large; it is estimated at no cost. For Markov chain draws, the required sample size grows in proportion to the sampler's relaxation time, unavoidably in the worst case. The estimated bridge, the fixed-point iteration and the fitted proposal matter only at second order, and we give free diagnostics for sampler failure and misestimated autocorrelation.
\end{abstract}

\begin{document}

\flushbottom
\maketitle
\thispagestyle{empty}

\section{Introduction}\label{sec:intro}

The marginal likelihood $\Z=p(\Data\mid\Model)$ is the basis of Bayesian model comparison \citep{jeffreys_probability_1961,kass_bayes_1995}, and bridge sampling \citep{bennett_estimation_1976,meng_bridge_1996,gelman_simulating_1998} is one of the few general-purpose methods for computing it from posterior output. The estimator of \citet{meng_bridge_1996} is the default in widely used software \citep{gronau_tutorial_2017,gronau_bridgesampling_2020}, and it is usually accompanied by a Monte Carlo standard error (MCSE) computed from a delta-method formula \citep{fruhwirth_estimating_2004,micaletto_diagnostics_2025}. The theory behind that MCSE is asymptotic, and cannot say whether the draws at hand are enough, and the question is most pressing exactly where bridge sampling is hardest. With high-dimensional posteriors and imperfect proposals, posterior and proposal overlap poorly, the importance ratios may have no finite moments, rare draws can dominate the estimate at practical sample sizes \citep{micaletto_diagnostics_2025}, and fitting the proposal on the same draws induces a bias that grows with dimension \citep{wong_properties_2020}. A practitioner who reports $\log\Zhat\pm\sighat$ therefore faces two questions that asymptotic theory alone cannot answer: is this run large enough for the estimate and its MCSE to be trusted, and if not, what went wrong?

Our answers rest on a classical observation: the optimal bridge estimator is the maximum-likelihood estimate of the intercept in a logistic regression that discriminates posterior draws from proposal draws \citep{geyer_estimating_1994,kong_theory_2003}, which is also the one-parameter case of noise-contrastive estimation \citep{gutmann_noise_2012}. Its estimating function $W(c)$, with $c=\log z$, is a sum of \emph{bounded}, monotone terms, and two consequences drive this paper. First, as for monotone location M-estimators \citep{huber_robust_1964}, $\{\log\Zhat\le\log\Z+t\}$ coincides with $\{W(\log\Z+t)\le0\}$, so the distribution of $\log\Zhat$ is that of the sign of a sum of independent bounded variables, at every sample size and for every target. Finite-sample control therefore requires no moment conditions on the importance ratios, in contrast to importance sampling, whose error depends on the distribution of unbounded ratios \citep{chatterjee_sample_2018}. Second, the population version of $W$ vanishes at $\log\Z$ for \emph{every} bridge function and \emph{every} proposal, which makes the estimating equation Neyman-orthogonal \citep{chernozhukov_double_2018} in all nuisance quantities.

Write $N=\Sone+\Stwo$ for the total number of posterior and proposal draws, $\sone=\Sone/N$ and $\stwo=\Stwo/N$, let $\mix=\sone p+\stwo g$ be the pooled law, and let $\pistar=\stwo g/\mix$ be the optimal classifier's probability that a draw came from the proposal. The central quantity is the overlap $\Hov=\EE_\mix[\pistar(1-\pistar)]$, which lies between one half of and the whole of the Bayes error $\varepsilon_B$ of this classification problem. Since $N\varepsilon_B$ is the expected number of draws that the optimal classifier assigns to the wrong sample, $N\Hov$ is that number up to a factor of two. The closely related \emph{effective overlap} $\neff=N\Hov(1-\Hov/(\sone\stwo))$ is the variance of $W$ at the truth, and $\neff\approx N\Hov$ when the overlap is poor. Our main message is that $\neff$ determines both how accurate $\log\Zhat$ is and how accurate its MCSE is, uniformly over targets, proposals and dimensions, and that it comes at no cost: its estimate $\neffhat$ is computed from the within-sample variances of the fitted classifier probabilities, which every bridge computation already produces.

\paragraph{Contributions.} Our main results are as follows.
\begin{enumerate}[leftmargin=*,label=(\roman*)]
\item \emph{The sample size required.} The error satisfies $\PP(|\log\Zhat-\log\Z|\ge t)\le 2\exp\{-N\Hov\kappa(t)\}$, where $\kappa(t)\sim t^2/2$, with no moment or tail assumptions (\cref{thm:concentration}). Conversely, for every $\Hov\in(0,1/8]$ there are targets and proposals for which $\PP(\Zhat\in\{0,\infty\})\ge e^{-4N\Hov}$ (\cref{thm:sharp}). Hence $N\asymp 1/\Hov$ draws are necessary and sufficient in the worst case. This inverse Bayes error law is the bridge-sampling counterpart of the result of \citet{chatterjee_sample_2018} that importance sampling needs about $\exp\{\KL\}$ draws, and for two Gaussians with a mean shift it improves the exponent by a factor of four (\cref{rem:ISvsbridge}).
\item \emph{A Berry--Esseen bound with an explicit constant.} Let $\sigN^2$ be the asymptotic variance of \citet{meng_bridge_1996}; the Kolmogorov distance between the law of $(\log\Zhat-\log\Z)/\sigN$ and $\mathcal N(0,1)$ is at most $3.1/\sqrt{\neff}$ (\cref{thm:BE}), and the rate cannot be improved (\cref{thm:sharp}). In particular, asymptotic normality holds whenever $\neff\to\infty$, with no further condition.
\item \emph{When the MCSE can be trusted.} With probability $1-\delta$, $|\log(\sighat/\sigN)|\le10\sqrt{\log(12/\delta)/\neff}$, and intervals built from the MCSE have coverage error $\mathcal O(\sqrt{\log\neff/\neff})$ uniformly over all targets and proposals (\cref{thm:mcse}). With the same probability, $\neffhat$ is within a factor $\exp\{12.1\sqrt{\log(12/\delta)/\neff}\}$ of $\neff$ (\cref{prop:certificate}). Conversely, a large $\neffhat$ certifies a large overlap: deterministically, $\neffhat$ is at most about twice the larger of two sums with mean $N\Hov$, so with independent draws a run with $N\Hov\le1$ passes the check $\neffhat\ge10$ with probability at most $0.04$ (\cref{prop:onesided}). In all our continuous examples, nominal $95\%$ intervals cover to within $0.01$ once $\neff\ge3.7$, which motivates the conservative rule of flagging $\neffhat<10$.
\item \emph{Markov chain draws.} If the posterior draws come from a reversible chain with spectral gap $\gamma$, the bound in (i) holds with $\kappa(t)$ replaced by $\kappa_P(t)\gtrsim\gamma t^2$, and the probability that $\Zhat\in\{0,\infty\}$ is at most $e^{-N\Hov}+e^{-\gamma N\Hov/2}$, and the inflation of the required sample size by $1/\gamma$ cannot be avoided (\cref{thm:mcmc-conc,thm:mcmc-sharp}). A Berry--Esseen bound holds whose main term is $C_P/\sqrt{v_P}$, where $v_P$ is the long-run variance of the estimating function and $C_P\le83/\gamma$ for positive chains, and no bound of this form can have $C_P$ of smaller order than $1/\gamma$ (\cref{thm:mcmc-BE,thm:mcmc-sharp}). Its proof rests on a Berry--Esseen bound for the sum of an independent part and a reversible-chain part (\cref{lem:be-chain}), a discrete-time analogue of a bound of \citet{lezaud_chernoff_2001} for continuous-time processes, with constant linear in $\|u\|_\infty/\sigma_P(u)$, rather than cubic \citep{kloeckner_effective_2019}, which is what the small-overlap regime requires; the lemma may be of independent interest. The MCMC version of the MCSE satisfies the analogue of (iii) up to the accuracy of its long-run variance estimate (\cref{thm:mcmc-mcse}), and in simulations coverage is a function of $\neffP=\neff/\tau_W$ alone, where $\tau_W$ is the integrated autocorrelation time of the estimating function.
\item \emph{Why nuisance matters only at second order.} Sharpening the convergence theorem of \citet{meng_bridge_1996}, the sample map is a global contraction on the log scale, and its derivative at the solution equals one minus the ratio of two estimates of $\Hov$, which is $\mathcal O_\PP(\neff^{-1/2})$; the iteration is therefore asymptotically perfectly conditioned (\cref{thm:iteration}). Estimating the proposal, even on the same draws, leaves the first-order behaviour of the estimator and the validity of the MCSE unchanged (\cref{thm:proposal}), with only a second-order bias remaining (if fitted on the same draws). An expansion one order further, checked by simulation, accounts for the correlation-induced bias documented by \citet{wong_properties_2020}, and sample splitting removes it (\cref{rem:secondorder}).
\item \emph{Diagnostics for what the MCSE cannot see.} The contraction factor $1-\Hhat/\Hzero$ doubles as a calibration test for sampler failure: it checks whether the residuals of the implicit classifier are uncorrelated with its fitted probabilities. Its studentized version is asymptotically $t_{G-1}$ under correct sampling (\cref{thm:calib-null}), and under sampler failure we derive its limit: to first order the test sees the part of the sampler's error that covaries with the classifier probabilities on the overlap region, the bias depends on the mean of the error there, and errors that are constant there cannot be detected from the bridge output by any method (\cref{prop:calib-alt}). In our experiments it detects posterior samplers that misweight modes, a failure that every variance-based diagnostic misses (\cref{sec:exp-failure}). For Markov chain output, a delete-a-group jackknife comparison with the spectral MCSE has fixed-$G$ calibration, $(G-1)\Vjack\Rightarrow\chi^2_{G-1}$, and pivotal thresholds for the influence of a single block (\cref{thm:jackknife}). None of these diagnostics requires additional evaluations of the posterior or the proposal density.
\end{enumerate}
Taken together, these results reduce the reliability of a bridge sampling estimate to a short list of conditions, each with its own diagnostic: insufficient overlap, which $\neffhat$ measures; a posterior sampler that does not sample from $p$, which the calibration test targets; and, for Markov chain output, misestimated autocorrelation, which the jackknife comparison targets. The experiments in \cref{sec:experiments} check each claim: finite-sample computations with independent and with Markov chain draws, null calibration, MCMC output, proposal fitting, sampler failure and metastability, and a universality study in which coverage curves from very different targets, proposals and dimensions collapse onto one curve in $\neff$. 

\subsection{Related work and what is new}\label{sec:related}

\paragraph{Bridge sampling and overlap.} Our results give finite-sample control of quantities that \citet{meng_bridge_1996} (henceforth MW) introduced and analysed asymptotically; see also \citet{chen_monte_2000}. MW write the optimal estimator as the unique root of a monotone ``score'' function, our $W$ (their (A.5)), prove that the iteration \cref{eq:MWupdate} converges to it from any starting value with monotonically decreasing error (their Theorem~2), and note that every iterate is consistent. Their minimal asymptotic relative mean squared error is expressed through $\int(\tilde p_1^{-1}+\tilde p_2^{-1})^{-1}$ with $\tilde p_i=s_ip_i$ (their (3.4)), which is our $\Hov$, so their error is our $\sigN^2$; the variance of their score at the true ratio (their (A.10)) is our $\neff$; and they observe that the numerator of the iteration at convergence gives a free consistent estimate of the error. Their Gaussian mean-shift example, in which the relative variance grows like $|\mu|e^{\mu^2/8}$ for the optimal bridge against $e^{\mu^2}$ for importance sampling (their (6.1) and (6.4)), already shows that overlap governs the asymptotic error. That dependence is therefore classical, and it motivates transformations that increase overlap \citep{meng_warp_2002,wang_warp_2022,ding_channelling_2024}. What this error analysis does not provide is control at a given sample size: how many draws are needed before the asymptotic description applies, and whether a reported MCSE is accurate for the run at hand. Supplying that control, together with a way to estimate it from the bridge output, is the aim of this paper.

Among later work, \citet{wang_warp_2022} and \citet{ding_channelling_2024} express the asymptotic variance through a sample-size-adjusted harmonic divergence, which is our $\beta$ up to the weighting convention, and \citet{xing_improving_2022} estimates the asymptotic relative mean squared error through this $f$-divergence. \citet{fruhwirth_estimating_2004} proposed the MCSE formula used in \citet{gronau_bridgesampling_2020}. \citet{micaletto_diagnostics_2025} derived its MCMC version and observed that, although the summands are bounded, rare draws can dominate at practical sample sizes when the overlap is poor, and \citet{wong_properties_2020} documented the bias caused by fitting the proposal on the same draws.

\paragraph{The same estimator in other fields.} The optimal bridge estimator is the maximum-likelihood estimator of reverse logistic regression \citep{geyer_estimating_1994,kong_theory_2003,tan_likelihood_2004}. It is also the Bennett acceptance ratio (BAR) of molecular simulation \citep{bennett_estimation_1976}, which \citet{shirts_equilibrium_2003} gave the same logistic-regression interpretation and MBAR extends to several states \citep{shirts_statistically_2008}, and the intercept-only case of noise-contrastive estimation (NCE) \citep{gutmann_noise_2012}; \citet{martino_unifying_2026} gives a recent unified account. In each of these literatures the error theory is asymptotic \citep[e.g.][]{gutmann_noise_2012,chehab_provable_2023,li_understanding_2023}, and work on multi-parameter NCE explains its failure under poor overlap through flat losses and ill-conditioned Hessians \citep{liu_analyzing_2022,lee_pitfalls_2023}. In free-energy computation, reliability is judged by a threshold of $0.03$ on the off-diagonal entries of the MBAR overlap matrix \citep{klimovich_guidelines_2015}. For two states those entries are $\Hhat/\sone$ and $\Hhat/\stwo$ (\cref{app:related}), so the threshold constrains the estimated overlap but not the sample size. \citet{reinhardt_small_2021} found numerically that Bennett's asymptotic formula underestimates the mean squared error when the overlap is small, and recommended judging BAR by the overlap together with the number of independent samples. The effective overlap $\neff$ combines the two, and the sample size at which BAR overtakes a simpler estimator in their experiments corresponds to $\neff$ of order one (\cref{rem:ISvsbridge}).

\paragraph{Non-asymptotic and Markov chain results.} Sample-size results exist for importance sampling \citep{chatterjee_sample_2018,block_partition_2026} and for Jarzynski-type and annealed importance sampling estimators \citep{guo_complexity_2025}. In theoretical computer science, $(\varepsilon,\delta)$-guarantees for product and paired-product estimators of Gibbs partition functions rest on Chebyshev bounds under bounded relative variances \citep{stefankovic_adaptive_2009,huber_approximation_2015,kolmogorov_faster_2018,guo_optimal_2026}; the paired-product estimator is a bridge-type estimator with the geometric bridge. For Markov chains we use the Bernstein inequality of \citet{jiang_bernstein_2025}, whose variance proxy is sharp, and bounds of \citet{paulin_concentration_2015}. Berry--Esseen bounds with explicit constants for Markov chains are rare \citep{mann_berry_1996,lezaud_chernoff_2001,kloeckner_effective_2019}. Our \cref{lem:be-chain} is a discrete-time analogue of Lezaud's bound for continuous-time processes. Its constant is linear in $\|f\|_\infty/\sigma_P(f)$, which the small-overlap regime requires (\cref{sec:mcmc-be}), whereas the general bound of \citet{kloeckner_effective_2019} is cubic in this ratio.

\paragraph{What is new.} To our knowledge, no non-asymptotic error bounds, Berry--Esseen bounds or guarantees for the MCSE have been given for the optimal bridge estimator or for its equivalents, BAR and intercept-only NCE. Relative to the work above, the new elements are the distributional reduction, which uses \citet{huber_robust_1964}; moment-free two-sided sample-size bounds in terms of $N\Hov$; a Berry--Esseen bound with a universal constant and a matching lower bound on the rate; guarantees for the delta-method MCSE and its intervals; non-asymptotic results for MCMC output, including \cref{lem:be-chain}; an explicit contraction factor for the MW iteration; the converse certificate of \cref{prop:onesided}; and the calibration test and the fixed-$G$ jackknife diagnostics, with their null calibration and, for the calibration test, a description of what it can detect. \Cref{sec:orth} makes explicit what is implicit in the invariance of the bridge identity and in Meng and Wong's observation that every iterate is consistent. In this light, the pre-asymptotic regime identified by \citet{micaletto_diagnostics_2025}, is the regime of small $\neff$, which $\neffhat$ measures directly.

\section{Setting}\label{sec:setting}

\subsection{Bridge sampling and the optimal bridge}

Let $\Data$ be the data and $\theta\in\Theta$ the parameter of a model $\Model$, with likelihood $p(\Data\mid\theta)$ and prior $p(\theta)$ (we suppress $\Model$). The unnormalized posterior is $f(\theta)=p(\Data\mid\theta)p(\theta)$, the evidence is $\Z=\int_\Theta f(\theta)\dd\theta$, and the posterior is $p=f/\Z$. All densities are taken with respect to a common $\sigma$-finite measure on $\Theta$. Let $g$ be a proposal density, which may depend on a parameter $\vartheta$ fitted to posterior draws. For any bridge function $h$ with $0<\EE_p[gh]<\infty$, the bridge identity \citep{bennett_estimation_1976,meng_bridge_1996} is $
\Z=\EE_g[f h]/\EE_p[g h]$, and it holds for every $h$, because both expectations equal $\int f g h$. Given posterior draws $\Xj1,\dots,\Xj{\Sone}$ and proposal draws $\Yi1,\dots,\Yi{\Stwo}$, write $N=\Sone+\Stwo$, $\sone=\Sone/N$ and $\stwo=\Stwo/N$. The asymptotically optimal bridge of MW is $h_z(\theta)=\{\sone f(\theta)+\stwo z\,g(\theta)\}^{-1}$ evaluated at $z=\Z$. Since $\Z$ is unknown, the estimator is defined as the solution of the fixed-point equation
\begin{equation}\label{eq:MWupdate}
z=T(z):=\frac{n(z)}{d(z)},\qquad
n(z)=\frac1\Stwo\sum_{i=1}^{\Stwo}\frac{\ell(\Yi i)}{\sone\ell(\Yi i)+\stwo z},\qquad
d(z)=\frac1\Sone\sum_{j=1}^{\Sone}\frac{1}{\sone\ell(\Xj j)+\stwo z},
\end{equation}
where $\ell=f/g$ is the importance ratio. In practice \cref{eq:MWupdate} is solved by iterating $z^{(t+1)}=T(z^{(t)})$.

\subsection{The classification representation}

For $z>0$ and $\theta$ with $f(\theta)+g(\theta)>0$ define the \emph{responsibility}
\begin{equation}\label{eq:pi}
\pi_z(\theta):=\frac{\stwo z\,g(\theta)}{\sone f(\theta)+\stwo z\,g(\theta)}\in[0,1].
\end{equation}
If $z=\Z$, this is the probability that $\theta$ was drawn from $g$ rather than from $p$ under $\mix=\sone p+\stwo g$. We write $\pistar:=\pi_\Z=\stwo g/\mix$, and with $c=\log z$ we abuse notation and write $\pi_c$. Since $\pi_c(\theta)=\{1+\exp(-c-\lambda(\theta))\}^{-1}$ with $\lambda=\log(\stwo g/(\sone f))$, we have $\partial_c\pi_c=\pi_c(1-\pi_c)$. Define the estimating function
\begin{equation}\label{eq:W}
W(c):=\sum_{i=1}^{\Stwo}\{1-\pi_c(\Yi i)\}-\sum_{j=1}^{\Sone}\pi_c(\Xj j).
\end{equation}

\begin{lemma}[Estimating equation]\label{lem:esteq}
{\rm (a)} For every $z>0$, $n(z)-z\,d(z)=W(\log z)/(N\sone\stwo)$. Hence, whenever $d(z)>0$ (i.e., some posterior draw lies in the set $A$ defined in {\rm (b)}), $z$ is a fixed point of $T$ if and only if $W(\log z)=0$.
{\rm (b)} $W$ is continuous and non-increasing, with $W'(c)=-\sum_{\text{all draws}}\pi_c(1-\pi_c)$. It is strictly decreasing as soon as one draw lies in $A=\{f>0,\,g>0\}$.
{\rm (c)} Define $\chat:=\sup\{c\in\RR:W(c)>0\}\in[-\infty,\infty]$ and $\Zhat=e^{\chat}$. If both samples meet $A$, then $\chat$ is the unique root of $W$ and $\Zhat$ is the unique positive fixed point of $T$.
{\rm (d)} $W$ is the derivative of the concave log-likelihood $c\mapsto\sum_i\log\pi_c(\Yi i)+\sum_j\log\{1-\pi_c(\Xj j)\}$ and $\chat$ is the maximum-likelihood estimate of the intercept of a logistic regression of the sample label (proposal $=1$) on the offset $\lambda$ \citep{geyer_estimating_1994,kong_theory_2003}.
\end{lemma}

For positive densities, parts (b) and (c) are the argument of the proof of Theorem~2 of MW for their score function. The convention in (c) extends it to samples that do not overlap, in which case $\chat=\pm\infty$; \cref{thm:sharp} shows that this is not a vacuous case. All our results are stated for $\Delta:=\chat-\log\Z=\log(\Zhat/\Z)$, the error on the scale on which evidence is reported.

\subsection{Overlap}

\begin{definition}[Overlap functionals]\label{def:overlap}
Let $\mix=\sone p+\stwo g$ and $\pistar=\stwo g/\mix$. Define
\[
\Hov:=\EE_\mix\big[\pistar(1-\pistar)\big]=\sone\stwo\int\frac{pg}{\sone p+\stwo g},\qquad
\Iov:=\frac{\Hov}{\sone\stwo}\in[0,1],\qquad
\beta:=1-\Iov,
\]
and
\[
\neff:=N\Hov\beta=N\sone\stwo\,\Iov(1-\Iov),\qquad \sigN^2:=\frac{\beta}{N\Hov}=\frac{1}{N\sone\stwo}\Big(\frac1\Iov-1\Big).
\]
\end{definition}
The quantity $\sigN^2$ is the asymptotic variance of $\log\Zhat$ found by MW for independent draws. The next lemma collects interpretations of $\Hov$ and $\beta$.

\begin{lemma}[Properties of the overlap]\label{lem:overlap}
Let $\varepsilon_B:=\int\min(\sone p,\stwo g)$ be the Bayes error for classifying a draw from $\mix$ as coming from $p$ or from $g$, and let $\chi^2(p\|g)=\int p^2/g-1$.
\begin{enumerate}[label={\rm (\alph*)}]
\item $\Hov=\sone\EE_p[\pistar]=\stwo\EE_g[1-\pistar]$.
\item $\tfrac12\varepsilon_B\le\Hov\le\varepsilon_B$.
\item $\Iov\ge\{1+\sone\chi^2(p\|g)\}^{-1}$ and $\Iov\ge\{1+\stwo\chi^2(g\|p)\}^{-1}$.
\item $\beta=\sone\stwo\int(p-g)^2/\mix$. In particular $\beta=0$ if and only if $p=g$ a.e.
\item $\Var(W(\log\Z))=\neff$ under independent sampling, and $\sigN^2=\neff/(N\Hov)^2$.
\end{enumerate}
\end{lemma}

Part (b) is the interpretation we use throughout: \emph{$N\Hov$ is, up to a factor of two, the expected number of draws that the optimal classifier cannot tell apart}. Part (c) recovers, non-asymptotically, the classical fact that the optimal bridge is never worse than the better of the two importance-sampling directions (MW). Part (d) identifies $\beta$ as a weighted triangular (Le~Cam) discrimination. The same quantity appears, as a sample-size-adjusted harmonic divergence, in the variance formulas of \citet{wang_warp_2022} and \citet{ding_channelling_2024}. Part (e) and the formula for $\sigN^2$ are due to MW. Their (3.4) is the functional $\int(\tilde p_1^{-1}+\tilde p_2^{-1})^{-1}=\Hov$, and $\neff$ is the variance of their score function, their (A.10). The effective size $\neff=N\sone\stwo\Iov(1-\Iov)$ is small either when the overlap is poor ($\Iov\to0$) or when the proposal is nearly perfect ($\Iov\to1$). Only the first case is a practical concern: in the second, the estimator is nearly exact (\cref{rem:nearperfect}).

\subsection{The delta-method MCSE}

Conditioning on the bridge $h_{\Zhat}$ and on the proposal, the delta method applied to $\Zhat=\bar{\mathscr N}/\bar{\mathscr D}$, with $\mathscr N_i=f(\Yi i)h_{\Zhat}(\Yi i)$ and $\mathscr D_j=g(\Xj j)h_{\Zhat}(\Xj j)$, gives the relative variance
\begin{equation}\label{eq:VarMC-def}
\sighat^2:=\frac{\widehat\Var(\bar{\mathscr N})}{\bar{\mathscr N}^2}+\frac{\widehat\Var(\bar{\mathscr D})}{\bar{\mathscr D}^2}.
\end{equation}
This is the standard error of $\log\Zhat$, equivalently the relative standard error of $\Zhat$ \citep{fruhwirth_estimating_2004,gronau_tutorial_2017,micaletto_diagnostics_2025}. For independent draws, $\widehat\Var(\bar{\mathscr N})$ and $\widehat\Var(\bar{\mathscr D})$ are sample variances divided by the sample sizes; for Markov chain output one uses spectral or batch-means variance estimates \citep{flegal_markov_2010,vats_strongly_2019}. Write $\widehat\pi_k=\pi_{\chat}(\cdot)$ for the fitted responsibility of the $k$-th draw, and let $\widehat V_{\mathrm{prop}}$ and $\widehat V_{\mathrm{post}}$ be the unbiased sample variances of $\{\widehat\pi_i\}$ over the proposal draws and of $\{\widehat\pi_j\}$ over the posterior draws.

\begin{lemma}[Structure of the MCSE]\label{lem:mcse-structure}
For independent draws,
\[
\sighat^2=\frac{\neffhat}{(N\Hzero)^2},\qquad
\neffhat:=\Stwo\widehat V_{\mathrm{prop}}+\Sone\widehat V_{\mathrm{post}},\qquad
N\Hzero:=\sum_{j=1}^{\Sone}\widehat\pi_j=\sum_{i=1}^{\Stwo}(1-\widehat\pi_i).
\]
Moreover $\Hzero$ and $\Hhat:=N^{-1}\sum_{\text{all draws}}\widehat\pi_k(1-\widehat\pi_k)$ both estimate $\Hov$.
\end{lemma}

The lemma shows that the MCSE is the plug-in version of the identity $\sigN^2=\neff/(N\Hov)^2$ from \cref{lem:overlap}(e). Everything the MCSE ``knows'' is contained in two scalars, the overlap and the variance of the estimating function. MW observed that the numerator of \cref{eq:MWupdate} at convergence, which equals $\Hzero/(\sone\stwo)$, yields the free estimate $\widehat\sigma^2_{\mathrm{MW}}=N^{-1}\{1/\Hzero-1/(\sone\stwo)\}$. With sample variances normalized by $1/S$, one checks that $N\sighat^2=(1+\Tp)/\Hzero-1/(\sone\stwo)$, where $\Tp=1-\Hhat/\Hzero$ is the contraction factor of \cref{thm:iteration}(c). Hence $\Tp=N\Hzero(\sighat^2-\widehat\sigma^2_{\mathrm{MW}})$: the delta-method MCSE and Meng and Wong's free estimate differ by $\Tp/(N\Hzero)$, which is negligible under correct sampling and is the basis of the calibration test of \cref{sec:calibration}. Under Markov chain sampling, $\Sone\widehat V_{\mathrm{post}}$ is replaced by $\Sone\widehat V_{\mathrm{post}}\widehat\tau$, where $\widehat\tau$ estimates the integrated autocorrelation time of $\{\pi_{\chat}(\Xj j)\}_j$.

\section{Non-asymptotic theory}\label{sec:nonasymp}

\begin{assumption}[Independent sampling]\label{asmp:iid}
$\Xj1,\dots,\Xj{\Sone}$ are i.i.d.\ from $p$, $\Yi1,\dots,\Yi\Stwo$ are i.i.d.\ from $g$, the two samples are independent, and $\Hov>0$.
\end{assumption}

No other condition is imposed. In particular, the importance ratios $\ell=f/g$ and $g/f$ may have no finite moments of any order, the supports of $p$ and $g$ need not be nested (only $\Hov>0$ is needed), and the dimension of $\Theta$ plays no role except through $\Hov$. Markov chain draws are treated in \cref{sec:mcmc}. Throughout, $\cstar=\log\Z$, $\Delta=\chat-\cstar$ and
\[
\kappa(t):=\frac{(1-e^{-t})^2}{2\cosh^2(t/2)+\tfrac23(1-e^{-t})},\qquad t\ge0.
\]
The function $\kappa(t)/t^2$ decreases from $1/2$ at $t=0$ to $0.259$ at $t=1/2$, and $\kappa(1)=0.135$.

\subsection{An exact reduction}

\begin{lemma}[Mean and variance of the shifted estimating function]\label{lem:meanvar}
Let $W_t:=W(\cstar+t)$, $\mu(t):=\EE W_t$ and $v(t):=\Var(W_t)$. Under \cref{asmp:iid}:
\begin{enumerate}[label={\rm (\alph*)}]
\item $\mu(t)=-N(e^t-1)\,\EE_\mix\!\big[\pistar(1-\pistar)/\{1+(e^t-1)\pistar\}\big]$. Hence $\operatorname{sign}\mu(t)=-\operatorname{sign}t$ and $N\Hov(1-e^{-|t|})\le|\mu(t)|\le N\Hov(e^{|t|}-1)$.
\item $|\mu(t)+N\Hov t|\le N\Hov(e^{|t|}-1-|t|)$.
\item $v(0)=\neff$, $v(t)\le N\Hov\cosh^2(t/2)$, and $e^{-2|t|}v(0)\le v(t)\le e^{2|t|}v(0)$.
\end{enumerate}
\end{lemma}

\begin{theorem}[Reduction and concentration]\label{thm:concentration}
(a) For every $t\in\RR$ and every joint law of the draws, $\{\Delta\le t\}=\{W(\cstar+t)\le0\}$, equivalently $\{\Delta>t\}=\{W(\cstar+t)>0\}$. Moreover $\{\Delta\ge t\}\subseteq\{W(\cstar+t)\ge0\}$, with equality whenever some draw lies in $A=\{f>0,g>0\}$.
(b) Under \cref{asmp:iid}, for all $t>0$,
\[
\PP(\Delta\ge t)\le e^{-N\Hov\kappa(t)},\qquad \PP(\Delta\le -t)\le e^{-N\Hov\kappa(t)}.
\]
\end{theorem}

Part (a) turns a question about the solution of a nonlinear equation into a question about a sum of independent bounded variables, exactly and at every sample size. Part (b) is the resulting Bernstein bound. Its exponent $N\Hov\kappa(t)\approx N\Hov t^2/2$ matches the Gaussian tail with variance $1/(N\Hov)\ge\sigN^2$ to leading order.

\begin{corollary}[Sample size sufficient for a given accuracy]\label{cor:samplesize}
Under \cref{asmp:iid}, if $N\Hov\ge16\log(2/\delta)$, then with probability at least $1-\delta$,
\[
|\log\Zhat-\log\Z|\le2\sqrt{\log(2/\delta)/(N\Hov)}.
\]
Consequently, relative accuracy $t\in(0,1/2]$ with confidence $1-\delta$ is guaranteed once $N\ge 8\log(2/\delta)/(t^2\varepsilon_B)$, where $\varepsilon_B$ is the Bayes error of \cref{lem:overlap}.
\end{corollary}

\subsection{A Berry--Esseen bound with an explicit constant}

\begin{theorem}[Berry--Esseen]\label{thm:BE}
Under \cref{asmp:iid},
\[
\sup_{x\in\RR}\Big|\PP\big(\log\Zhat-\log\Z\le x\,\sigN\big)-\Phi(x)\Big|\le\frac{3.1}{\sqrt{\neff}}.
\]
\end{theorem}

The bound is universal: it holds for every target, proposal and dimension, including cases where the ratios $\ell$ have infinite variance. Letting $\neff\to\infty$ gives asymptotic normality with the MW variance under no condition beyond $\neff\to\infty$. By \cref{thm:sharp}(b) the rate $\neff^{-1/2}$ cannot be improved.

\begin{remark}[Near-perfect proposals]\label{rem:nearperfect}
If $g\approx p$, then $\beta\to0$ and $\neff\to0$ even though the estimator is nearly exact ($\sigN\to0$). The theorem then says little, because it measures the error in units of the vanishing $\sigN$. The bound uses $|\pi-\EE\pi|\le1$ to control third moments, which is loose when the responsibilities barely fluctuate. The regime of practical concern is poor overlap, where $\beta\approx1$ and $\neff\approx N\Hov$.
\end{remark}

\subsection{When can the MCSE be trusted?}

\begin{theorem}[Reliability of the delta-method MCSE]\label{thm:mcse}
Assume \cref{asmp:iid} with $\Sone,\Stwo\ge2$, and $\delta\in(0,1)$ with $\neff\ge64\log(12/\delta)$, then with probability at least $1-\delta$,
\[
\Big|\log\frac{\sighat}{\sigN}\Big|\le10\sqrt{\frac{\log(12/\delta)}{\neff}}.
\]
Let $\mathcal C_\alpha=[\log\Zhat-z_{1-\alpha/2}\sighat,\ \log\Zhat+z_{1-\alpha/2}\sighat]$. If $\neff\ge350$, then
\[
\big|\PP(\log\Z\in\mathcal C_\alpha)-(1-\alpha)\big|\le\frac{7.2}{\sqrt{\neff}}+2z_{1-\alpha/2}(e^{r}-1)\,\phi\big(z_{1-\alpha/2}e^{-r}\big),\qquad r=10\sqrt{\frac{\log(12\sqrt{\neff})}{\neff}}.
\]
In particular, the coverage error is $O(\sqrt{\log\neff/\neff})$ uniformly over all $(p,g)$.
\end{theorem}

\begin{proposition}[A free certificate]\label{prop:certificate}
Under the assumptions of \cref{thm:mcse}, with probability at least $1-\delta$,
\[
\big|\log(\neffhat/\neff)\big|\le12.1\sqrt{\log(12/\delta)/\neff}\quad\text{and}\quad\big|\log(\Hzero/\Hov)\big|\le 3.5\sqrt{\log(12/\delta)/\neff}.
\]
\end{proposition}

\Cref{thm:mcse,prop:certificate} answer the question posed in the introduction. \emph{The delta-method MCSE is trustworthy when the effective overlap $\neff$ is large, and $\neff$ is estimated at no cost by $\neffhat=(N\Hzero\sighat)^2$.} The explicit constants are conservative. The experiments in \cref{sec:experiments} show that coverage of nominal $95\%$ intervals is within $0.01$ of nominal for $\neff\ge3.7$ and degrades appreciably only below $\neff\approx1$. This suggests reporting $\neffhat$ next to every bridge estimate and flagging $\neffhat<10$. \Cref{prop:certificate} guarantees that $\neffhat$ is accurate when $\neff$ is large. The next result covers the other direction: $\neffhat$ cannot be large when the overlap is poor.

\begin{proposition}[A large certificate certifies overlap]\label{prop:onesided}
Let $M_{\mathrm{post}}:=\sum_{j=1}^{\Sone}\pistar(\Xj j)$, $M_{\mathrm{prop}}:=\sum_{i=1}^{\Stwo}\{1-\pistar(\Yi i)\}$ and $\varsigma:=\Sone/(\Sone-1)+\Stwo/(\Stwo-1)$.
\begin{enumerate}[label={\rm (\alph*)}]
\item For any draws, independent or not, $\neffhat\le\varsigma\,N\Hzero\le\varsigma\max\{M_{\mathrm{post}},M_{\mathrm{prop}}\}$.
\item Under \cref{asmp:iid}, $M_{\mathrm{post}}$ and $M_{\mathrm{prop}}$ have mean $N\Hov$, and for every $x\ge N\Hov$,
\[
\PP(\neffhat\ge\varsigma x)\le2e^{-N\Hov}\Big(\frac{eN\Hov}{x}\Big)^{x}.
\]
Moreover, for every $\delta\in(0,1)$, with probability at least $1-\delta$,
\[
\neffhat\le\varsigma\Big(\sqrt{N\Hov}+\sqrt{\tfrac23\log(2/\delta)}\Big)^2.
\]
\item If $\Iov\le1/2$, then $N\Hov\le2\neff$. If $N\Hov\le1$ and $\Sone,\Stwo\ge100$, then $\PP(\neffhat\ge10)\le0.04$.
\end{enumerate}
\end{proposition}

Part (a) needs no probabilistic assumption, and it is nearly sharp: in the simulations of \cref{app:repro} the ratio $\neffhat/(\varsigma\max\{M_{\mathrm{post}},M_{\mathrm{prop}}\})$ reaches $0.97$. Together, \cref{prop:certificate,prop:onesided} make $\neffhat$ a two-sided certificate in the regime that matters, and when the overlap is poor ($\Iov\le1/2$), $N\Hov$ and $\neff$ agree within a factor of two, so a large $\neffhat$ certifies a large $\neff$ and \cref{thm:BE,thm:mcse} apply. When $\Iov>1/2$, $N\Hov>N\sone\stwo/2$, and \cref{thm:concentration} makes $\log\Zhat$ accurate to order $(N\sone\stwo)^{-1/2}$ whatever the value of $\neff$; this is the near-perfect regime of \cref{rem:nearperfect}. The constant in (c) is conservative. In six configurations with $\neff$ between $0.36$ and $3.2$, two of them with importance ratios of infinite variance, $\neffhat$ never reached $10$ in $10^4$ replications each, and its $99\%$ quantile stayed below $5.6$ (\cref{app:repro}).

\subsection{Sharpness}

\begin{theorem}[Lower bounds]\label{thm:sharp}
Let $\varepsilon\in(0,1/2]$ and $S\ge1$. Let $\Theta=\{0,1,2\}$ with $p=(1-\varepsilon)\delta_0+\varepsilon\delta_1$ and $g=\varepsilon\delta_1+(1-\varepsilon)\delta_2$, and take $\Sone=\Stwo=S$. Then $\Hov=\varepsilon/4$, $\beta=1-\varepsilon$, $\neff=S\varepsilon(1-\varepsilon)/2$, and $\Delta=\log(n_{\mathrm{prop}}/n_{\mathrm{post}})$ whenever both counts are positive, where $n_{\mathrm{post}},n_{\mathrm{prop}}\sim\mathrm{Bin}(S,\varepsilon)$ are the independent numbers of draws landing on the overlap point. Moreover:
\begin{enumerate}[label={\rm (\alph*)}]
\item $\PP(\Zhat\in\{0,\infty\})\ge e^{-4N\Hov}$;
\item if $\neff\ge2$, then $\sup_x|\PP(\Delta\le x\sigN)-\Phi(x)|\ge0.04/\sqrt{\neff}$.
\end{enumerate}
\end{theorem}

Together with \cref{thm:concentration}, part (a) shows that $\log\PP(\text{failure})\asymp -N\Hov$ over the class of problems with overlap $\Hov\in(0,1/8]$. In this example the failure probability is $2p_0-p_0^2$ with $p_0=(1-\varepsilon)^S\approx e^{-2N\Hov}$, so the constant $4$ in the exponent is not sharp. No bound that depends only on $N$, or on moments of the importance ratios, can replace the overlap. This is a worst-case statement about the optimal bridge estimator. A minimax lower bound over all estimators that use both samples remains open; \citet{block_partition_2026} treat the setting with proposal samples only. Part (b) shows that the Berry--Esseen rate of \cref{thm:BE} is attained. In this example the Kolmogorov distance is $\approx0.1/\sqrt{\neff}$ (\cref{sec:exp-sharp}).

\begin{remark}[Comparison with importance sampling]\label{rem:ISvsbridge}
\citet{chatterjee_sample_2018} show that importance sampling from $g$ to $p$ needs about $\exp\{\KL(p\|g)\}$ draws (see \citet{jarzynski_rare_2006} for the analogous law for exponential work averages). For the optimal bridge the corresponding requirement is $N\asymp1/\Hov\asymp1/\varepsilon_B$. For two Gaussians with a mean shift $\|\mu\|$, $\KL=\|\mu\|^2/2$ while $\varepsilon_B\approx e^{-\|\mu\|^2/8}$ up to polynomial factors, so the exponent improves by a factor of four. The corresponding asymptotic comparison of relative variances, $e^{\|\mu\|^2}$ for importance sampling against $\|\mu\|e^{\|\mu\|^2/8}$ for the optimal bridge, is due to MW, Equation (6.1), (6.4). Since $\Hov\le\varepsilon_B$ (\cref{lem:overlap}(b)), the requirement grows exponentially in the dimension whenever the Bayes error decays exponentially, which quantifies the known curse of dimensionality of bridge sampling \citep{wang_warp_2022}. For BAR, \citet{reinhardt_small_2021} found empirically that the sample size per state at which BAR's mean squared error falls below that of the linear estimator (with the constant in BAR set to its exact value) follows $n\approx b\,\Omega^{-a}$, with $a=0.77$--$0.83$ and $b=6.0$--$8.5$. Here $\Omega$ is our $\Iov$, and their test systems have $\sone=\stwo=1/2$. Translated into our units, $\neff=(n/2)\Omega(1-\Omega)$ along their fitted crossovers lies between about $0.5$ and $2.5$ for all four systems and all $\Omega$ between $4\times10^{-4}$ and $1/2$, while $n$ itself varies by a factor of $200$--$300$ over that range. By our reading of their Figure~2(b,c), the actual mean squared error of BAR also departs from Bennett's asymptotic formula at $\neff\approx2$, for both $n=20$ and $n=1000$. These observations are consistent with our finding (\cref{sec:exp-universality}) that the transition to asymptotic behaviour is governed by $\neff$ rather than by $n$ or $\Iov$ separately. Constant $\neff$ would appear over their range of $\Omega$ as a power law with exponent about $0.9$. The slightly smaller fitted exponents indicate that their crossover value of $\neff$ decreases slowly as the overlap decreases, which is plausible since their criterion is a comparison with another estimator rather than a reliability criterion.
\end{remark}

\section{Markov chain draws}\label{sec:mcmc}

In practice the posterior draws are sampled via MCMC. \Cref{thm:concentration}(a) holds for every joint law of the draws, so the distribution of $\Delta$ is still that of the sign of $W(\cstar+t)$, which is now the sum of an additive functional of a Markov chain and an independent sum. This section gives the MC equivalents of \cref{thm:concentration,thm:BE,thm:mcse,thm:sharp}, and the proposal draws remain independent, as they are in the known implementations.

\begin{assumption}[Reversible posterior sampler]\label{asmp:mcmc}
The proposal draws are i.i.d.\ from $g$ and independent of the posterior draws. The posterior draws $\Xj1,\dots,\Xj{\Sone}$ are consecutive states of a Markov chain with transition kernel $P$ that is reversible with respect to $p$ and started at stationarity, $\Xj1\sim p$. As an operator on $L^2_0(p)=\{u\in L^2(p):\EE_pu=0\}$, $P$ has spectrum in $[-1+\gamma_-,1-\gamma]$ for some $\gamma\in(0,1]$ and $\gamma_-\in(0,2-\gamma]$. We write $\gs=\min(\gamma,\gamma_-)$ and assume $\Hov>0$.
\end{assumption}

Thus $\gamma$ is a lower bound on the spectral gap of the posterior sampler and $\gs$ on its absolute spectral gap; independent draws have $\gamma=\gamma_-=1$. We call the chain \emph{positive} if one can take $\gamma_-=1$, as for lazy chains; then $\gs=\gamma$. Chains that are not started at stationarity are covered by \cref{thm:mcmc-conc}(c). For bounded $u$, let $\sigma_P^2(u)=\lim_{S\to\infty}S^{-1}\Var\{\sum_{j\le S}u(\Xj j)\}$ be the asymptotic variance and $\tau_P(u)=\sigma^2_P(u)/\Var_p(u)$ the integrated autocorrelation time. By the spectral theorem (\cref{lem:mcmc-basic}),
\begin{equation}\label{eq:sandwich}
\cminus\le\tau_P(u)\le\cplus,\qquad \cminus:=\frac{\gamma_-}{2-\gamma_-},\qquad \cplus:=\frac{2-\gamma}{\gamma},
\end{equation}
and $\cminus\ge1$ for positive chains.

\begin{definition}[Overlap functionals under MCMC]\label{def:mcmc-overlap}
Let
\[
v_P:=\Stwo\Var_g(\pistar)+\Sone\sigma_P^2(\pistar),\qquad \sigNP^2:=\frac{v_P}{(N\Hov)^2},\qquad \tau_W:=\frac{v_P}{\neff},\qquad \neffP:=\frac{\neff}{\tau_W}=\frac{\neff^2}{v_P}.
\]
\end{definition}

Here $v_P$ is the long-run variance of $W(\cstar)$, and it replaces $\neff=\Var W(\cstar)$ of \cref{lem:overlap}(e), and $\sigNP^2$ is the asymptotic variance of $\log\Zhat$ \citep{micaletto_diagnostics_2025}. The ratio $\tau_W\in[\min(1,\cminus),\cplus]$ is the integrated autocorrelation time of the estimating function: a weighted average of $1$ for the proposal draws and of $\tau_P(\pistar)$ for the posterior draws. Hence $\sigNP^2=\tau_W\sigN^2$, and an MCSE that ignores autocorrelation is too small by the factor $\sqrt{\tau_W}$. We call $\neffP$ the \emph{effective overlap under MCMC}: the number of ambiguous draws divided by the autocorrelation time. Let $\widehat\omega^2$ be an estimate of $\sigma^2_P(\pistar)$ from the posterior draws, such as $\widehat V_{\mathrm{post}}\widehat\tau$ in \cref{sec:setting}, and let $\sighatP^2:=(\Stwo\widehat V_{\mathrm{prop}}+\Sone\widehat\omega^2)/(N\Hzero)^2$ be the MCMC version of the MCSE. Then $\widehat n_P:=(N\Hzero\sighatP)^2=\Stwo\widehat V_{\mathrm{prop}}+\Sone\widehat\omega^2$ estimates $v_P$, and \(\neffPhat:=\neffhat^2/\widehat n_P\) estimates $\neffP$ at no cost.

\subsection{Concentration and the failure probability}

\begin{theorem}[Concentration under MCMC]\label{thm:mcmc-conc}
Assume \cref{asmp:mcmc}. Let $b_P:=5/\gamma$ if $\gamma<1$ and $b_P:=1/3$ if $\gamma=1$, and
\[
\kappa_P(t):=\frac{(1-e^{-t})^2}{2\cplus\cosh^2(t/2)+2b_P(1-e^{-t})},\qquad t\ge0.
\]
\begin{enumerate}[label={\rm (\alph*)}]
\item For all $t>0$, $\PP(\Delta\ge t)\le e^{-N\Hov\kappa_P(t)}$ and $\PP(\Delta\le-t)\le e^{-N\Hov\kappa_P(t)}$.
\item $\PP(\Zhat\in\{0,\infty\})\le e^{-N\Hov}+e^{-\gamma N\Hov(1-1/\Sone)}$.
\item Suppose instead that the chain is started from a law $q$ with $\chi^2(q\|p)<\infty$, and that its first $t_0$ states are discarded. Then every event $E$ determined by the draws satisfies $\PP_q(E)\le\PP_p(E)+\tfrac12(1-\gs)^{t_0}\chi^2(q\|p)^{1/2}$ and $\PP_q(E)\le\{1+(1-\gs)^{2t_0}\chi^2(q\|p)\}^{1/2}\PP_p(E)^{1/2}$, where $\PP_p$ refers to the stationary chain.
\end{enumerate}
\end{theorem}

For $\gamma=1$, and in particular for independent draws, $\kappa_P=\kappa$ and (a) is \cref{thm:concentration}(b). For $\gamma<1$ and $t\le1/2$ we have $\kappa_P(t)\ge0.075\,\gamma t^2$, so $|\log\Zhat-\log\Z|\le t$ with probability at least $1-\delta$ once $\gamma N\Hov\ge15\log(2/\delta)/t^2$. The sample size required in bridge sampling from MCMC output is therefore inflated by a factor of order $1/\gamma$, the relaxation time of the posterior sampler. By (b) and \cref{thm:mcmc-sharp}(a) below, the worst-case failure probability over reversible chains with spectral gap $\gamma$ is $\exp\{-\Theta(\gamma N\Hov)\}$, so this inflation cannot be avoided. Part (c), a direct consequence of \citet[Propositions~3.15 and 3.16]{paulin_concentration_2015}, extends all results of this section to warm starts with burn-in.

\subsection{A Berry--Esseen bound}\label{sec:mcmc-be}

\begin{theorem}[Berry--Esseen under MCMC]\label{thm:mcmc-BE}
Assume \cref{asmp:mcmc} with $\Sone\ge2$, and let $\zeta_P:=(\cplus/\cminus)^{1/2}$. Then
\[
\sup_{x\in\RR}\big|\PP(\Delta\le x\,\sigNP)-\Phi(x)\big|\le\frac{C_P}{\sqrt{v_P}}+0.63(1+\zeta_P)\,\sigNP+e^{-N\Hov\kappa_P(1/(4+4\zeta_P))}+\frac{0.12}{\gs}\,e^{-\gs\Sone/2},
\]
where $C_P$ depends only on $(\gamma,\gs,\gamma_-)$ and is given in \cref{eq:CP}. If the chain is positive, then $C_P\le83/\gamma$.
\end{theorem}

The main term has the form of \cref{thm:BE}, with $\neff=\Var W(\cstar)$ replaced by the long-run variance $v_P$. The second term is of the order of the MCSE itself, and the third is exponentially small in $N\Hov/\{\cplus(1+\zeta_P)^2\}$. Since $\sigNP\le\cplus/\sqrt{v_P}$ and $N\Hov\ge v_P/\cplus$ (\cref{lem:mcmc-basic}(c)), and $\sup_{y>0}\sqrt ye^{-y}=(2e)^{-1/2}$, the right-hand side is at most $C'_P/\sqrt{v_P}+(0.12/\gs)e^{-\gs\Sone/2}$, with
\[
C'_P:=C_P+0.63(1+\zeta_P)\cplus+\big\{\cplus/(2e\,\kappa_P(\tau_P))\big\}^{1/2},\qquad\tau_P:=1/(4+4\zeta_P).
\]
For the chain of \cref{thm:mcmc-sharp} below, $\zeta_P=1$ and $C'_P\le90/\gamma$; for every positive chain $\zeta_P\le\cplus^{1/2}$ and $C'_P\le83/\gamma+11\gamma^{-3/2}$. Whether the order $\gamma^{-3/2}$ can be improved to $\gamma^{-1}$ for all positive chains is open. The factor $\zeta_P$ enters because the long-run variance of $W(\cstar+t)$ may change quickly with $t$ when the spectrum of $P$ is spread out (\cref{lem:mcmc-basic}(b)). The proof rests on a Berry--Esseen bound for the sum of an independent part and a reversible-chain part (\cref{lem:be-chain}). We derive it from the perturbation approach of \citet{mann_berry_1996} and \citet{lezaud_chernoff_2001}, and we control the leading eigenvalue of the perturbed transition operator with the coefficient bounds of \citet{paulin_concentration_2015}. The constant of \cref{lem:be-chain} is linear in $\|u\|_\infty/\sigma_P(u)$, and this is what makes it usable here. For the responsibilities, $\sigma_P^2(\pistar)\le\cplus\Var_p(\pistar)\le\cplus\|\pistar\|_\infty\Hov/\sone$, so the ratio $\|\pistar\|_\infty/\sigma_P(\pistar)$ is at least of order $\Hov^{-1/2}$ when the overlap is poor and $\|\pistar\|_\infty$ is of order one (as it is whenever $g/p$ is unbounded on the support of $p$). A bound that is cubic in this ratio, such as the general bound of \citet[Theorem~C]{kloeckner_effective_2019}, would then lose a factor of order $1/\Hov$. As in \cref{thm:BE}, the constant $83/\gamma$ is conservative. The next result shows that the order $1/\gamma$ of the bound cannot be improved in general.

\begin{theorem}[Lower bounds under MCMC]\label{thm:mcmc-sharp}
In the two-point family of \cref{thm:sharp}, let the posterior draws be the stationary Markov chain on $\{0,1\}$ with $P(0,1)=\varepsilon\gamma$ and $P(1,0)=(1-\varepsilon)\gamma$, where $\gamma\in(0,1]$. This chain is positive and satisfies \cref{asmp:mcmc} with right and absolute spectral gap $\gamma$. Moreover $\tau_W=1/\gamma$, $v_P=\neff/\gamma$ and $\neffP=\gamma\neff$, and:
\begin{enumerate}[label={\rm (\alph*)}]
\item $\PP(\Zhat\in\{0,\infty\})\ge(1-\varepsilon)(1-\varepsilon\gamma)^{S-1}\ge\tfrac12e^{-4\gamma N\Hov}$;
\item if $\varepsilon\le1/4$ and $S=\lceil1/(\varepsilon\gamma)\rceil$, then $\sup_x|\PP(\Delta\le x\sigNP)-\Phi(x)|\ge0.15$, while $v_P\ge3/(8\gamma^2)$. Consequently no bound of the form $C/\sqrt{v_P}$ holds for all reversible chains with spectral gap $\gamma$ unless $C\ge0.09/\gamma$.
\end{enumerate}
\end{theorem}

Together with \cref{thm:mcmc-conc}(b), part (a) shows that the worst-case failure probability is $\exp\{-\Theta(\gamma N\Hov)\}$. Part (b) shows that no bound of the form $C/\sqrt{v_P}$ can have $C$ of smaller order than $1/\gamma$. For this chain $\zeta_P=1$, and \cref{thm:mcmc-BE} gives such a bound with $C=C'_P\le90/\gamma$, so the order $1/\gamma$ is attained. In computations in this family (\cref{sec:exp-mcmc-exact}), $\gamma\sqrt{v_P}$ times the Kolmogorov distance lies between $0.095$ and $0.17$ for all $\gamma$ and all $\neffP\ge1$, whereas $\gamma C_P\le83$. So the order in $\gamma$ is right and the constant is conservative. Moreover, coverage in that family is, to within $0.02$ for $\neffP\ge2$, a function of $\neffP$ alone, over a twenty-fold range of $\gamma$.

\subsection{The MCSE under MCMC}

\begin{theorem}[MCSE under MCMC]\label{thm:mcmc-mcse}
Assume \cref{asmp:mcmc} with $\Sone,\Stwo\ge2$. Let $\widehat\omega^2$ be any estimate of $\sigma^2_P(\pistar)$, and let $\varepsilon_\omega\in[0,1/2]$ and $\delta_\omega\in[0,1]$ be such that $\PP(|\widehat\omega^2/\sigma_P^2(\pistar)-1|>\varepsilon_\omega)\le\delta_\omega$. Let $\delta\in(0,1)$, $L=\log(12/\delta)$, $\ell_P=\cplus L/(N\Hov)$ and $r_P=2\sqrt{L/v_P}+\varepsilon_\omega$. If $\ell_P\le1/64$ and $r_P\le1/2$, then with probability at least $1-\delta-\delta_\omega$,
\[
\Big|\log\frac{\sighatP}{\sigNP}\Big|\le8.4\sqrt{\ell_P}+1.4\,r_P=:r .
\]
Consequently the intervals $\mathcal C^P_\alpha=[\log\Zhat-z_{1-\alpha/2}\sighatP,\ \log\Zhat+z_{1-\alpha/2}\sighatP]$ satisfy
\[
\big|\PP(\log\Z\in\mathcal C^P_\alpha)-(1-\alpha)\big|\le2B_P+\delta+\delta_\omega+2z_{1-\alpha/2}(e^{r}-1)\,\phi\big(z_{1-\alpha/2}e^{-r}\big),
\]
where $B_P$ is the right-hand side of \cref{thm:mcmc-BE}.
\end{theorem}

The theorem separates the two sources of error of an MCSE computed from MCMC output. The first is the overlap, which enters through $N\Hov/\cplus$ and $v_P$ as in the independent case. The second is the accuracy $\varepsilon_\omega$ of the long-run variance estimate, which no property of the bridge estimator can control. Spectral and batch-means estimators are consistent for geometrically ergodic chains \citep{flegal_markov_2010}, but finite-sample guarantees of the required form are not available in this generality. The comparison $\Vjack$ of \cref{sec:jackknife} is an empirical check of the event $|\widehat\omega^2/\sigma^2_P(\pistar)-1|\le\varepsilon_\omega$, through a second, batch-type estimate of the same variance. In the experiments of \cref{sec:exp-mcmc-universality} the second error dominates. With Geyer's estimator, coverage is within $0.025$ of nominal once $\neffP\ge19$, while intervals based on the true $\sigNP$ are accurate throughout.

\section{Orthogonality, the fixed-point iteration, and proposal fitting}\label{sec:orth}

This section explains why three natural suspects for unreliable bridge-sampling MCSEs, the estimated bridge, the fixed-point iteration and the fitted proposal, are second order.

\begin{proposition}[Exact orthogonality]\label{prop:orth}
Let $h_{\zeta,\vartheta}=\{\sone f+\stwo\zeta g_\vartheta\}^{-1}$ for $\zeta>0$ and a proposal $g_\vartheta$ from any family.
\begin{enumerate}[label={\rm (\alph*)}]
\item For all $(\zeta,\vartheta)$ with $\Hov(\vartheta)>0$: $\EE_{g_\vartheta}[f h_{\zeta,\vartheta}]-\Z\,\EE_p[g_\vartheta h_{\zeta,\vartheta}]=0$. Hence every derivative of this population moment with respect to the nuisance quantities $(\zeta,\vartheta)$ vanishes identically.
\item The population fixed-point map is constant: $T_{\mathrm{pop}}(\zeta):=\EE_g[f h_\zeta]/\EE_p[g h_\zeta]=\Z$ for all $\zeta>0$. The population MW iteration therefore terminates after one step from any starting value.
\item The moment map $\Psi(z)=\EE_g[fh_z]-z\EE_p[gh_z]$, in which $z$ appears both as the target and inside the bridge, factorizes as $\Psi(z)=(1-z/\Z)K(z)$ with $K(z)=\int fg/(\sone f+\stwo zg)>0$. It has the unique root $\Z$, and $\Psi'(\Z)=-K(\Z)/\Z=-\EE_p[g h_\Z]$.
\end{enumerate}
\end{proposition}

Part (a) is Neyman orthogonality \citep{chernozhukov_double_2018} in its strongest, global form: the moment condition holds at every value of the nuisance parameters, not only to first order at the truth. Part (b) is the population counterpart of the observation of Section 4 of MW that every iterate of \cref{eq:MWupdate} is consistent, whatever the starting value. The rest of this section draws out its consequences.

\subsection{The Meng--Wong iteration}

Write $\varphi(x):=\log T(e^x)$ for the sample map \cref{eq:MWupdate} on the log scale, and let $L_k\in\{0,1\}$ be the label of draw $k$ (with $1$ for proposal draws).

\begin{theorem}[Global contraction and the contraction factor]\label{thm:iteration}
Suppose both samples meet $A=\{f>0,g>0\}$.
\begin{enumerate}[label={\rm (\alph*)}]
\item $\varphi'(x)=\mathsf A(x)-\mathsf B(x)$, where
\[
\mathsf A(x)=\frac{\sum_j\pi_x(\Xj j)^2}{\sum_j\pi_x(\Xj j)}\qquad\text{and}\qquad \mathsf B(x)=\frac{\sum_i\pi_x(\Yi i)\{1-\pi_x(\Yi i)\}}{\sum_i\{1-\pi_x(\Yi i)\}}
\]
are weighted averages of responsibilities in $(0,1)$. The weights are those of the bridge-weighted posterior and proposal samples, whose population versions coincide.
\item $L:=\sup_x|\varphi'(x)|<1$. The iteration $x_{t+1}=\varphi(x_t)$ converges to $\chat$ from every starting value, with $|x_t-\chat|\le L^t|x_0-\chat|$ and $|x_t-\chat|\le|x_{t+1}-x_t|/(1-L)$.
\item At the solution,
\[
T'(\Zhat)=\varphi'(\chat)=1-\frac{\Hhat}{\Hzero}=-\frac{\sum_{k}(L_k-\widehat\pi_k)\,\widehat\pi_k}{\sum_{j}\widehat\pi_j}.
\]
\item Under \cref{asmp:iid}, if $N\Hov\ge64\log(6/\delta)$, then with probability at least $1-\delta$,
\[
|\log\{1-T'(\Zhat)\}|\le5.9\sqrt{\log(6/\delta)/(N\Hov)}.
\]
\end{enumerate}
\end{theorem}

Theorem 2 of MW proved that the iteration converges to the unique root from any starting value, with $|r^{(t+1)}-\widehat r|<|r^{(t)}-\widehat r|$. Their proof shows that $rM(r)$ is increasing and $M(r)/r$ is decreasing, which is the pointwise bound $|\varphi'|<1$ of (b). They report fast convergence empirically. \Cref{thm:iteration} adds the explicit form (a) of $\varphi'$, the uniform bound $L<1$, which gives a geometric rate and an a posteriori error bound, the identity (c) for the contraction factor at the solution, and its size (d). An ``iteration conditioning'' diagnostic would flag a large contraction factor $C=|T'(\Zhat)|$ on the grounds that perturbations of the map are amplified by $1/(1-C)$. By (c)--(d), $C$ is the discrepancy between two consistent estimators of the same overlap and is $O_\PP(\neff^{-1/2})$. The MW fixed point is asymptotically \emph{perfectly} conditioned, and iteration error is controlled by the stopping tolerance alone. In our experiments the iteration reaches a tolerance of $10^{-13}$ on the log scale in $7$--$13$ steps, from starting values that differ from the solution by a factor $e^{\pm200}$ (\cref{sec:exp-identities}). The contraction factor is still useful, but as a calibration statistic (\cref{sec:calibration}) rather than a conditioning number.

\subsection{Proposal fitting}

Let $\{g_\vartheta:\vartheta\in\mathsf\Theta\subseteq\RR^k\}$ be a proposal family generated by a reparametrization $\tilde\theta=\tau_\vartheta(\epsilon)$, with $\epsilon\sim\nu$ drawn independently of everything else. Let $\chat(\vartheta)$ denote the estimator computed with proposal $g_\vartheta$ and proposal draws $\tau_\vartheta(\epsilon_i)$, using common base variates $\epsilon_1,\dots,\epsilon_{\Stwo}$.

\begin{theorem}[First-order invariance to the proposal fit]\label{thm:proposal}
Assume that the posterior draws are i.i.d.\ from $p$, that $\sone$ is fixed as $N\to\infty$, that $\widehat\vartheta\xrightarrow{\,\Pr\,}\vartheta_0$ ($\widehat\vartheta$ may be computed from the same posterior draws), and:
\begin{enumerate}[label={\rm(P\arabic*)}]
    \item On a compact neighbourhood of $\vartheta_0$ the maps $\vartheta\mapsto\log g_\vartheta(\theta)$ and $\vartheta\mapsto\log\{g_\vartheta(\tau_\vartheta(\epsilon))/f(\tau_\vartheta(\epsilon))\}$ are finite and Lipschitz, with Lipschitz constants square-integrable under $p$ and $\nu$ respectively (this requires $g_\vartheta>0$ $p$-a.s.\ and $f>0$ on the support of $g_\vartheta$);
    \item $\Hov(\vartheta_0)>0$ and $\vartheta\mapsto\Hov(\vartheta)$ is continuous at $\vartheta_0$.
\end{enumerate}
Then $\sqrt N\{\chat(\widehat\vartheta)-\chat(\vartheta_0)\}\xrightarrow{\,\Pr\,}0$. If moreover $\beta(\vartheta_0)>0$ (i.e., $g_{\vartheta_0}\neq p$), then $(\chat(\widehat\vartheta)-\log\Z)/\sigN(\vartheta_0)\Rightarrow\mathcal N(0,1)$ and the delta-method MCSE computed with $g_{\widehat\vartheta}$ is consistent for $\sigN(\vartheta_0)$, whether or not the sample is split. If instead $g_{\vartheta_0}=p$, then $\chat(\vartheta_0)=\log\Z$ and the error of $\chat(\widehat\vartheta)$ is $o_\PP(N^{-1/2})$.
\end{theorem}

\begin{remark}[What remains: a second-order bias; and a pitfall]\label{rem:secondorder}
(i) When $\widehat\vartheta$ is computed from the same posterior draws, the $o_\PP(N^{-1/2})$ remainder in \cref{thm:proposal} has non-zero mean. Expanding one order further, this mean is $(N\Hov)^{-1}\EE[\partial_\vartheta\{W_N-\EE W_N\}(\cstar,\vartheta_0)\cdot(\widehat\vartheta-\vartheta_0)]-\Hov^{-1}\nabla_\vartheta\Hov(\vartheta_0)\cdot\EE[(\chat-\cstar)(\widehat\vartheta-\vartheta_0)]$, up to smaller terms; the first term is typically the dominant one. For fixed $k$ this bias is $O(N^{-1})$, so its ratio to the MCSE is $O(N^{-1/2})$, with a constant that grows with $k$. This is the ``correlation-induced bias'' found by simulation in \citet{wong_properties_2020}, including its growth with dimension: a Gaussian proposal with full covariance has $k=d+d(d+1)/2$ parameters. With splitting or cross-fitting, $\widehat\vartheta$ is independent of the draws used in $W_N$, and both terms have mean zero at this order. Splitting is therefore advisable when the proposal has many parameters relative to the effective overlap; otherwise it only wastes draws. \Cref{sec:exp-proposal} quantifies this: with $2d$ fitted parameters and $\Sone=\Stwo=2000$, the same-draw bias grows from $0.3$ to $1.1$ MCSEs as $d$ increases from $2$ to $32$, while splitting removes it.
(ii) A natural proposal-fitting diagnostic is $R_g^2=(\partial_\vartheta\Zhat)\Sigma_\vartheta(\partial_\vartheta\Zhat)^\top/\sighat^2$, where $\Sigma_\vartheta$ is the covariance of $\widehat\vartheta$ and $\partial_\vartheta\Zhat$ is computed by differentiating $T$ while \emph{holding the proposal draws fixed}. That derivative omits the score term $\EE_{g_\vartheta}[(1-\pi)\partial_\vartheta\log g_\vartheta]$ generated by the dependence of the proposal draws on $\vartheta$. Its expectation is therefore of order one instead of zero, and $R_g$ converges to a positive constant unrelated to the variance of $\Zhat$. The pathwise derivative, computed with common random numbers, is $O_\PP((N\Hov)^{-1/2})$, and the implied variance contribution is of smaller order than $\sigN^2$ (\cref{sec:exp-proposal}). The regime is different if the proposal draws are held fixed and $\vartheta$ is re-estimated from those same draws. First-order effects are then real, and they can even reduce the variance, as for importance sampling with an estimated sampler \citep{henmi_importance_2007} and for NCE \citep{uehara_analysis_2018}. This may be a route to improving bridge sampling, but it does not measure the variance of the standard estimator.
(iii) For the same reason, a jackknife that refits $\vartheta$ within each deletion must regenerate the proposal draws from the same base variates. Refitting while holding the proposal draws fixed inflates the jackknife variance spuriously (\cref{sec:exp-proposal}).
\end{remark}

\begin{corollary}[The conditional MCSE misses nothing at first order]\label{cor:nothingmissed}
Under the conditions of \cref{thm:BE} or \cref{thm:proposal}, the asymptotic variance of $\log\Zhat$ is $\sigN^2$: the conditional delta-method variance \cref{eq:VarMC-def} evaluated at the final bridge and proposal. The dependence of the bridge on $\Zhat$, the fixed-point iteration and the proposal fit add nothing at order $N^{-1}$. In other words, when the MCSE is estimated consistently, $(G-1)\Vjack\Rightarrow\chi^2_{G-1}$ for fixed $G$ (\cref{thm:jackknife}) and $\Vjack\to1$ when $G\to\infty$ (\cref{rem:thresholds}(iv)).
\end{corollary}

\section{Diagnostics}\label{sec:diagnostics}

The previous sections show that, with independent draws, the delta-method MCSE fails only when the effective overlap is small, and that this is visible in $\neffhat$. With MCMC draws the same holds for $\neffP$, provided the long-run variance of the posterior stream is estimated accurately (\cref{thm:mcmc-mcse}). Three failure modes remain that no variance formula can reveal. The overlap may be too small to trust the normal approximation. The posterior sampler may not sample from $p$. And, for Markov chains, the long-run variance may be misestimated. We give one diagnostic for each. All three are computed from quantities that the estimator already produces.

\subsection{Overlap certificate}\label{sec:certificate}

Report $\neffhat=(N\Hzero\sighat)^2=\Stwo\widehat V_{\mathrm{prop}}+\Sone\widehat V_{\mathrm{post}}$ alongside $\log\Zhat$ and $\sighat$; for MCMC output, $\neffhat$ is computed from the sample variances and $\sighat$ uses the long-run variance, as in \cref{alg:diag}. By \cref{thm:BE,thm:mcse,prop:certificate}, $\neffhat$ controls the accuracy of the normal approximation and of the MCSE itself, uniformly over targets and proposals. The experiments suggest flagging $\neffhat<10$, which is conservative, and by \cref{prop:onesided} a run with poor overlap passes this check only with small probability. Coverage of nominal $95\%$ intervals is within $0.01$ of nominal in all our continuous examples with $\neff\ge3.7$, and it deteriorates only below $\neff\approx1$ (\cref{sec:exp-universality}). A small $\neffhat$ should be read as ``the two samples barely overlap''. The remedy is a better proposal \citep{meng_warp_2002,wang_warp_2022,micaletto_diagnostics_2025}, or more draws in proportion to $1/\Hov$. In free-energy practice the same information is summarized by the MBAR overlap matrix \citep{klimovich_guidelines_2015}, whose two-state overlap scalar equals $\Hhat/(\sone\stwo)$ (\cref{sec:related}), and a common rule requires this overlap to exceed $0.03$. Our results show that a threshold on the overlap alone cannot be right, because what matters is $\neff=N\sone\stwo\Iov(1-\Iov)$. This agrees with the empirical recommendation of \citet{reinhardt_small_2021} to judge BAR by the overlap together with the number of independent samples. With $\sone=\stwo=1/2$, an overlap of $0.03$ gives $\neff\approx730$ with $N=10^5$ draws, which is ample, but only $\neff\approx3.6$ with $N=500$. For MCMC output, report in addition $\neffPhat=\neffhat^2/\widehat n_P$ with $\widehat n_P=(N\Hzero\sighatP)^2$, the effective overlap divided by the estimated autocorrelation time of the estimating function. \Cref{thm:mcmc-BE,thm:mcmc-mcse} and the experiments of \cref{sec:exp-mcmc-exact,sec:exp-mcmc-universality} suggest flagging $\neffPhat<30$. The threshold is higher than for independent draws because the long-run variance estimate is itself unreliable when $\neffP$ is small, and it then biases $\neffPhat$ upwards.

\subsection{A calibration test for sampler failure}\label{sec:calibration}

By \cref{thm:iteration}(c), the contraction factor of the MW map at the solution is
\[
\Tp:=T'(\Zhat)=1-\frac{\Hhat}{\Hzero}=-\frac{\sum_k(L_k-\widehat\pi_k)\widehat\pi_k}{\sum_j\widehat\pi_j}.
\]
The estimating equation forces the implicit classifier to be calibrated in the large, $\sum_k(L_k-\widehat\pi_k)=0$. The statistic $\Tp$ checks the next moment: whether the residuals are uncorrelated with the fitted responsibilities, as in a calibration-slope test for logistic regression. We studentize $\Tp$ with the same delete-a-group jackknife used below, $t_{\Tp}:=\Tp/\widehat{\mathrm{se}}_{\mathrm{jack}}(\Tp)$, and compare it with $t_{G-1}$ quantiles. The test is free and needs no extra model evaluations. It complements multi-run checks such as bidirectional Monte Carlo \citep{grosse_sandwiching_2015}, which sandwiches $\log\Z$ between stochastic bounds but requires posterior samples and additional annealing runs. Let $\mix_\star(\dd\theta):=\pistar(\theta)\{1-\pistar(\theta)\}\,\mix(\dd\theta)/\Hov$ denote the pooled law weighted by the variance of the label, which we call the overlap-weighted law.

\begin{theorem}[Null calibration of the calibration test]\label{thm:calib-null}
Assume \cref{asmp:iid} with $p$, $g$, $\sone$ and $\stwo$ fixed as $N\to\infty$, and $\Var_{\mix_\star}(\pistar)>0$.
\begin{enumerate}[label={\rm(\alph*)}]
\item $\sqrt{N\Hov}\,\Tp\Rightarrow\mathcal N\big(0,\Var_{\mix_\star}(\pistar)\big)$.
\item With the blocks of \cref{sec:jackknife}, $G$ fixed, and $\widehat{\mathrm{se}}_{\mathrm{jack}}^2(\Tp):=\frac{G-1}{G}\sum_{g=1}^G(\Tp_{(-g)}-\bar\Tp_{(-)})^2$, where $\Tp_{(-g)}$ is computed without block $g$ and $\bar\Tp_{(-)}$ is their mean, $t_\Tp\Rightarrow t_{G-1}$.
\end{enumerate}
Part {\rm (b)} extends to Markov chain output under the analogue of \cref{asmp:fixedG} for the functions $\pistar$ and ${\pistar}^2$.
\end{theorem}

Part (a) makes \cref{thm:iteration}(d) precise and identifies the constant measured in \cref{tab:E1}: for F2 with $\delta=2$, $\sqrt{\neff}\,\mathrm{sd}(\Tp)\to\sqrt\beta\,\mathrm{sd}_{\mix_\star}(\pistar)=0.202$, against $0.197$--$0.206$ observed. Since $\pistar\in[0,1]$, $\mathrm{sd}(\Tp)$ is asymptotically at most $(4N\Hov)^{-1/2}$. The condition $\Var_{\mix_\star}(\pistar)>0$ excludes only problems in which $p/g$ is constant on the overlap; then $\Tp\equiv0$, as in the two-point family of \cref{thm:sharp}. If the posterior draws come from some $\tilde p\neq p$, for instance due to the sampler under-representing a mode, the estimating equation still has a root, but the implicit classifier is miscalibrated, and the next result describes what $\Tp$ measures.

\begin{proposition}[Power and blind spot of the calibration test]\label{prop:calib-alt}
Suppose that the posterior draws are i.i.d.\ from $\tilde p\ll p$ instead of $p$, that the proposal draws are i.i.d.\ from $g$ and independent of them, and that $\tilde p(A)>0$. Let $\tilde c$ be the root of $\widetilde F(c):=\stwo\EE_g[1-\pi_c]-\sone\EE_{\tilde p}[\pi_c]$, $\tilde\pi:=\pi_{\tilde c}$ and $\widetilde H_0:=\sone\EE_{\tilde p}[\tilde\pi]$.
\begin{enumerate}[label={\rm (\alph*)}]
\item As $N\to\infty$ with $\sone$ and $\stwo$ fixed, $\chat\to\tilde c$ and $\Tp\to\Tp_\infty$ almost surely, where
\[
\Tp_\infty:=\frac{\sone\EE_{\tilde p}[\tilde\pi^2]-\stwo\EE_g[\tilde\pi(1-\tilde\pi)]}{\widetilde H_0}=-\frac{\Cov\big(L-\tilde\pi(\theta),\,\tilde\pi(\theta)\big)}{\widetilde H_0},
\]
the covariance being taken under the pooled law of the label $L$ and the draw $\theta$. If $\Tp_\infty\ne0$, then $\PP(|t_\Tp|>t_{G-1,1-\alpha/2})\to1$.
\item If $\tilde p=p(1+\epsilon h)$ with $h$ bounded and $\EE_ph=0$, then, as $\epsilon\to0$,
\[
\tilde c-\cstar=-\epsilon\,\EE_{\mix_\star}[h]+O(\epsilon^2),\qquad \Tp_\infty=\epsilon\,\Cov_{\mix_\star}(\pistar,h)+O(\epsilon^2).
\]
\item If $\EE_p[\tilde p/p\mid f/g]=\lambda$ on $A$ for a constant $\lambda$ (in particular, if $\tilde p=\lambda p$ on $A$), then $\lambda=\tilde p(A)/p(A)$, and for every $N$ the responsibilities of all draws, as functions of $c$, have the same joint law as in a correctly specified problem whose unnormalized posterior equals $f$ on $A$ and whose evidence is $\Z/\lambda$. Consequently $\Zhat\to\Z\,p(A)/\tilde p(A)$, and every statistic computed from the responsibilities, including $\neffhat$, $\sighat$, $t_\Tp$ and the jackknife diagnostics, has the same law as in that correctly specified problem.
\end{enumerate}
\end{proposition}

\begin{remark}[What the test can and cannot see]\label{rem:calib-power}
Part (b) separates the two effects of a sampler error $h$ on the overlap region. The bias of $\log\Zhat$ is governed by the mean $\EE_{\mix_\star}[h]$, the test by the covariance $\Cov_{\mix_\star}(\pistar,h)$. Under local alternatives $\tilde p=p(1+h/\sqrt N)$, the expansions in the proofs show that $t_\Tp$ converges to a noncentral $t_{G-1}$ law with noncentrality $\sqrt{\Hov}\,\Cov_{\mix_\star}(\pistar,h)/\mathrm{sd}_{\mix_\star}(\pistar)$, while the bias is $-\EE_{\mix_\star}[h]\sqrt{\Hov/\beta}$ MCSEs. The test therefore detects samplers whose error varies with $\pistar$ on the overlap region, such as samplers that misweight modes that the proposal covers unequally (\cref{sec:exp-failure}). It is blind to errors that are nearly constant there, and part (c) shows that this is not a weakness of the test: the responsibilities depend on the draws only through $f/g$, and an error whose average over the level sets of $f/g$ is constant on $A$ leaves no trace in them, although it biases $\log\Zhat$ by $\log\{p(A)/\tilde p(A)\}$. The typical case is a mode that the sampler misses and the proposal does not cover. In \cref{app:repro}, such a mode carrying $30\%$ of the posterior mass produces a bias of $21$ MCSEs and zero interval coverage, while $t_\Tp$ rejects at the nominal rate. Errors confined to $A^c$ that leave $\tilde p(A)=p(A)$ do not affect $\Zhat$ at all.
\end{remark}

In \cref{sec:exp-failure} the test detects biases that are invisible to every variance-based diagnostic. The biases in question are several MCSEs large and drive interval coverage to zero.

\subsection{Delete-a-group jackknife diagnostics for Markov chain output}\label{sec:jackknife}

For MCMC output, $\sighat$ requires a long-run variance estimate for the posterior stream, and misestimating it is the most common failure of MCSEs in practice. Partition both samples into $G$ contiguous blocks $\PostBlock g$ and $\PropBlock g$ of sizes $\Sone/G$ and $\Stwo/G$, aligned with the sampling order. Deleting block $g$ from both samples leaves $\sone$ and $\stwo$ unchanged. Let $\chat_{(-g)}$ be the estimate without block $g$, let $\bar c_{(-)}=G^{-1}\sum_g\chat_{(-g)}$, and let $R_g=G\chat-(G-1)\chat_{(-g)}$ be the pseudo-values. Define
\begin{gather*}
\JackVar:=\frac{G-1}{G}\sum_{g=1}^G(\chat_{(-g)}-\bar c_{(-)})^2,\qquad
\Vjack:=\frac{\JackVar}{\sighat^2},\\
\Imax:=\max_g\frac{|\chat_{(-g)}-\chat|}{\sighat},\qquad
\Gr:=\max_g\frac{|R_g-\bar R|}{s_R},
\end{gather*}
where $s_R$ is the sample standard deviation of the pseudo-values. We work on the log scale; versions on the $\Zhat$ scale agree to first order. Instead of an asymptotic linearization we use the following identity.

\begin{lemma}[Block influence]\label{lem:block}
Let $W_g(c)=\sum_{i\in\PropBlock g}\{1-\pi_c(\Yi i)\}-\sum_{j\in\PostBlock g}\pi_c(\Xj j)$ be the contribution of block $g$ to $W$, and let $\widehat{\mathcal I}_{(-g)}(c)=\sum_{k\notin\text{block }g}\pi_c(1-\pi_c)$ be the observed information without block $g$. If the reduced samples meet $A$, then for some $\tilde c_g$ between $\chat$ and $\chat_{(-g)}$,
\[
\chat_{(-g)}-\chat=-\frac{W_g(\chat)}{\widehat{\mathcal I}_{(-g)}(\tilde c_g)},\qquad\text{and}\qquad \sum_{g=1}^GW_g(\chat)=0.
\]
\end{lemma}

The influence of a block is therefore its own contribution to the estimating equation, evaluated at the full-sample solution, divided by the leave-block-out information. The block contributions sum to zero.

\begin{assumption}[Fixed-$G$ regime]\label{asmp:fixedG}
The proposal draws are i.i.d.\ from $g$ and independent of the stationary posterior sequence. The partial sums of $\{\pistar(\Xj j)-\EE_p\pistar\}$ satisfy a functional CLT with long-run variance $\omega^2$, and the ergodic theorem holds for $\pi_c(1-\pi_c)$ uniformly in $c$ near $\cstar$. $G$ is fixed and divides $\Sone$ and $\Stwo$. Finally, the MCSE is consistent: $\sighat^2/\sigma_{\mathrm{as}}^2\xrightarrow{\,\Pr\,}1$, where $\sigma_{\mathrm{as}}^2=(\Stwo\Var_g\pistar+\Sone\omega^2)/(N\Hov)^2$.
\end{assumption}

The functional CLT holds, for example, for geometrically ergodic chains, because $\pistar$ is bounded \citep{jones_markov_2004}. Consistency of spectral variance estimators is discussed by \citet{flegal_markov_2010}.

\begin{theorem}[Fixed-$G$ calibration]\label{thm:jackknife}
Under \cref{asmp:fixedG}, let $\zeta_1,\dots,\zeta_G$ be i.i.d.\ $\mathcal N(0,1)$ with mean $\bar\zeta$ and standard deviation $s_\zeta$. As $N\to\infty$:
\begin{enumerate}[label={\rm (\alph*)}]
\item $(G-1)\Vjack\Rightarrow\chi^2_{G-1}$;
\item $\Imax\Rightarrow\frac{\sqrt G}{G-1}\max_g|\zeta_g-\bar\zeta|$, where each coordinate $\frac{\sqrt G}{G-1}(\zeta_g-\bar\zeta)$ is $\mathcal N(0,1/(G-1))$;
\item $\Gr\Rightarrow\max_g|\zeta_g-\bar\zeta|/s_\zeta$, the Grubbs statistic \citep{grubbs_sample_1950}, whose law does not depend on any unknown quantity.
\end{enumerate}
\end{theorem}

\begin{remark}[Calibration pitfalls]\label{rem:thresholds}
(i) The threshold $\chi^2_{G-1,1-\alpha}/(G-1)$ for $\Vjack$ is correct in the fixed-$G$ regime. If instead $G\to\infty$, then $\Vjack\to1$ and the threshold degenerates to $1$. (ii) The standardized block effects have variance $1/(G-1)$, and the Bonferroni threshold for $\Imax$ is $\Phi^{-1}(1-\alpha/(2G))/\sqrt{G-1}$. Since the limit in (b) is pivotal, exact quantiles are cheap to simulate, and we use them. (iii) Because both samples are split into $G$ blocks, the block sizes satisfy $m_1/\Sone=m_2/\Stwo=1/G$, so weights that apportion the variance between the posterior and proposal contributions cancel out of every threshold. (iv) If $G\to\infty$ with $\Sone/G\to\infty$, then $\JackVar/\sigma_{\mathrm{as}}^2\to1$ \citep{kunsch_jackknife_1989}.
\end{remark}

$\Vjack$ compares a batch-type variance estimate with the spectral one used in $\sighat$, so it flags underestimated autocorrelation. $\Imax$ and $\Gr$ flag a single dominating block, such as a transient excursion or a brief visit to a rare mode. No within-run diagnostic can detect metastability on time scales longer than the run, and \cref{sec:exp-failure} shows that all of them lose power as the switching time approaches the block length.

\subsection{What to report}\label{sec:report}

We recommend reporting $\log\Zhat$, $\sighat$ and $\neffhat$ for every bridge estimate, together with the $p$-value of the calibration test $t_\Tp$. For MCMC output, also report $\Vjack$ and $\Imax$ (or $\Gr$) with their fixed-$G$ thresholds; we use $G=20$ \citep{schmeiser_batch_1982}. \Cref{alg:diag} summarizes the computation. Its cost is dominated by $G$ root solves, each of which takes a few Newton steps from $\chat$, and it needs no additional evaluations of $f$ or $g$.

\begin{algorithm}
\caption{Bridge sampling with diagnostics (log scale)}\label{alg:diag}
\small
\textbf{Input:} log ratios $a_j=\log\{f(\Xj j)/g(\Xj j)\}$ and $a_i=\log\{f(\Yi i)/g(\Yi i)\}$; number of blocks $G$; level $\alpha$.
\begin{enumerate}[leftmargin=2em,itemsep=1pt,topsep=3pt]
\item Set $\sone=\Sone/N$, $\stwo=\Stwo/N$ and $\pi_c(a):=\{1+\exp(\log(\sone/\stwo)+a-c)\}^{-1}$.
\item Solve $W(c)=\sum_i\{1-\pi_c(a_i)\}-\sum_j\pi_c(a_j)=0$ by safeguarded Newton steps, using $W'=-\sum\pi_c(1-\pi_c)$; call the root $\chat$.
\item Compute $\widehat\pi_k=\pi_{\chat}(a_k)$, $N\Hzero=\sum_j\widehat\pi_j$, $N\Hhat=\sum_k\widehat\pi_k(1-\widehat\pi_k)$ and $\Tp=1-\Hhat/\Hzero$.
\item Compute $\neffhat=\Stwo\widehat V_{\mathrm{prop}}+\Sone\widehat V_{\mathrm{post}}$, $\widehat n_P=\Stwo\widehat V_{\mathrm{prop}}+\Sone\widehat V_{\mathrm{post}}\widehat\tau$ (with $\widehat\tau=1$ for independent draws), the MCSE of $\log\Zhat$, $\sighat=\sqrt{\widehat n_P}/(N\Hzero)$, and $\neffPhat=\neffhat^2/\widehat n_P$.
\item For $g=1,\dots,G$: compute $\chat_{(-g)}$, the root of $W-W_g$, by Newton steps started at $\chat$, and $\Tp_{(-g)}$, the statistic $\Tp$ without block $g$.
\item Compute $\Vjack$, $\Imax$ and $\Gr$ as in \cref{sec:jackknife}, and $t_\Tp=\Tp/\widehat{\mathrm{se}}_{\mathrm{jack}}(\Tp)$.
\item Raise a flag if $\neffhat<10$, or $\neffPhat<30$ for MCMC output (overlap); if $|t_\Tp|>t_{G-1,1-\alpha/2}$ (sampler failure); if $\Vjack>\chi^2_{G-1,1-\alpha}/(G-1)$ (MCSE too small); or if $\Imax>q_{G,1-\alpha}$ or $\Gr>\mathrm{Grubbs}_{G,1-\alpha}$ (influential block).
\end{enumerate}
\textbf{Output:} $\chat$, $\sighat$, $\neffhat$, $\neffPhat$, $\Tp$, $t_\Tp$, $\Vjack$, $\Imax$, $\Gr$ and the flags.
\end{algorithm}

\section{Numerical experiments}\label{sec:experiments}

\paragraph{Implementation.} All computations work on the log scale. The root of $W$ is found by safeguarded Newton iterations in $c=\log z$, and the jackknife roots are obtained by Newton iterations started at $\chat$. All targets are normalized densities multiplied by a known constant, so the true $\log\Z$ is known. Unless stated otherwise, $\Sone=\Stwo$, $G=20$ and $\alpha=0.05$. Null quantiles of $\Imax$ and $\Gr$ are simulated once from the pivotal limits of \cref{thm:jackknife}, using $4\times10^5$ draws. With $R=5000$ replications, the Monte Carlo standard error of a rejection rate near $0.05$ is $0.003$. 

\subsection{Identities of Sections~\ref{sec:setting} and~\ref{sec:orth}}\label{sec:exp-identities}

For three problems we compute the MW iteration from starting values $\chat\pm20$ and $\chat\pm200$, the supremum of $|\varphi'|$ over a grid of $4001$ points in $[\chat-60,\chat+60]$, and the contraction factor at the solution by three routes: formula (a) of \cref{thm:iteration}, a finite difference, and $1-\Hhat/\Hzero$. The problems are F1, $p=\mathcal N(0,I_{10})$ with $g=\mathcal N(0,1.5^2I_{10})$; F2, $p=\mathcal N(0,1)$ with $g=\mathcal N(4,1)$; and F3, $p=t_3(0,I_{20})$ with $g=\mathcal N(0,3I_{20})$. \Cref{tab:E1} confirms the identity and the global contraction. The map is nearly flat everywhere, not only near the solution, and $\sqrt{\neff}\,\mathrm{sd}(\Tp)$ is constant across sample sizes, as \cref{thm:iteration}(d) predicts; its value agrees with the limit $\sqrt\beta\,\mathrm{sd}_{\mix_\star}(\pistar)=0.202$ given by \cref{thm:calib-null}(a).

\begin{table}
\centering\small
\caption{Identities of \cref{lem:esteq,thm:iteration} ($\Sone=\Stwo=2000$). Iterations of the MW map needed to reach $|x_{t+1}-x_t|\le10^{-13}$ from $\chat\pm20$ and $\chat\pm200$ (worst case over the two signs), the resulting error, the supremum of the contraction factor over $[\chat-60,\chat+60]$, and $T'(\Zhat)$ computed from \cref{thm:iteration}(a), by finite differences, and as $1-\Hhat/\Hzero$. Bottom: $T'(\Zhat)$ over $1000$ replications for F2 with $\delta=2$; $\sqrt{\neff}\,\mathrm{sd}(T')$ is constant.}\label{tab:E1}
\setlength{\tabcolsep}{4pt}
\resizebox{\textwidth}{!}{%
\begin{tabular}{lccccccc}
\toprule
Problem & iter.\ ($\pm20$) & iter.\ ($\pm200$) & final error & $\sup_x|\varphi'(x)|$ & $T'$ (formula) & $T'$ (fin.\ diff.) & $1-\Hhat/\Hzero$\\
\midrule
F1: $\mathcal N(0,I_{10})$ vs $\mathcal N(0,1.5^2I_{10})$ & 8 & 8 & 9e-16 & 0.083 & 0.007805 & 0.007805 & 0.007805\\
F2: $\mathcal N(0,1)$ vs $\mathcal N(4,1)$ & 13 & 13 & 9e-14 & 0.213 & -0.060488 & -0.060488 & -0.060488\\
F3: $t_3(0,I_{20})$ vs $\mathcal N(0,3I_{20})$ & 7 & 7 & 4e-16 & 0.037 & 0.002837 & 0.002837 & 0.002837\\
\bottomrule
\end{tabular}}

\vspace{6pt}
\begin{tabular}{rrrrr}
\toprule
$\Sone=\Stwo$ & $\neff$ & mean $T'$ & sd $T'$ & $\sqrt{\neff}\,$sd $T'$\\
\midrule
250 & 30.9 & -0.00211 & 0.03614 & 0.201\\
1000 & 123.7 & -0.00088 & 0.01855 & 0.206\\
4000 & 494.9 & -0.00047 & 0.00886 & 0.197\\
16000 & 1979.7 & +0.00002 & 0.00448 & 0.199\\
\bottomrule
\end{tabular}
\end{table}

\subsection{Finite-sample behaviour in the two-point family}\label{sec:exp-sharp}

For the family of \cref{thm:sharp}, the law of $\Delta$ is that of $\log(n_{\mathrm{prop}}/n_{\mathrm{post}})$ for independent binomial counts, so every quantity can be computed exactly by enumeration. \Cref{tab:E7} reports the Kolmogorov distance to $\mathcal N(0,\sigN^2)$, the probability of a non-finite estimate, tail probabilities, and the coverage of the nominal $95\%$ delta-method interval. The Kolmogorov distance equals half the atom at zero and behaves as $0.1/\sqrt{\neff}$. It lies between the lower bound $0.04/\sqrt{\neff}$ and the upper bound $3.1/\sqrt{\neff}$, so the Berry--Esseen rate is attained. Coverage is close to nominal once $\neff\ge2$ and collapses below $\neff\approx1$, where non-overlapping samples become likely.

\begin{table}
\centering\small
\caption{Computations in the two-point family of \cref{thm:sharp} ($\Sone=\Stwo=S$). Kolmogorov distance between the law of $\Delta/\sigN$ and $\mathcal N(0,1)$, with the lower bound of \cref{thm:sharp}(b) (valid for $\neff\ge2$) and the upper bound of \cref{thm:BE}; $\log_{10}$ of the probability that $\Zhat\in\{0,\infty\}$ and of the lower bound $e^{-4N\Hov}$ of \cref{thm:sharp}(a); coverage of the nominal $95\%$ delta-method interval; and $\PP(|\Delta|\ge0.25)$ with the bound $2e^{-N\Hov\kappa(0.25)}$ of \cref{thm:concentration}.}\label{tab:E7}
\setlength{\tabcolsep}{3.5pt}
\resizebox{\textwidth}{!}{%
\begin{tabular}{rrrcccccccc}
\toprule
$S$ & $\varepsilon$ & $\neff$ & Kolmogorov & $0.04/\sqrt{\neff}$ & $3.1/\sqrt{\neff}$ & $\log_{10}\PP(\Zhat\in\{0,\infty\})$ & $\log_{10}e^{-4N\Hov}$ & coverage & $\PP(|\Delta|\ge0.25)$ & bound\\
\midrule
50 & 0.02 & 0.49 & 0.3403 & -- & 1.000 & -0.2 & -0.9 & 0.4043 & 0.8234 & 1.0000\\
100 & 0.02 & 0.98 & 0.1316 & -- & 1.000 & -0.6 & -1.7 & 0.7518 & 0.8020 & 1.0000\\
200 & 0.02 & 1.96 & 0.0723 & -- & 1.000 & -1.5 & -3.5 & 0.9548 & 0.7432 & 1.0000\\
400 & 0.02 & 3.92 & 0.0507 & 0.0202 & 1.000 & -3.2 & -6.9 & 0.9647 & 0.6399 & 1.0000\\
1{,}000 & 0.01 & 4.95 & 0.0451 & 0.0180 & 1.000 & -4.1 & -8.7 & 0.9595 & 0.5941 & 1.0000\\
2{,}000 & 0.01 & 9.90 & 0.0318 & 0.0127 & 0.985 & -8.4 & -17.4 & 0.9547 & 0.4371 & 1.0000\\
4{,}000 & 0.01 & 19.80 & 0.0224 & 0.0090 & 0.697 & -17.2 & -34.7 & 0.9529 & 0.2668 & 1.0000\\
10{,}000 & 0.005 & 24.88 & 0.0200 & 0.0080 & 0.622 & -21.5 & -43.4 & 0.9515 & 0.2145 & 1.0000\\
20{,}000 & 0.005 & 49.75 & 0.0141 & 0.0057 & 0.440 & -43.2 & -86.9 & 0.9510 & 0.0788 & 0.6507\\
50{,}000 & 0.004 & 99.60 & 0.0100 & 0.0040 & 0.311 & -86.7 & -173.7 & 0.9505 & 0.0128 & 0.2117\\
100{,}000 & 0.004 & 199.20 & 0.0071 & 0.0028 & 0.220 & -173.8 & -347.4 & 0.9501 & 0.0004 & 0.0224\\
1{,}000{,}000 & 0.001 & 499.50 & 0.0045 & 0.0018 & 0.139 & -434.2 & -868.6 & 0.9501 & 0.0000 & 0.0000\\
1{,}000{,}000 & 0.004 & 1992.00 & 0.0022 & 0.0009 & 0.069 & -1740.4 & -3474.4 & 0.9500 & 0.0000 & 0.0000\\
\bottomrule
\end{tabular}}
\end{table}

\subsection{Universality of the effective overlap}\label{sec:exp-universality}

\begin{figure}
\centering
\includegraphics[width=\textwidth]{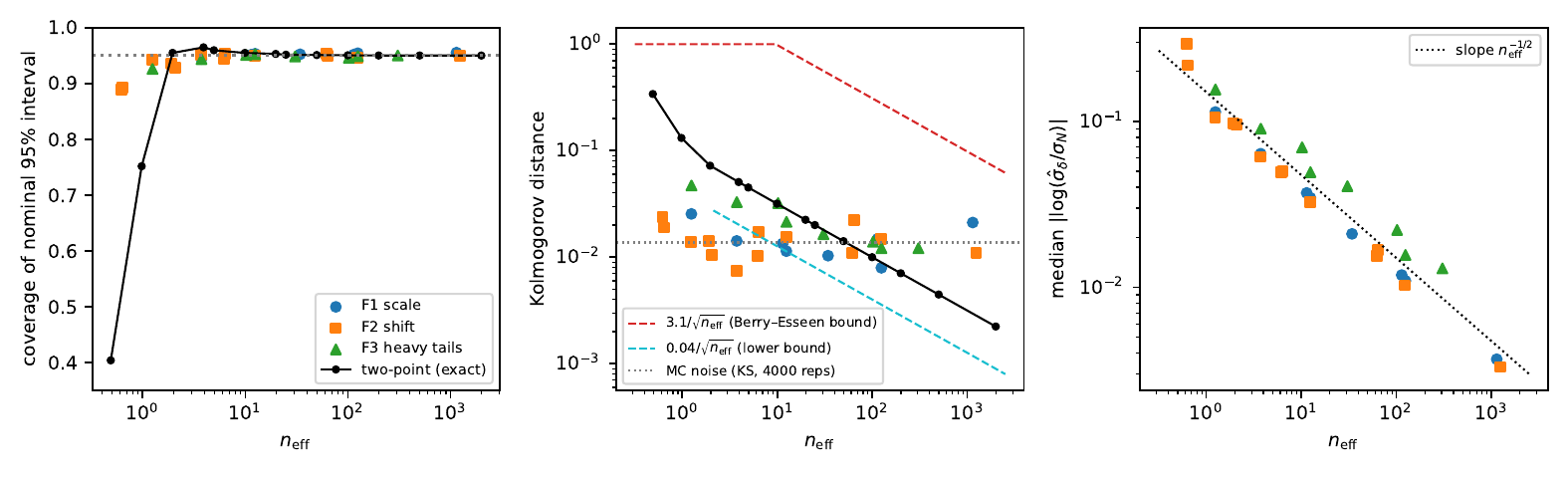}
\caption{Universality of $\neff$. Left: coverage of nominal $95\%$ delta-method intervals for $\log\Z$ against $\neff$, for three families and the two-point family of \cref{thm:sharp}. Middle: Kolmogorov distance between the law of $\Delta/\sigN$ and $\mathcal N(0,1)$; the two-point values are exact, the continuous-family values are Kolmogorov--Smirnov statistics from $4000$ replications (the dotted line marks their Monte Carlo noise level), and the dashed lines are the bounds $3.1/\sqrt{\neff}$ (\cref{thm:BE}) and $0.04/\sqrt{\neff}$ (\cref{thm:sharp}). Right: median of $|\log(\sighat/\sigN)|$, with reference slope $\neff^{-1/2}$.}\label{fig:universality}
\end{figure}

We consider three independent-sampling families with true overlap $\Hov$ computed by one-dimensional quadrature and validated against Monte Carlo with $2\times10^6$ draws. F1 is a scale mismatch, $p=\mathcal N(0,I_d)$ with $g=\mathcal N(0,\sigma^2I_d)$, for $(d,\sigma)\in\{(10,1.5),(40,1.3)\}$. F2 is a mean shift, $p=\mathcal N(0,1)$ with $g=\mathcal N(\delta,1)$, for $\delta\in\{2,4,6\}$. F3 is a heavy-tailed target, $p=t_3(0,I_d)$ with the moment-matched $g=\mathcal N(0,3I_d)$, for $d\in\{20,100\}$. Sample sizes range from $\Sone=\Stwo=10$ to $30{,}000$, giving $\neff$ from $0.6$ to $1.2\times10^3$, and each configuration uses $4000$ replications. \Cref{fig:universality} and \cref{tab:E6} show that coverage, the ratio of the true to the predicted standard deviation, and the accuracy of both $\sighat$ and $\neffhat$ are functions of $\neff$ alone. The curves collapse across families that differ in dimension, in tail behaviour (for F3 the ratios $g/f$ have infinite variance under $g$) and in the type of mismatch. Coverage is within $0.01$ of nominal for all configurations with $\neff\ge3.7$. It is $0.93$--$0.94$ for $\neff\approx1$--$2$ and falls to about $0.89$ only at $\neff\approx0.6$. The median relative error of the MCSE decays like $\neff^{-1/2}$, as \cref{thm:mcse} predicts, with a family-dependent constant between about $0.1$ and $0.25$ instead of the conservative $10$. The same transition at $\neff$ of order one is visible in the independent BAR experiments of \citet{reinhardt_small_2021}, as discussed in \cref{rem:ISvsbridge}.

\subsection{Null calibration of the jackknife diagnostics}\label{sec:exp-null}

\Cref{tab:E2} reports two settings. In the first, $p=\mathcal N(0,I_5)$ with the fixed proposal $g=\mathcal N(0.2\cdot\mathbf 1,1.3^2I_5)$, for $\Sone\in\{1000,4000,16000\}$ and $G\in\{10,20\}$. In the second, the Gaussian proposal is fitted on an independent posterior sample of the same size, as in the \texttt{bridgesampling} package; this is done for heavy-tailed, multimodal and high-dimensional targets with $G=20$. Every diagnostic holds its nominal level. In the first setting, the corrected Bonferroni threshold for $\Imax$ and the pivotal threshold give $4.3$--$5.1\%$ rejections, while the naive threshold with $\sqrt G$ gives $5.4$--$7.2\%$. The only visible departures, rejection rates of about $7\%$ for $\Imax$ and $\Gr$ (and up to $9\%$ with the naive threshold), occur in the two high-dimensional settings with poor overlap, where $\neffhat/G$ is of order one, so that each block contains only one or two ``ambiguous'' draws. The small $\neffhat$ in the $16$-dimensional Gaussian case has the opposite cause: the fitted proposal is nearly perfect ($\beta\approx0.01$), and there the diagnostics are calibrated. As predicted by \cref{cor:nothingmissed}, the ratio of the true variance to the average squared MCSE is within Monte Carlo error of one throughout.

\begin{table}
\centering\small
\caption{Null calibration with independent draws ($5000$ replications per row; rejection rates in \% at nominal $5\%$). ``Var/MCSE$^2$'' is the empirical variance of $\log\Zhat$ divided by the mean of $\sighat^2$. $\Imax$ is tested with the corrected Bonferroni threshold $\Phi^{-1}(1-\alpha/2G)/\sqrt{G-1}$, the naive threshold $\Phi^{-1}(1-\alpha/2G)/\sqrt G$, and the pivotal quantile; $\Gr$ with the Grubbs quantile; $t_\Tp$ with $t_{G-1}$.}\label{tab:E2}
\setlength{\tabcolsep}{3pt}
\resizebox{\textwidth}{!}{%
\begin{tabular}{lrrrcccccccccc}
\toprule
Setting & $S$ & $G$ & med $\neffhat$ & Var/MCSE$^2$ & mean $\Vjack$ & $\Vjack$ & $\Imax$ (Bonf.) & $\Imax$ (naive) & $\Imax$ (exact) & $\Gr$ & $t_\Tp$ & coverage\\
\midrule
$\mathcal N(0,I_5)$, fixed $g$ & 1000 & 10 & 70 & 0.984 & 0.988 & 5.0 & 4.8 & 7.0 & 5.0 & 5.2 & 5.5 & 0.954\\
$\mathcal N(0,I_5)$, fixed $g$ & 1000 & 20 & 70 & 1.020 & 0.987 & 4.5 & 4.3 & 5.4 & 4.5 & 4.6 & 4.6 & 0.947\\
$\mathcal N(0,I_5)$, fixed $g$ & 4000 & 10 & 280 & 1.027 & 0.998 & 4.8 & 4.7 & 7.2 & 4.9 & 5.0 & 5.1 & 0.947\\
$\mathcal N(0,I_5)$, fixed $g$ & 4000 & 20 & 279 & 1.005 & 1.001 & 4.9 & 4.9 & 6.3 & 5.1 & 5.0 & 4.5 & 0.949\\
$\mathcal N(0,I_5)$, fixed $g$ & 16000 & 10 & 1118 & 1.009 & 0.998 & 4.9 & 4.5 & 6.8 & 4.7 & 4.8 & 5.1 & 0.950\\
$\mathcal N(0,I_5)$, fixed $g$ & 16000 & 20 & 1118 & 1.010 & 0.999 & 5.1 & 4.6 & 5.8 & 4.8 & 4.5 & 5.4 & 0.949\\
\midrule
$\mathcal N(0,I_{16})$, split & 4000 & 20 & 19 & 1.019 & 1.000 & 5.1 & 5.4 & 6.7 & 5.5 & 5.3 & 5.1 & 0.945\\
$t_3(0,I_{8})$, split & 4000 & 20 & 433 & 1.020 & 1.000 & 5.0 & 4.6 & 5.9 & 4.8 & 4.7 & 4.6 & 0.945\\
$t_3(0,I_{32})$, split & 4000 & 20 & 469 & 0.943 & 1.004 & 4.6 & 4.9 & 6.4 & 5.1 & 5.1 & 5.5 & 0.958\\
$t_3(0,I_{64})$, split & 1000 & 20 & 45 & 1.011 & 1.007 & 5.5 & 7.3 & 9.0 & 7.5 & 6.8 & 4.4 & 0.947\\
mixture ($\pm3.5e_1$), $d=8$, split & 4000 & 20 & 410 & 1.013 & 0.994 & 4.6 & 4.7 & 6.2 & 4.8 & 5.2 & 4.9 & 0.948\\
$\mathcal N(0,I_{64})$, split & 300 & 20 & 26 & 1.000 & 1.003 & 5.6 & 6.9 & 8.5 & 7.0 & 7.2 & 5.3 & 0.948\\
\bottomrule
\end{tabular}}
\end{table}

\subsection{Jackknife diagnostics for Markov chain output}\label{sec:exp-mcmc}

Posterior draws are generated from a stationary Gaussian AR(1) chain, independently in each coordinate, with coefficient $\phi\in\{0.5,0.9\}$ and marginal $\mathcal N(0,I_5)$; $\Sone=\Stwo=8000$ and the proposal is as in \cref{sec:exp-null}. The MCSE $\sighat$ uses Geyer's initial monotone sequence estimator of the integrated autocorrelation time of $\{\widehat\pi_j\}$ \citep{geyer_practical_1992}. The ``naive'' MCSE ignores autocorrelation. \Cref{tab:E3} shows that with the spectral MCSE all diagnostics are calibrated and the intervals cover, while the naive MCSE undercovers badly ($0.67$ at $\phi=0.9$). $\Vjack$ detects this in $99.5\%$ of replications at $\phi=0.9$ and in $22\%$ at $\phi=0.5$, where the naive variance is too small by a factor of about $1.25$.

\begin{table}
\centering\small
\caption{Markov chain output: AR(1) posterior chains, $\Sone=\Stwo=8000$, $G=20$, $5000$ replications. Left: spectral MCSE (Geyer's initial monotone sequence). Right: naive MCSE that ignores autocorrelation. Rejection rates in \% at nominal $5\%$.}\label{tab:E3}
\setlength{\tabcolsep}{4pt}
\resizebox{\textwidth}{!}{%
\begin{tabular}{c|ccccccc|cccc}
\toprule
& \multicolumn{7}{c|}{spectral MCSE} & \multicolumn{4}{c}{naive MCSE}\\
$\phi$ & Var/MCSE$^2$ & coverage & mean $\Vjack$ & $\Vjack$ & $\Imax$ & $\Gr$ & $t_\Tp$ & Var/MCSE$^2$ & coverage & mean $\Vjack$ & $\Vjack$\\
\midrule
0.5 & 0.962 & 0.959 & 0.991 & 4.3 & 4.6 & 5.7 & 5.0 & 1.247 & 0.923 & 1.286 & 22.2\\
0.9 & 0.983 & 0.949 & 0.977 & 3.4 & 4.6 & 5.4 & 5.2 & 4.283 & 0.666 & 4.259 & 99.5\\
\bottomrule
\end{tabular}}
\end{table}

\subsection{Exact finite-sample behaviour with Markov chain draws}\label{sec:exp-mcmc-exact}

For the two-state chain of \cref{thm:mcmc-sharp}, the number of posterior draws at the overlap point is the occupation time of a two-state chain. Its law is computed by dynamic programming, so $\Delta=\log(n_{\mathrm{prop}}/n_{\mathrm{post}})$ can again be analysed. \Cref{tab:E8} reports $\varepsilon=0.05$, $\gamma\in\{1,0.5,0.2,0.05\}$ and sample sizes chosen so that $\neffP=\gamma\neff\in\{1,2,4,8,16,32\}$; see also the top row of \cref{fig:mcmc}. There are four findings. (i) For $\gamma<1$, the probability of a non-finite estimate agrees with the lower bound of \cref{thm:mcmc-sharp}(a) to the displayed precision, and it is a function of $\gamma N\Hov$: at fixed $\neffP$ it does not depend on $\gamma$. (ii) The Kolmogorov distance between the law of $\Delta/\sigNP$ and $\mathcal N(0,1)$ behaves like $0.1/(\gamma\sqrt{v_P})=0.1/\sqrt{\neffP}$. At fixed $v_P$ it therefore grows like $1/\gamma$. This is the order of the bound of \cref{thm:mcmc-BE} for this chain, and \cref{thm:mcmc-sharp}(b) shows that it cannot be improved. (iii) The coverage of the nominal $95\%$ interval based on the correct asymptotic variance $\sigNP^2$ is, to within $0.016$ for $\neffP\ge2$ and within $0.003$ for $\neffP\ge16$, a function of $\neffP$ alone, over a twenty-fold range of $\gamma$. (iv) The interval that ignores autocorrelation covers with probability close to $2\Phi(1.96\sqrt\gamma)-1$, for example $0.34$ at $\gamma=0.05$, whatever the sample size.

\begin{table}
\centering\small
\caption{Computations in the two-point family of \cref{thm:sharp} with $\varepsilon=0.05$, $\Sone=\Stwo=S$ and posterior draws from the stationary two-state chain of \cref{thm:mcmc-sharp} with spectral gap $\gamma$; for $\gamma=1$ the draws are independent. Here $\tau_W=1/\gamma$, so $\neff^P=\gamma\neff$. Columns: $\log_{10}\PP(\Zhat\in\{0,\infty\})$ and its lower bound $\log_{10}\{(1-\varepsilon)(1-\varepsilon\gamma)^{S-1}\}$ from \cref{thm:mcmc-sharp}(a); the Kolmogorov distance between the law of $\Delta/\sigma_{N,P}$ and $\mathcal N(0,1)$, and the same distance multiplied by $\gamma\sqrt{v_P}$, to be compared with $\gamma C'_P\le90$ in the bound of \cref{thm:mcmc-BE} for this chain; exact coverage of nominal $95\%$ intervals $\log\Zhat\pm1.96\,\sigma_{N,P}$ and $\log\Zhat\pm1.96\,\sigN$, the latter ignoring autocorrelation.}\label{tab:E8}
\setlength{\tabcolsep}{4pt}
\resizebox{\textwidth}{!}{%
\begin{tabular}{rrrrrrcccc}
\toprule
$\gamma$ & $S$ & $\neff$ & $\neff^P$ & $\log_{10}\PP(\Zhat\in\{0,\infty\})$ & lower bound & Kolmogorov & $\gamma\sqrt{v_P}\times$Kolm. & coverage ($\sigma_{N,P}$) & coverage ($\sigN$)\\
\midrule
1 & 42 & 1.0 & 1 & -0.7 & -0.9 & 0.1160 & 0.116 & 0.7792 & 0.7792\\
1 & 84 & 2.0 & 2 & -1.6 & -1.9 & 0.0715 & 0.101 & 0.8850 & 0.8850\\
1 & 168 & 4.0 & 4 & -3.4 & -3.7 & 0.0502 & 0.100 & 0.9249 & 0.9249\\
1 & 337 & 8.0 & 8 & -7.2 & -7.5 & 0.0354 & 0.100 & 0.9408 & 0.9408\\
1 & 674 & 16.0 & 16 & -14.7 & -15.0 & 0.0250 & 0.100 & 0.9447 & 0.9447\\
1 & 1{,}347 & 32.0 & 32 & -29.7 & -30.0 & 0.0176 & 0.100 & 0.9473 & 0.9473\\
\midrule
0.5 & 84 & 2.0 & 1 & -0.9 & -0.9 & 0.1262 & 0.126 & 0.8394 & 0.7276\\
0.5 & 168 & 4.0 & 2 & -1.9 & -1.9 & 0.0788 & 0.111 & 0.8999 & 0.7906\\
0.5 & 337 & 8.0 & 4 & -3.7 & -3.7 & 0.0477 & 0.095 & 0.9257 & 0.8153\\
0.5 & 674 & 16.0 & 8 & -7.4 & -7.4 & 0.0336 & 0.095 & 0.9379 & 0.8241\\
0.5 & 1{,}347 & 32.0 & 16 & -14.8 & -14.8 & 0.0237 & 0.095 & 0.9444 & 0.8290\\
0.5 & 2{,}695 & 64.0 & 32 & -29.6 & -29.6 & 0.0168 & 0.095 & 0.9471 & 0.8314\\
\midrule
0.2 & 211 & 5.0 & 1 & -0.9 & -0.9 & 0.1555 & 0.156 & 0.8229 & 0.5578\\
0.2 & 421 & 10.0 & 2 & -1.9 & -1.9 & 0.0921 & 0.130 & 0.8918 & 0.5925\\
0.2 & 842 & 20.0 & 4 & -3.7 & -3.7 & 0.0587 & 0.117 & 0.9220 & 0.6053\\
0.2 & 1{,}684 & 40.0 & 8 & -7.4 & -7.4 & 0.0390 & 0.110 & 0.9359 & 0.6120\\
0.2 & 3{,}368 & 80.0 & 16 & -14.7 & -14.7 & 0.0260 & 0.104 & 0.9430 & 0.6160\\
0.2 & 6{,}737 & 160.0 & 32 & -29.4 & -29.4 & 0.0178 & 0.100 & 0.9465 & 0.6176\\
\midrule
0.05 & 842 & 20.0 & 1 & -0.9 & -0.9 & 0.1698 & 0.170 & 0.8094 & 0.3083\\
0.05 & 1{,}684 & 40.0 & 2 & -1.9 & -1.9 & 0.1009 & 0.143 & 0.8841 & 0.3244\\
0.05 & 3{,}368 & 80.0 & 4 & -3.7 & -3.7 & 0.0650 & 0.130 & 0.9175 & 0.3319\\
0.05 & 6{,}737 & 160.0 & 8 & -7.3 & -7.3 & 0.0438 & 0.124 & 0.9339 & 0.3354\\
0.05 & 13{,}474 & 320.0 & 16 & -14.7 & -14.7 & 0.0297 & 0.119 & 0.9420 & 0.3371\\
0.05 & 26{,}947 & 640.0 & 32 & -29.3 & -29.3 & 0.0205 & 0.116 & 0.9460 & 0.3379\\
\bottomrule
\end{tabular}}
\end{table}

\subsection{Universality under MCMC}\label{sec:exp-mcmc-universality}

\begin{figure}
\centering
\includegraphics[width=0.92\textwidth]{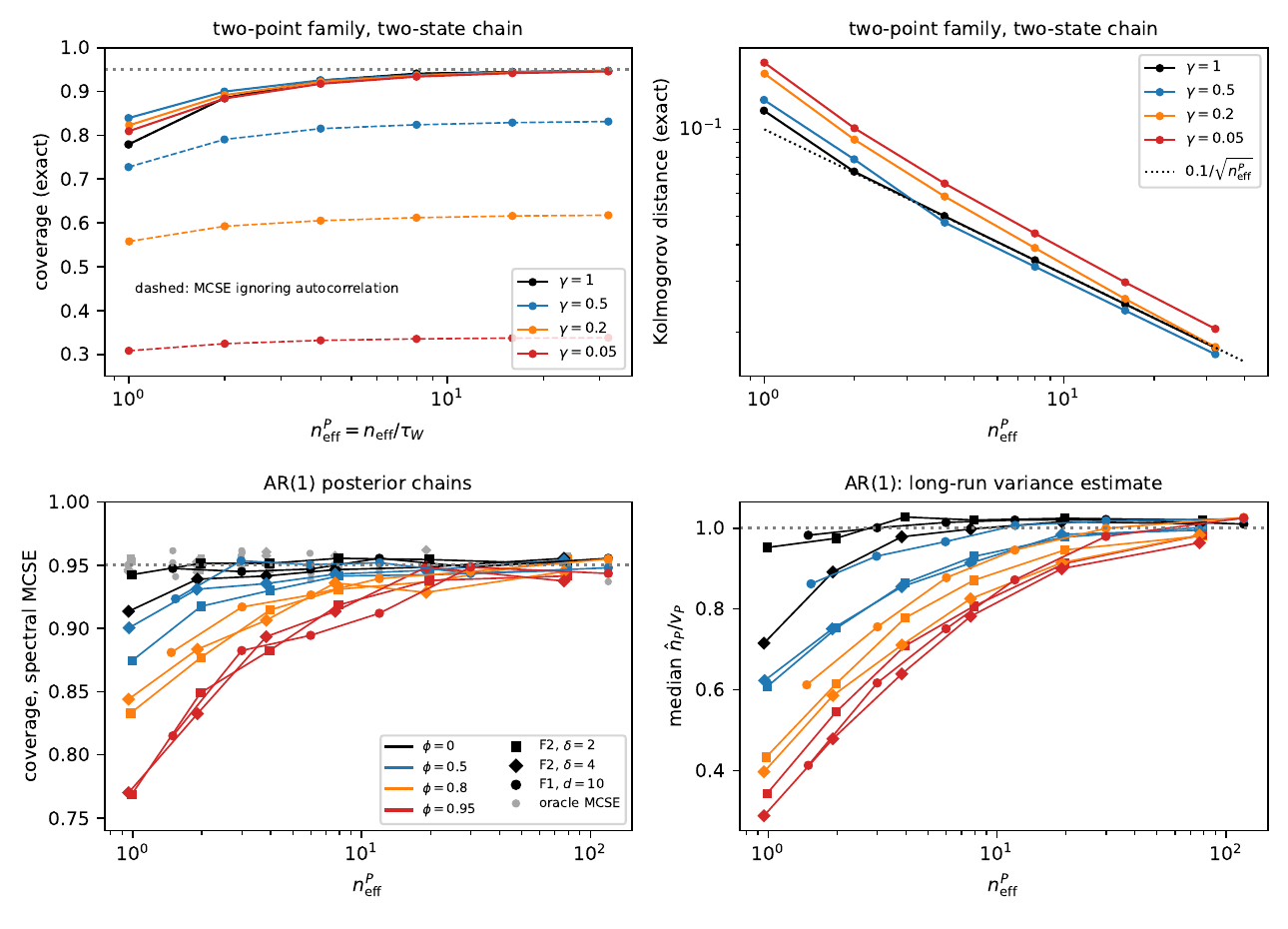}
\caption{Markov chain draws. Top: exact computations for the two-point family with the two-state posterior chain of \cref{thm:mcmc-sharp} (\cref{tab:E8}). Left, coverage of nominal $95\%$ intervals based on $\sigNP$ (solid) and on $\sigN$, which ignores autocorrelation (dashed); right, Kolmogorov distance of $\Delta/\sigNP$ to $\mathcal N(0,1)$. Both are plotted against $\neffP=\neff/\tau_W$ for four spectral gaps $\gamma$. Bottom: AR(1) posterior chains (\cref{tab:E9}). Left, coverage of intervals based on the spectral MCSE (coloured by $\phi$) and on the true $\sigNP$ (grey); right, median of $\widehat n_P/v_P$, the accuracy of the long-run variance estimate.}\label{fig:mcmc}
\end{figure}

We return to families F1 ($d=10$, $\sigma=1.5$) and F2 ($\delta\in\{2,4\}$) of \cref{sec:exp-universality}, now with posterior draws from stationary Gaussian AR(1) chains with coefficient $\phi\in\{0,0.5,0.8,0.95\}$ in each coordinate. These chains are positive with $\gamma=1-\phi$, and the proposal draws are independent. The asymptotic variance $\sigma^2_P(\pistar)$ is computed by Gaussian quadrature over the bivariate normal law of $(\Xj1,\Xj{1+k})$, summed over lags $k$, and we checked it against batch means over runs of length $2\times10^6$. For each $(\phi,\text{family})$ the sample size is chosen so that $\neffP$ takes six values between $1$ and $120$, and each configuration uses $2000$ replications. \Cref{fig:mcmc} (bottom row) and \cref{tab:E9} in \cref{app:repro} show three things. First, intervals based on the true $\sigNP$ cover at $0.937$--$0.962$ in all $72$ configurations, even at $\neffP\approx1$. In these smooth examples the normal approximation with the correct variance is accurate well before \cref{thm:mcmc-BE} guarantees it; the slow approach of the two-state example requires a chain that rarely visits the overlap region. Second, the interval that ignores autocorrelation covers with probability close to $2\Phi(1.96/\sqrt{\tau_W})-1$, between $0.36$ and $0.92$ for $\phi>0$. Third, with the spectral MCSE the coverage is governed by $\neffP$. For $\phi>0$ it is $0.77$--$0.92$ at $\neffP\approx1$, within $0.025$ of nominal for $\neffP\ge19$ and within $0.012$ for $\neffP\ge76$. The undercoverage comes entirely from the long-run variance estimate: the median of $\widehat n_P/v_P$ is $0.3$--$0.9$ at $\neffP\approx1$ and at least $0.9$ once $\neffP\ge19$. This is the term $\varepsilon_\omega$ of \cref{thm:mcmc-mcse}. Because $\widehat\omega^2$ is then biased downwards, the certificate $\neffPhat$ overstates small values of $\neffP$ by a factor of up to about $2$, but it is within $20\%$ of $\neffP$ once $\neffP\ge8$. These findings motivate the threshold $\neffPhat<30$ of \cref{sec:certificate}.

\subsection{Proposal fitting}\label{sec:exp-proposal}

The target has independent coordinates $0.5+s_kt_7$ with $s_k$ equally spaced in $[1,2]$, so it is not in the proposal family. The proposal is a diagonal Gaussian fitted by moments, with $k=2d$ parameters. It is used in three ways: fixed at its limit $\vartheta_0$, fitted on the same posterior draws, or fitted on an independent sample (split). \Cref{tab:E4} confirms \cref{thm:proposal,rem:secondorder}. The standard deviations of the three estimators agree, so there is no first-order variance from fitting. Fitting on the same draws produces a bias that grows with $d$ and reaches about one MCSE at $d=32$, and splitting removes it. The diagnostic $R_g$ of \cref{rem:secondorder}(ii), computed with the proposal draws held fixed, stays near $0.4$ as $S$ quadruples, whereas the pathwise version with common random numbers halves, as an $O(\neff^{-1/2})$ quantity should. A jackknife that refits the proposal while holding the proposal draws fixed inflates $\Vjack$ by about $25\%$ at every sample size. Refitting with common random numbers inflates it by about $10\%$ at $S=2000$ and not at all at $S=8000$, consistent with the second-order size of the fitting effect.

\begin{table}
\centering\small
\caption{Proposal fitting ($t_7$ target with independent coordinates, diagonal Gaussian proposal with $2d$ fitted parameters, $G=20$). Standard deviations of $\log\Zhat$ with the proposal fixed at its limit, fitted on the same draws, and fitted on an independent sample (split). Bias in units of the MCSE (Monte Carlo standard error in parentheses). Median of $R_g$ computed with the proposal draws held fixed and of its pathwise version with common random numbers (not computed for $d=32$). Mean $\Vjack$ for three jackknife variants: no refitting, refitting with the proposal draws held fixed, and refitting with common random numbers.}\label{tab:E4}
\setlength{\tabcolsep}{3pt}
\resizebox{\textwidth}{!}{%
\begin{tabular}{rrrcccccccccc}
\toprule
& & & \multicolumn{3}{c}{sd of $\log\Zhat$} & \multicolumn{2}{c}{bias / MCSE} & \multicolumn{2}{c}{$R_g$} & \multicolumn{3}{c}{mean $\Vjack$}\\
\cmidrule(lr){4-6}\cmidrule(lr){7-8}\cmidrule(lr){9-10}\cmidrule(lr){11-13}
$d$ & $S$ & reps & fixed & same & split & same & split & draws fixed & pathwise & no refit & refit, draws fixed & refit, CRN\\
\midrule
2 & 2000 & 2000 & 0.0035 & 0.0037 & 0.0036 & $-0.26\,(0.02)$ & $+0.00\,(0.02)$ & 0.384 & 0.204 & 1.006 & 1.272 & 1.109\\
8 & 2000 & 2000 & 0.0071 & 0.0072 & 0.0072 & $-0.60\,(0.02)$ & $-0.01\,(0.02)$ & 0.402 & 0.223 & 1.011 & 1.269 & 1.113\\
32 & 2000 & 1000 & 0.0140 & 0.0147 & 0.0152 & $-1.11\,(0.03)$ & $-0.02\,(0.03)$ & -- & -- & 0.996 & 1.242 & 1.086\\
2 & 8000 & 600 & 0.0018 & 0.0018 & 0.0019 & $-0.13\,(0.04)$ & $-0.01\,(0.04)$ & 0.368 & 0.105 & 0.968 & 1.235 & 0.995\\
\bottomrule
\end{tabular}}
\end{table}

\subsection{Sampler failure and metastability}\label{sec:exp-failure}

The target is the asymmetric mixture $\tfrac12\mathcal N(-3e_1,I_2)+\tfrac12\mathcal N(3e_1,\mathrm{diag}(1/4,1))$ and the fixed proposal is $\mathcal N(0,\mathrm{diag}(10,1))$. In scenario (a) the posterior ``sampler'' draws independently but puts weight $w\in\{0.5,0.6,0.7,0.9\}$ on the narrow component ($w=0.5$ is correct), with $\Sone=\Stwo=4000$. In scenario (b) the posterior draws come from a metastable chain whose stationary law is exactly the target. The component label flips with probability $p_{\mathrm{sw}}\in\{10^{-2},10^{-3},2\times10^{-4}\}$ per step, and there is AR(1) dependence with $\phi=0.5$ within components; $\Sone=\Stwo=8000$ and the MCSE is spectral. \Cref{tab:E5} shows the division of labour predicted in \cref{sec:diagnostics}. Under mode imbalance the estimator is biased by $2.4$--$9.7$ MCSEs and interval coverage collapses. Variance-based diagnostics ($\Vjack$, $\Imax$, $\Gr$) stay at their nominal level, as they must, while the calibration test rejects in $46\%$, $97\%$ and $100\%$ of replications. Under metastability the true variance exceeds the spectral MCSE by factors up to $4.7$. For the two slowest chains the jackknife underestimates it as well ($\Vjack\approx0.63$), because the blocks are shorter than the switching time. The calibration test retains partial power ($23\%$ and $60\%$), because a finite run spends unequal time in the two modes.

\begin{table}
\centering\small
\caption{Sampler failure and metastability ($5000$ replications per row; rejection rates in \% at nominal $5\%$). Bias of $\log\Zhat$ in MCSE units, ratio of the true variance to the mean squared MCSE, coverage of nominal $95\%$ intervals, and the diagnostics of \cref{sec:diagnostics}.}\label{tab:E5}
\setlength{\tabcolsep}{4pt}
\resizebox{\textwidth}{!}{%
\begin{tabular}{lrccccccccc}
\toprule
Scenario & med $\neffhat$ & bias/MCSE & Var/MCSE$^2$ & coverage & mean $\Vjack$ & $\Vjack$ & $\Imax$ & $\Gr$ & $t_\Tp$\\
\midrule
(a) imbalance, $w=0.5$ (correct) & 458 & -0.01 & 1.02 & 0.948 & 1.005 & 5.5 & 5.6 & 5.3 & 5.1\\
(a) imbalance, $w=0.6$ & 450 & +2.42 & 0.98 & 0.324 & 0.986 & 4.6 & 4.6 & 5.2 & 45.9\\
(a) imbalance, $w=0.7$ & 440 & +4.80 & 0.94 & 0.003 & 0.957 & 3.7 & 4.6 & 5.4 & 96.9\\
(a) imbalance, $w=0.9$ & 417 & +9.72 & 0.91 & 0.000 & 0.911 & 2.0 & 3.5 & 5.6 & 100.0\\
\midrule
(b) metastable, $p_{\mathrm{sw}}=0.01$ & 914 & +0.02 & 1.11 & 0.932 & 1.041 & 3.0 & 1.8 & 1.8 & 5.5\\
(b) metastable, $p_{\mathrm{sw}}=0.001$ & 915 & +0.01 & 1.63 & 0.800 & 0.636 & 0.5 & 1.5 & 2.2 & 23.5\\
(b) metastable, $p_{\mathrm{sw}}=0.0002$ & 915 & +0.08 & 4.72 & 0.496 & 0.631 & 4.4 & 9.5 & 13.8 & 59.8\\
\bottomrule
\end{tabular}}
\end{table}

\section{Discussion}\label{sec:discussion}

The optimal bridge sampling estimator is better behaved than its reputation suggests, and its failures are simpler to characterize. With independent draws, one number, the effective overlap $\neff$, determines both how accurate the estimator is and how accurate its MCSE is. The results hold uniformly over targets, proposals and dimensions, and require no moment conditions. The sample size required is of order $1/\Hov$, the inverse Bayes error of discriminating the posterior from the proposal, and this is sharp in the worst case. With draws from a reversible Markov chain, the picture is the same once the overlap is measured in effectively independent units. The sample size required grows by a factor of order $1/\gamma$, the relaxation time of the sampler, and this is again sharp in the worst case. The relevant certificate is $\neffP=\neff/\tau_W$, the number of ambiguous draws divided by the autocorrelation time of the estimating function. The remaining failures are sampler failure and misestimated autocorrelation, which call for different tools: a calibration test and jackknife comparisons with fixed-$G$ calibration.

Several limitations should be stated. Our Markov chain results assume a reversible posterior sampler with a spectral gap. Non-reversible samplers can be handled in the same way through pseudo-spectral gaps \citep{paulin_concentration_2015}, at the cost of worse constants. The MCMC guarantee for the MCSE is conditional on the accuracy of the long-run variance estimate, for which we know of no finite-sample theory in the required generality, and the certificate threshold $\neffPhat\ge30$ is calibrated empirically. For independent draws, \cref{prop:onesided} backs the threshold $\neffhat\ge10$ with an explicit guarantee only when $N\Hov\lesssim1$; between that and $10$ it rests on simulation. The calibration test cannot detect sampler errors that are constant on the overlap region, but by \cref{prop:calib-alt}(c) neither can any other method that uses only the bridge output. The constants in \cref{thm:mcse} are conservative by one to two orders of magnitude relative to what we observe, and $C_P$ in \cref{thm:mcmc-BE} by more than two. The second-order bias of \cref{rem:secondorder} is described heuristically and by simulation, not by a theorem with explicit constants. Finally, no within-run diagnostic can detect metastability on time scales longer than the run. Natural extensions include multiple bridge sampling and MBAR \citep{kong_theory_2003,tan_likelihood_2004,shirts_statistically_2008}, where the overlap becomes the MBAR overlap matrix and its spectral gap is the natural analogue of $\Iov$; a minimax lower bound over all estimators that use both samples; warp transformations chosen to maximize $\Hov$ \citep{meng_warp_2002,wang_warp_2022}; an omnibus calibration test, since $\Tp$ is sensitive to one direction of error and a test of the whole relation that correct sampling implies between the laws of $f/g$ in the two samples should detect every error that the bridge output reveals; and noise-contrastive estimation with more than one parameter, where our reduction no longer applies but orthogonality and overlap arguments should.

\section*{Acknowledgments}
The author thanks Professor Aki Vehtari for his mentorship during the Aalto Science Institute International Summer Research Programme. The block reshuffling approach that inspired the jackknife diagnostics in this work originated from research conducted under his supervision, which appeared in an early preprint of \cite{micaletto_diagnostics_2025}.

\newpage
\bibliography{references}

\appendix

\section{Proofs for Section~\ref{sec:setting}}\label{app:proofs-setting}

\begin{proof}[Proof of \cref{lem:esteq}]
(a) For $\theta$ with $g(\theta)>0$, $\ell/(\sone\ell+\stwo z)=f/(\sone f+\stwo zg)=(1-\pi_z)/\sone$, and this also holds when $f(\theta)=0$, since then both sides vanish. For $\theta$ with $f(\theta)>0$, $1/(\sone\ell+\stwo z)=g/(\sone f+\stwo zg)=\pi_z/(\stwo z)$, which also holds when $g(\theta)=0$. Hence $n(z)=(\sone\Stwo)^{-1}\sum_i\{1-\pi_z(\Yi i)\}$ and $z\,d(z)=(\stwo\Sone)^{-1}\sum_j\pi_z(\Xj j)$. Since $\sone\Stwo=\stwo\Sone=N\sone\stwo$, we get $n(z)-z\,d(z)=W(\log z)/(N\sone\stwo)$. Whenever $d(z)>0$, $z=n(z)/d(z)$ is equivalent to $n(z)-zd(z)=0$.
(b) We have $\pi_c=\{1+e^{-c-\lambda}\}^{-1}$ with $\lambda=\log\{\stwo g/(\sone f)\}\in[-\infty,\infty]$, so $\partial_c\pi_c=\pi_c(1-\pi_c)$, which is positive on $A$. Summing gives the formula for $W'$.
(c) As $c\to-\infty$, $\pi_c\to0$ on $A$; proposal draws outside $A$ have $f=0$, hence $\pi_c\equiv1$, and contribute nothing. Thus $W(c)\to\#\{i:\Yi i\in A\}>0$. Symmetrically, $W(c)\to-\#\{j:\Xj j\in A\}<0$ as $c\to\infty$. Continuity and strict monotonicity give a unique root, which equals $\sup\{c:W(c)>0\}$.
(d) $\partial_c\log\pi_c=1-\pi_c$ and $\partial_c\log(1-\pi_c)=-\pi_c$. The second derivative is $-\sum\pi_c(1-\pi_c)\le0$.
\end{proof}

\begin{proof}[Proof of \cref{lem:overlap}]
(a) We have $\sone\EE_p\pistar=\sone\stwo\int pg/\mix$, $\stwo\EE_g(1-\pistar)=\stwo\sone\int gp/\mix$, and $\EE_\mix[\pistar(1-\pistar)]=\int\mix\,(\stwo g/\mix)(\sone p/\mix)$.
(b) For $x\in[0,1]$, $x(1-x)=\min(x,1-x)\max(x,1-x)$ and $\max(x,1-x)\in[1/2,1]$. Moreover $\EE_\mix\min(\pistar,1-\pistar)=\int\min(\stwo g,\sone p)=\varepsilon_B$.
(c) $\Iov=\int pg/\mix=\EE_p[1/(\sone r+\stwo)]$ with $r=p/g\in[0,\infty]$. The map $x\mapsto(\sone x+\stwo)^{-1}$ is convex, so Jensen's inequality gives $\Iov\ge(\sone\EE_pr+\stwo)^{-1}$ with $\EE_pr=1+\chi^2(p\|g)$. The bound is trivial if $\chi^2=\infty$. The second inequality follows in the same way from $\Iov=\EE_g[1/(\sone+\stwo g/p)]$.
(d) From $\sone\int p^2/\mix+\stwo\int pg/\mix=\int p=1$ we get $\sone\int p^2/\mix=1-\stwo\Iov$, and likewise $\stwo\int g^2/\mix=1-\sone\Iov$. Hence $\sone\stwo\int(p-g)^2/\mix=\stwo(1-\stwo\Iov)+\sone(1-\sone\Iov)-2\sone\stwo\Iov=1-\Iov$.
(e) Under independent sampling, $\Var W(\cstar)=\Stwo\Var_g(\pistar)+\Sone\Var_p(\pistar)$. Now $\Stwo\EE_g(1-\pistar)^2+\Sone\EE_p(\pistar)^2=N\int\{\stwo g(1-\pistar)^2+\sone p(\pistar)^2\}=N\int\mix\{\pistar(1-\pistar)^2+(1-\pistar)(\pistar)^2\}=N\Hov$. By (a) the squared means contribute $\Stwo(\Hov/\stwo)^2+\Sone(\Hov/\sone)^2=N\Hov^2/(\sone\stwo)$. Hence $\Var W(\cstar)=N\Hov(1-\Iov)=\neff$, and $\sigN^2=\beta/(N\Hov)=\neff/(N\Hov)^2$.
\end{proof}

\begin{proof}[Proof of \cref{lem:mcse-structure}]
With $\mathscr N_i=f h_{\Zhat}(\Yi i)=(1-\widehat\pi_i)/\sone$ and $\mathscr D_j=gh_{\Zhat}(\Xj j)=\widehat\pi_j/(\stwo\Zhat)$, the relative variances in \cref{eq:VarMC-def} are scale-free. They equal $\widehat V_{\mathrm{prop}}/\{\Stwo(\Stwo^{-1}\sum_i(1-\widehat\pi_i))^2\}$ and $\widehat V_{\mathrm{post}}/\{\Sone(\Sone^{-1}\sum_j\widehat\pi_j)^2\}$. At the root, $\sum_i(1-\widehat\pi_i)=\sum_j\widehat\pi_j=N\Hzero$ by \cref{lem:esteq}, so the sum equals $(\Stwo\widehat V_{\mathrm{prop}}+\Sone\widehat V_{\mathrm{post}})/(N\Hzero)^2$. That $\Hzero$ and $\Hhat$ estimate $\Hov$ follows from \cref{lem:overlap}(a) and the definition of $\Hov$; see \cref{prop:certificate} and \cref{thm:iteration}(d) for quantitative statements.
\end{proof}

\section{Proofs for Section~\ref{sec:nonasymp}}\label{app:proofs-nonasymp}

Throughout, $\pi_t:=\pi_{\cstar+t}$. Since $\pi_c=\sigma(c+\lambda)$ with $\sigma$ the logistic function, and $\pistar=\sigma(\cstar+\lambda)$,
\begin{equation}\label{eq:shift}
\pi_t=\frac{e^t\pistar}{D_t},\qquad 1-\pi_t=\frac{1-\pistar}{D_t},\qquad D_t:=1-\pistar+e^t\pistar=1+(e^t-1)\pistar\in[\min(1,e^t),\max(1,e^t)].
\end{equation}
Two elementary facts are used repeatedly. First, for every $\theta$ and every $\Delta\in\RR$,
\begin{equation}\label{eq:ratio}
e^{-|\Delta|}\le\frac{\pi_{c+\Delta}(\theta)}{\pi_c(\theta)},\ \frac{1-\pi_{c+\Delta}(\theta)}{1-\pi_c(\theta)},\ \frac{\pi_{c+\Delta}(1-\pi_{c+\Delta})(\theta)}{\pi_c(1-\pi_c)(\theta)}\le e^{|\Delta|},
\end{equation}
which follows from $\sigma(x+\Delta)/\sigma(x)=(1+e^{-x})/(1+e^{-x-\Delta})$. Second, for any probability measure $Q$ (population or empirical), $\partial_c\Var_Q(\pi_c)=2\Cov_Q(\pi_c,\pi_c(1-\pi_c))$. Since $x\mapsto x(1-x)$ is $1$-Lipschitz on $[0,1]$, $\Var_Q(\pi_c(1-\pi_c))\le\Var_Q(\pi_c)$, and Cauchy--Schwarz gives
\begin{equation}\label{eq:varlip}
|\partial_c\log\Var_Q(\pi_c)|\le2,\qquad\text{so}\qquad e^{-2|\Delta|}\le\Var_Q(\pi_{c+\Delta})/\Var_Q(\pi_c)\le e^{2|\Delta|}.
\end{equation}

\begin{proof}[Proof of \cref{lem:meanvar}]
(a) By \cref{eq:shift} and $\stwo g=\mix\pistar$, $\sone p=\mix(1-\pistar)$,
\[
\mu(t)=N\!\int\!\Big\{\stwo g\frac{1-\pistar}{D_t}-\sone p\frac{e^t\pistar}{D_t}\Big\}=N\!\int\!\mix\,\frac{\pistar(1-\pistar)(1-e^t)}{D_t}.
\]
The bounds follow from the range of $D_t$ in \cref{eq:shift}: for $t>0$, $|\mu(t)|\in[N(e^t-1)\Hov/e^t,N(e^t-1)\Hov]$, and for $t<0$, $|\mu(t)|\in[N(1-e^t)\Hov,N(1-e^t)\Hov/e^t]$.
(b) Differentiating, $\partial_t\{(e^t-1)/D_t\}=e^t/D_t^2\in[e^{-|t|},e^{|t|}]$. Hence $\mu'(s)=-N\EE_\mix[\pistar(1-\pistar)e^s/D_s^2]$ and $|\mu'(s)+N\Hov|\le N\Hov(e^{|s|}-1)$. Integrating from $0$ to $t$ gives the claim.
(c) $v(0)=\neff$ is \cref{lem:overlap}(e). Next, $v(t)\le\Stwo\EE_g(1-\pi_t)^2+\Sone\EE_p\pi_t^2=N\int\mix\,\pistar(1-\pistar)\,\phi_t(\pistar)$ with $\phi_t(q)=\{1+(e^{2t}-1)q\}/\{1+(e^t-1)q\}^2$. Writing $r=e^t$ and $x=(r-1)q$, the function $x\mapsto(1+(r+1)x)/(1+x)^2$ is maximized at $x=(r-1)/(r+1)$, which lies in the admissible range. Its maximum is $(r+1)^2/(4r)=\cosh^2(t/2)$. The last claim follows from \cref{eq:varlip} applied to $\Var_g$ and $\Var_p$ separately.
\end{proof}

\begin{proof}[Proof of \cref{thm:concentration}]
(a) $W$ is continuous and non-increasing. If $W(\cstar+t)\le0$, then $W(c)\le0$ for all $c\ge\cstar+t$, so $\chat=\sup\{c:W(c)>0\}\le\cstar+t$. Conversely, if $\chat\le\cstar+t$, then $W(c)\le0$ for all $c>\cstar+t$, and $W(\cstar+t)\le0$ by continuity. For the inclusion, suppose $\chat\ge\cstar+t$ but $W(\cstar+t)<0$; by continuity and monotonicity $W<0$ on $[\cstar+t-\eta,\infty)$ for some $\eta>0$, so $\chat\le\cstar+t-\eta$, a contradiction. If some draw lies in $A$, then $W$ is strictly decreasing and $W(\cstar+t)\ge0$ forces $\chat\ge\cstar+t$. If no draw lies in $A$, then $W\equiv0$ and $\chat=-\infty$, so the inclusion can be strict. No probabilistic assumption is used.
(b) Fix $t>0$. $W_t-\mu(t)$ is a sum of $N$ independent centred terms bounded by $1$ in absolute value, with total variance $v(t)\le N\Hov\cosh^2(t/2)$ (\cref{lem:meanvar}(c)). By the inclusion in (a) and Bernstein's inequality, since $-\mu(t)\ge\lambda_0:=N\Hov(1-e^{-t})$ and $\lambda\mapsto\lambda^2/\{2(v+\lambda/3)\}$ is increasing,
\[
\PP(\Delta\ge t)\le\PP(W_t-\mu(t)\ge-\mu(t))\le\exp\Big\{-\frac{\lambda_0^2}{2(v(t)+\lambda_0/3)}\Big\}\le\exp\{-N\Hov\kappa(t)\}.
\]
The lower tail is identical, using $\{\Delta\le-t\}=\{W_{-t}\le0\}$ and $\mu(-t)\ge\lambda_0$.
\end{proof}

\begin{proof}[Proof of \cref{cor:samplesize}]
Let $L=\log(2/\delta)$ and $t=2\sqrt{L/(N\Hov)}\le1/2$. On $(0,1/2]$, $(1-e^{-t})/t$ is decreasing and the denominator of $\kappa$ is increasing. Hence $\kappa(t)/t^2\ge\kappa(1/2)/(1/4)=0.259\ge1/4$, so $N\Hov\kappa(t)\ge L$, and \cref{thm:concentration} gives $\PP(|\Delta|\ge t)\le2e^{-L}=\delta$. For the second claim, $N\Hov\ge N\varepsilon_B/2$ by \cref{lem:overlap}(b). The condition $N\Hov\ge4L/t^2$ implies both $N\Hov\ge16L$ and $2\sqrt{L/(N\Hov)}\le t$.
\end{proof}

\begin{proof}[Proof of \cref{thm:BE}]
Let $F(t)=\PP(\Delta\le t)=\PP(W_t\le0)$ (\cref{thm:concentration}(a)) and $a(t)=-\mu(t)/\sqrt{v(t)}$. Fix $\tau\in(0,1/2)$; we take $\tau=0.2312$.

\emph{Central region $|t|\le\tau$.} $W_t$ is a sum of independent, non-identically distributed variables whose centred versions are bounded by $1$, so the sum of their absolute third moments is at most $v(t)$. The Berry--Esseen theorem with the constant $C_0=0.5600$ of \citet{shevtsova_improvement_2010} gives $|F(t)-\Phi(a(t))|\le C_0/\sqrt{v(t)}\le C_0e^{\tau}/\sqrt{\neff}$, by \cref{lem:meanvar}(c). Next, write $y=t/\sigN=N\Hov t/\sqrt{\neff}$ and $a(t)=\rho(t)y$. By \cref{lem:meanvar}(a),(c), and since $(1-e^{-x})/x\ge e^{-x}$ and $(e^x-1)/x\le e^x$, we have $\rho(t)=\{-\mu(t)/(N\Hov t)\}\sqrt{\neff/v(t)}\in[e^{-2|t|},e^{2|t|}]$. The mean value theorem gives
\[
|\Phi(\rho y)-\Phi(y)|\le|y|\,|\rho-1|\,\phi(|y|e^{-2\tau})\le2\sigN e^{2\tau}\,y^2\phi(ye^{-2\tau})\le2\sigN e^{6\tau}\sup_ww^2\phi(w)=0.587\,e^{6\tau}\sigN,
\]
using $e^{2|t|}-1\le2|t|e^{2\tau}$ and $|t|=\sigN|y|$. Since $\sigN=\beta/\sqrt{\neff}\le1/\sqrt\neff$, the central region contributes at most $(C_0e^\tau+0.587e^{6\tau})/\sqrt{\neff}$.

\emph{Tails $|t|>\tau$.} For $t>\tau$, both $1-F(t)$ and $1-\Phi(t/\sigN)$ lie in $[0,\max\{1-F(\tau),1-\Phi(\tau/\sigN)\}]$. By \cref{thm:concentration}, $1-F(\tau)\le e^{-N\Hov\kappa(\tau)}$. Also $1-\Phi(\tau/\sigN)\le e^{-\tau^2N\Hov/(2\beta)}\le e^{-N\Hov\kappa(\tau)}$, since $\kappa(\tau)\le\tau^2/2$. As $N\Hov\ge\neff$ and $\sup_{x>0}\sqrt xe^{-\kappa x}=(2e\kappa)^{-1/2}$, the tails contribute at most $(2e\kappa(\tau))^{-1/2}/\sqrt{\neff}$. The case $t<-\tau$ is symmetric.

With $\tau=0.2312$ both constants are at most $3.06$, which proves the claim.
\end{proof}

\begin{proof}[Proof of \cref{thm:mcse}]
Let $L=\log(12/\delta)$ and $\ell=L/\neff\le1/64$. By \cref{lem:mcse-structure}, and since $\sigN=\sqrt{\neff}/(N\Hov)$,
\begin{equation}\label{eq:ratiodecomp}
\frac{\sighat}{\sigN}=\sqrt{\frac{\neffhat(\chat)}{\neff}}\cdot\frac{\Hov}{\Hzero(\chat)},
\end{equation}
where $\neffhat(c)$ and $\Hzero(c)$ denote the statistics evaluated at $c$. Consider three events.
\emph{E1}: $|\Delta|\le t_1:=1.642\sqrt\ell$. Since $\kappa(t)/t^2$ is decreasing, $\kappa(t)\ge k_0t^2$ on $[0,t_0]$ with $t_0=0.2256$ and $k_0=\kappa(t_0)/t_0^2=0.3710\ge1.642^{-2}$. As $t_1\le0.2053\le t_0$, \cref{thm:concentration} gives $\PP(E1^c)\le2e^{-N\Hov\kappa(t_1)}\le2e^{-L}=\delta/6$.
\emph{E2}: $|\Hzero(\cstar)/\Hov-1|\le\eta:=1.4606\sqrt\ell\le0.1826$. The variable $N\Hzero(\cstar)=\sum_j\pistar(\Xj j)$ has mean $N\Hov$ and variance at most $N\Hov$, with summands in $[0,1]$. Bernstein's inequality gives $\PP(E2^c)\le2\exp\{-\eta^2N\Hov/(2+2\eta/3)\}\le2e^{-L}=\delta/6$.
\emph{E3}: $|\sqrt{\neffhat(\cstar)}-\sqrt{\neff}|\le\sqrt{8L}$. The empirical Bernstein bound for the sample variance of $[0,1]$-valued i.i.d.\ variables \citep[Theorem~10]{maurer_empirical_2009} gives, with probability at least $1-\delta/6$ each, $|\sqrt{\widehat V_{\mathrm{post}}(\cstar)}-\sqrt{\Var_p\pistar}|\le\sqrt{2L/(\Sone-1)}$, and similarly for the proposal sample. Multiplying by $\sqrt{\Sone}$ and $\sqrt{\Stwo}$ and using $S/(S-1)\le2$, each coordinate of the vector $(\sqrt{\Sone\widehat V_{\mathrm{post}}},\sqrt{\Stwo\widehat V_{\mathrm{prop}}})$ is within $2\sqrt L$ of its population counterpart. The reverse triangle inequality for the Euclidean norm then gives E3.

On $E1\cap E2\cap E3$, which has probability at least $1-2\delta/3\ge1-\delta$: by \cref{eq:varlip} applied to the two empirical measures, $\sqrt{\neffhat(\chat)/\neffhat(\cstar)}\in[e^{-|\Delta|},e^{|\Delta|}]$, and by \cref{eq:ratio}, $\Hzero(\chat)/\Hzero(\cstar)\in[e^{-|\Delta|},e^{|\Delta|}]$. With $\epsilon_1=\sqrt{8\ell}\le0.3536$, \cref{eq:ratiodecomp} gives
\[
\Big|\log\frac{\sighat}{\sigN}\Big|\le2t_1-\log(1-\epsilon_1)-\log(1-\eta)\le\Big(3.284+\frac{2.828}{1-0.3536}+\frac{1.4606}{1-0.1826}\Big)\sqrt\ell\le9.45\sqrt\ell.
\]
For coverage, take $\delta=\neff^{-1/2}$; then $\neff\ge350$ implies $\neff\ge64\log(12\sqrt{\neff})$. On the event above, $\sighat\in\sigN[e^{-r},e^r]$. By \cref{thm:BE}, applied at $\pm x\sigN$ (the bound also holds for left limits), $|\PP(|\Delta|\le x\sigN)-(2\Phi(x)-1)|\le2\cdot3.1/\sqrt{\neff}$ for every $x$. Hence
\[
|\PP(\log\Z\in\mathcal C_\alpha)-(1-\alpha)|\le\frac{6.2}{\sqrt\neff}+\delta+\max_\pm|2\Phi(ze^{\pm r})-2\Phi(z)|,
\]
and the last term is at most $2z(e^r-1)\phi(ze^{-r})$ by the mean value theorem.
\end{proof}

\begin{proof}[Proof of \cref{prop:certificate}]
On $E1\cap E2\cap E3$ of the previous proof, $|\log(\neffhat/\neff)|\le2|\Delta|-2\log(1-\epsilon_1)\le2(1.642+4.375)\sqrt\ell\le12.1\sqrt\ell$. Similarly, $|\log(\Hzero/\Hov)|\le|\Delta|-\log(1-\eta)\le(1.642+1.787)\sqrt\ell\le3.5\sqrt\ell$.
\end{proof}

\begin{proof}[Proof of \cref{thm:sharp}]
Here $\pistar(0)=0$, $\pistar(2)=1$ and $\pistar(1)=1/2$, so only point $1$ carries overlap. This gives $\Hov=\mix(1)/4=\varepsilon/4$, $\Iov=\varepsilon$ and $\neff=2S(\varepsilon/4)(1-\varepsilon)$. Draws at $0$ or $2$ contribute nothing to $W$, so $W(c)=n_{\mathrm{prop}}\{1-\sigma(c-\cstar)\}-n_{\mathrm{post}}\sigma(c-\cstar)$ and $\Delta=\log(n_{\mathrm{prop}}/n_{\mathrm{post}})$ when both counts are positive.
(a) If $n_{\mathrm{post}}=0$, then $W\ge0$ everywhere, so $\chat=+\infty$ if $n_{\mathrm{prop}}>0$ and $\chat=-\infty$ otherwise. Hence $\PP(\Zhat\in\{0,\infty\})\ge(1-\varepsilon)^S\ge e^{-S\varepsilon/(1-\varepsilon)}\ge e^{-2S\varepsilon}=e^{-4N\Hov}$ for $\varepsilon\le1/2$.
(b) Let $b(k)=\binom Sk\varepsilon^k(1-\varepsilon)^{S-k}$ and $\sigma_b^2=S\varepsilon(1-\varepsilon)=2\neff$. The atom $\PP(\Delta=0)=\sum_{k\ge1}b(k)^2$ produces a jump of that size in the distribution function of $\Delta/\sigN$, whereas $\Phi$ is continuous. Hence the Kolmogorov distance is at least $\frac12\PP(\Delta=0)$. Let $K=\{k:|k-S\varepsilon|\le2\sigma_b\}$. By Chebyshev's inequality $\sum_{k\in K}b(k)\ge3/4$. If $\neff\ge2$, then $S\varepsilon\ge4/(1-\varepsilon)>2\sigma_b$, so $0\notin K$, and $|K|\le4\sigma_b+1\le(4\sqrt2+1/\sqrt2)\sqrt{\neff}$. By Cauchy--Schwarz, $\sum_{k\ge1}b(k)^2\ge(\sum_Kb)^2/|K|\ge0.0884/\sqrt{\neff}$, so the Kolmogorov distance is at least $0.0442/\sqrt{\neff}\ge0.04/\sqrt{\neff}$.
\end{proof}

\section{Proofs for Section~\ref{sec:mcmc}}\label{app:proofs-mcmc}

Throughout, $\langle a,b\rangle=\EE_p[\bar ab]$ is the inner product of the complex space $L^2(p)$, $\|\cdot\|$ denotes its norm and the corresponding operator norm, and $\Pi u=\EE_pu$ (a constant function). Under \cref{asmp:mcmc}, $P$ is a self-adjoint contraction with $P1=1$, and $Z:=(I-P+\Pi)^{-1}$ exists. On $L^2_0(p)$ the spectrum of $Z$ is $\{1/(1-\lambda):\lambda\in\operatorname{spec}(P|_{L^2_0})\}\subset[1/(2-\gamma_-),1/\gamma]$. For bounded $u$ with $\bar u=u-\EE_pu$, $\sigma_P^2(u)=\langle\bar u,(2Z-I)\bar u\rangle$ \citep[Theorem~3.5 and Remark~3.6]{paulin_concentration_2015}. We write $R(z,M)=(z-M)^{-1}$ for resolvents.

\begin{lemma}\label{lem:mcmc-basic}
Assume \cref{asmp:mcmc}.
\begin{enumerate}[label=(\alph*)]
\item For bounded $u$, $\cminus\Var_p(u)\le\sigma_P^2(u)\le\cplus\Var_p(u)$; and $\cminus\ge1$ if $\gamma_-\ge1$.
\item Let $s(t)^2:=\Stwo\Var_g(\pi_t)+\Sone\sigma_P^2(\pi_t)$, so that $s(0)^2=v_P$. Then $e^{-2\zeta_P|t|}v_P\le s(t)^2\le e^{2\zeta_P|t|}v_P$ for all $t$, and $\sigma_P^2(\pi_t)/\sigma_P^2(\pistar)$ obeys the same bounds when $\Var_p(\pistar)>0$.
\item $\min(1,\cminus)\,\neff\le v_P\le\cplus\neff\le\cplus N\Hov$.
\end{enumerate}
\end{lemma}

\begin{proof}
(a) We have $\langle\bar u,(2Z-I)\bar u\rangle=\int\frac{1+\lambda}{1-\lambda}\,\mu_{\bar u}(\dd\lambda)$, where the spectral measure $\mu_{\bar u}$ of $\bar u$ has mass $\Var_p(u)$ and is supported in $[-1+\gamma_-,1-\gamma]$. The integrand is increasing in $\lambda$ and equals $\cminus$ and $\cplus$ at the endpoints.
(b) Let $G=2Z-I$ on $L^2_0(p)$, a positive operator with spectrum in $[\cminus,\cplus]$, and let $f_t=\pi_t-\EE_p\pi_t$. Since $\partial_t\pi_t=\pi_t(1-\pi_t)=:h_t$, we have $\partial_tf_t=h_t-\EE_ph_t$ and $\partial_t\sigma^2_P(\pi_t)=2\langle\partial_tf_t,Gf_t\rangle$. The map $x\mapsto x(1-x)$ is $1$-Lipschitz on $[0,1]$, so $\Var_p(h_t)\le\Var_p(\pi_t)$. By the Cauchy--Schwarz inequality for the form $\langle\cdot,G\cdot\rangle$ and part (a),
\[
|\partial_t\sigma_P^2(\pi_t)|\le2\sigma_P(h_t)\sigma_P(\pi_t)\le2\{\cplus\Var_p(\pi_t)\}^{1/2}\sigma_P(\pi_t)\le2\zeta_P\,\sigma_P^2(\pi_t).
\]
Integrating gives the claim for $\sigma^2_P(\pi_t)$. If $\Var_p(\pistar)=0$, then $\pi_t$ is $p$-a.s.\ constant for every $t$, since it is a function of $\pistar$. The bound for $s(t)^2$ follows by combining this with $e^{-2|t|}\le\Var_g(\pi_t)/\Var_g(\pistar)\le e^{2|t|}$ from \cref{eq:varlip}, and with $\zeta_P\ge1$.
(c) By (a), $v_P$ lies between $\min(1,\cminus)$ and $\max(1,\cplus)=\cplus$ times $\Stwo\Var_g(\pistar)+\Sone\Var_p(\pistar)=\neff$ (\cref{lem:overlap}(e)), and $\neff=N\Hov\beta\le N\Hov$.
\end{proof}

\begin{proof}[Proof of \cref{thm:mcmc-conc}]
(a) Fix $t>0$ and write $W_t-\mu(t)=U+V$, with $U=\sum_i\{1-\pi_t(\Yi i)-\EE_g(1-\pi_t)\}$ and $V=-\sum_j\{\pi_t(\Xj j)-\EE_p\pi_t\}$. By stationarity, $\mu(t)$ is the function of \cref{lem:meanvar}, and $U$ and $V$ are independent. The summands of $U$ are independent, centred and bounded by $1$, so $\log\EE e^{\theta U}\le\Stwo\Var_g(\pi_t)(e^\theta-1-\theta)$ for $\theta\ge0$. Let $f=-(\pi_t-\EE_p\pi_t)$, which satisfies $|f|\le1$ and $\EE_pf^2=\Var_p(\pi_t)$, and let $\Sigma_t=\Stwo\Var_g(\pi_t)+\Sone\Var_p(\pi_t)$. If $\gamma=1$, then $\sup\operatorname{spec}(P|_{L^2_0})\le0$, and Lemma~2 of \citet{jiang_bernstein_2025} gives $\EE e^{\theta V}\le\|E^{\theta f/2}\Pi E^{\theta f/2}\|^{\Sone}=(\EE_pe^{\theta f})^{\Sone}$ for every $\theta\ge0$, where $E^{\theta f/2}$ is multiplication by $e^{\theta f/2}$; hence $\log\EE e^{\theta V}\le\Sone\Var_p(\pi_t)(e^\theta-1-\theta)$, as for independent draws. Then $e^\theta-1-\theta\le\theta^2/\{2(1-\theta/3)\}$ gives $\log\EE e^{\theta(U+V)}\le\Sigma_t\theta^2/\{2(1-b_P\theta)\}$ for $0\le\theta<3$. If $\gamma<1$, let $\lambda_+=1-\gamma\ge\sup\operatorname{spec}(P|_{L^2_0})$. Theorem~2 of \citet{jiang_bernstein_2025} gives
\[
\log\EE e^{\theta V}\le\Sone\Var_p(\pi_t)\Big\{e^\theta-1-\theta+\frac{\lambda_+\theta^2}{1-\lambda_+-5\theta}\Big\},\qquad0\le\theta<\frac{1-\lambda_+}{5}=\frac1{b_P}
\]
(their bound is increasing in $\lambda_+$). Since $1-\theta/3\ge1-5\theta/\gamma$,
\[
\log\EE e^{\theta(U+V)}\le\Sigma_t\theta^2\Big(\frac12+\frac{1-\gamma}{\gamma}\Big)\Big/\Big(1-\frac{5\theta}{\gamma}\Big)=\frac{\cplus\Sigma_t\theta^2}{2(1-b_P\theta)} .
\]
In both cases the bound $\cplus\Sigma_t\theta^2/\{2(1-b_P\theta)\}$ holds for $0\le\theta<1/b_P$ (with $\cplus=1$ if $\gamma=1$), and $\cplus\Sigma_t\le v:=\cplus N\Hov\cosh^2(t/2)$ by the proof of \cref{lem:meanvar}(c). The choice $\theta=x/(v+b_Px)<1/b_P$ in Chernoff's bound gives $\PP(U+V\ge x)\le\exp\{-x^2/(2(v+b_Px))\}$. Since $-\mu(t)\ge\lambda_0:=N\Hov(1-e^{-t})$ by \cref{lem:meanvar}(a), and $\{\Delta\ge t\}\subseteq\{W_t\ge0\}$ by \cref{thm:concentration}(a),
\[
\PP(\Delta\ge t)\le\PP(U+V\ge\lambda_0)\le\exp\Big\{-\frac{\lambda_0^2}{2(v+b_P\lambda_0)}\Big\}=e^{-N\Hov\kappa_P(t)}.
\]
The lower tail is identical, with $f$ replaced by $-f$, $\{\Delta\le-t\}=\{W_{-t}\le0\}$ and $\mu(-t)\ge\lambda_0$.

(b) By the proof of \cref{lem:esteq}(c), $\chat=+\infty$ requires that no posterior draw lies in $A=\{f>0,g>0\}$, and $\chat=-\infty$ requires that no proposal draw does. Since $1-\pistar\le\ind_A$ $g$-a.e., $\Stwo g(A)\ge\Stwo\EE_g(1-\pistar)=N\Hov$ by \cref{lem:overlap}(a), so the second event has probability $\{1-g(A)\}^{\Stwo}\le e^{-N\Hov}$. Similarly $\Sone p(A)\ge N\Hov$. Let $D$ be multiplication by $\ind_{A^c}$. Then
\[
\PP(\Xj1,\dots,\Xj{\Sone}\notin A)=\langle\ind_{A^c},(DPD)^{\Sone-1}\ind_{A^c}\rangle\le\|DPD\|^{\Sone-1}.
\]
The operator $DPD$ is self-adjoint and positivity preserving, so $|\langle u,DPDu\rangle|\le\langle|u|,DPD|u|\rangle$ and $\|DPD\|=\sup\{\langle u,DPDu\rangle:u\ge0,\ \|u\|=1\}$. If $u$ vanishes on $A$, then $(\EE_pu)^2\le p(A^c)\|u\|^2$ by the Cauchy--Schwarz inequality, so $\Var_p(u)\ge p(A)\|u\|^2$ and $\langle u,Pu\rangle\le\|u\|^2-\gamma\Var_p(u)\le\{1-\gamma p(A)\}\|u\|^2$. Applying this to $Du$ gives $\|DPD\|\le1-\gamma p(A)$, and the first event has probability at most $\exp\{-\gamma p(A)(\Sone-1)\}\le e^{-\gamma N\Hov(1-1/\Sone)}$.

(c) Discarding the first $t_0$ states leaves a chain started from $qP^{t_0}$. By \citet[Proposition~3.16]{paulin_concentration_2015}, $d_{\mathrm{TV}}(qP^{t_0},p)\le\frac12(1-\gs)^{t_0}\chi^2(q\|p)^{1/2}$ and $1+\chi^2(qP^{t_0}\|p)\le1+(1-\gs)^{2t_0}\chi^2(q\|p)$. The first bound follows because the joint law of the retained states depends on the start only through the law of the first retained state. For the second, apply \citet[Proposition~3.15]{paulin_concentration_2015} conditionally on the independent proposal draws, and then Jensen's inequality.
\end{proof}

\paragraph{Remark on the constant in \cref{thm:mcmc-conc}.} For $\gamma<1$ and $0<t\le1/2$, $0.7869\,t\le1-e^{-t}\le0.3935$, $\cosh^2(t/2)\le1.0639$ and $\cplus<2/\gamma$, so the denominator of $\kappa_P(t)$ is at most $(4.2556+3.9347)/\gamma$ and $\kappa_P(t)\ge0.6192\,\gamma t^2/8.1903\ge0.075\,\gamma t^2$. Hence $N\Hov\kappa_P(t)\ge\log(2/\delta)$ when $\gamma N\Hov\ge15\log(2/\delta)/t^2$.

\begin{lemma}[Characteristic function of a reversible chain]\label{lem:chf}
Assume \cref{asmp:mcmc}. Let $y:\Theta\to[0,1]$ be measurable with $V:=\Var_p(y)>0$, and let $\eta:=y-\EE_py$, $\sigma^2:=\sigma^2_P(y)$, $\rho:=V/\sigma^2$ and $a:=(1.64+2.58\rho)/\gamma$. For real $u$ let $E_u$ be multiplication by $e^{iu\eta/2}$ and $K(u):=E_uPE_u$. If $|u|\le\gs/16$, then $K(u)$ has exactly one eigenvalue $\lambda(u)$ with $|\lambda(u)-1|<\gamma/2$. This eigenvalue is simple, and for every $n\ge1$,
\[
\EE_p\exp\Big\{iu\sum_{j=1}^n\eta(\Xj j)\Big\}=\lambda(u)^{n-1}A(u)+B_n(u),
\]
where
\begin{enumerate}[label=(\roman*)]
\item $|\lambda(u)-1+\sigma^2u^2/2|\le a\,\sigma^2|u|^3$;
\item $|A(u)-1|\le16|u|/(7\gamma)+u^2V/4+u^4V^2/64$;
\item $|B_n(u)|\le\frac{16}{7\gs}\big(\frac2{\gs}+\frac12\big)\sqrt V\,|u|\,(1-\gs/2)^n$.
\end{enumerate}
\end{lemma}

\begin{proof}
Let $e:=e^{iu\eta/2}$ and $h:=e-1$, so that $|e|=1$, $\|E_u-I\|\le|u|/2$ and $\|h\|\le|u|\sqrt V/2$. By the Markov property, the characteristic function equals $\EE_p[e\,K(u)^{n-1}e]=\langle\bar e,K(u)^{n-1}e\rangle$.

\emph{Spectral separation.} We have $\|K-P\|\le\|(E_u-I)PE_u\|+\|P(E_u-I)\|\le|u|$. As $P$ is self-adjoint, $\|R(z,P)\|=1/d(z)$ with $d(z)=\operatorname{dist}(z,\operatorname{spec}P)$ and $\operatorname{spec}P\subset\{1\}\cup[-1+\gamma_-,1-\gamma]$. If $d(z)>|u|$, a Neumann series gives $\|R(z,K)\|\le1/\{d(z)-|u|\}$, so $\operatorname{spec}K\subset\{d\le|u|\}$. Let $\Gamma_1=\{|z-1|=\gamma/2\}$ and $\Gamma_2=\{|z|=1-\gs/2\}$. Then $d\ge\gamma/2$ on $\Gamma_1$ and $d\ge\gs/2$ on $\Gamma_2$, so $\|R(z,K)\|\le16/(7\gamma)$ on $\Gamma_1$ and $\|R(z,K)\|\le16/(7\gs)$ on $\Gamma_2$, and $\{d\le|u|\}$ lies inside $\Gamma_1\cup\Gamma_2$. Let $\Pi(u)$ and $Q(u)$ be the Riesz projections $(2\pi i)^{-1}\oint_{\Gamma_k}R(z,K)\dd z$. By the bound in (ii) below, which uses only the resolvent identity, $\|\Pi(u)-\Pi\|\le16|u|/(7\gamma)\le1/7<1$, so $\Pi(u)$ has the same rank as $\Pi$, namely one \citep[Section~I.4.6]{kato_perturbation_1995}; moreover $\Pi(u)+Q(u)=I$. Hence $\lambda(u)$ exists and is simple, $K^{n-1}=\lambda^{n-1}\Pi(u)+K^{n-1}Q(u)$ with $K^{n-1}Q(u)=(2\pi i)^{-1}\oint_{\Gamma_2}z^{n-1}R(z,K)\dd z$, and the decomposition holds with $A=\langle\bar e,\Pi(u)e\rangle$ and $B_n=\langle\bar e,K^{n-1}Q(u)e\rangle$.

(i) Let $D_\eta$ be multiplication by $\eta$. Since $K(u)=E_u(Pe^{iuD_\eta})E_u^{-1}$, $\lambda(u)=\Lambda(iu)$, where $\Lambda(r)$ is the eigenvalue near $1$ of $Pe^{rD_\eta}$. By the argument above with $\|Pe^{rD_\eta}-P\|\le e^{|r|}-1$, $\Lambda$ is holomorphic for complex $|r|<\log(1+\gamma/2)$, which exceeds $\gamma/5$, and its Taylor coefficients $\beta_k$ are those of Kato's series \citep[II.\S2 and VII.\S1]{kato_perturbation_1995}, whose validity on general state spaces is shown in the appendix of \citet{lezaud_chernoff_2001}. \citet[proof of Lemma~5.3]{paulin_concentration_2015} shows that $\beta_1=0$, that $\beta_2=\sigma^2/2$ (with the normalization $\langle\eta,Z\eta\rangle=(\sigma^2+V)/2$ above), that $\beta_3=\langle\eta,ZPD_\eta ZP\eta\rangle+\langle\eta,ZP\eta^2\rangle+\EE_p\eta^3/6$, and that $|\beta_k|\le(5/\gamma)^{k-2}(\sigma^2+0.8V)/2$ for $k\ge4$ (his (5.12); see the remark following this proof). In particular the Taylor series converges for $|r|<\gamma/5$. Now $|\langle\eta,ZPD_\eta ZP\eta\rangle|=|\EE_p[\eta(PZ\eta)^2]|\le\|PZ\eta\|^2\le\gamma^{-1}\langle\eta,Z\eta\rangle=(\sigma^2+V)/(2\gamma)$, because $\lambda^2/(1-\lambda)^2\le\gamma^{-1}(1-\lambda)^{-1}$ on $[-1,1-\gamma]$. Next, $|\langle\eta,ZP\eta^2\rangle|=|\langle PZ\eta,\eta^2-V\rangle|\le\gamma^{-1}\sqrt V\cdot\sqrt V$, since $|\lambda|/(1-\lambda)\le\gamma^{-1}$ on $[-1,1-\gamma]$ and $\Var_p(\eta^2)\le\EE_p\eta^4\le V$. Finally $|\EE_p\eta^3|\le V$. Hence $|\beta_3|\le\sigma^2/(2\gamma)+3V/(2\gamma)+V/6$. For $|u|\le\gamma/16$,
\[
\sum_{k\ge4}|\beta_k||u|^k\le\frac{\sigma^2+0.8V}{2}\cdot\frac{25u^4/\gamma^2}{1-5/16}\le\frac{1.137}{\gamma}(\sigma^2+0.8V)|u|^3 .
\]
Adding up, and using $\gamma\le1$, gives $|\lambda-1+\sigma^2u^2/2|\le\gamma^{-1}(1.637\sigma^2+2.576V)|u|^3\le a\sigma^2|u|^3$.

(ii) Since $\Pi e=(\EE_pe)1$ and $\langle\bar e,1\rangle=\EE_pe$, we have $A-1=\langle\bar e,\{\Pi(u)-\Pi\}e\rangle+\{(\EE_pe)^2-1\}$. The resolvent identity $R(z,K)-R(z,P)=R(z,K)(K-P)R(z,P)$ gives $\|\Pi(u)-\Pi\|\le\frac\gamma2\cdot\frac{16}{7\gamma}\cdot|u|\cdot\frac2\gamma=\frac{16|u|}{7\gamma}$. Moreover $\EE_pe=1+\varepsilon_0$ with $|\varepsilon_0|=|\EE_p(e^{iu\eta/2}-1-iu\eta/2)|\le u^2V/8$, so $|(\EE_pe)^2-1|\le u^2V/4+u^4V^2/64$.

(iii) Write $e=1+h$. Since $R(z,K)1=(z-1)^{-1}\{1+R(z,K)(K-P)1\}$, and $z\mapsto z^{n-1}/(z-1)$ is holomorphic inside $\Gamma_2$,
\[
K^{n-1}Q(u)1=\frac1{2\pi i}\oint_{\Gamma_2}z^{n-1}R(z,K)(K-P)1\,\frac{\dd z}{z-1}.
\]
Here $(K-P)1=h+E_uPh$, so $\|(K-P)1\|\le|u|\sqrt V$, and $|z-1|\ge\gs/2$ on $\Gamma_2$. Hence $\|K^{n-1}Q(u)1\|\le(1-\gs/2)^n\frac{16}{7\gs}|u|\sqrt V\frac2{\gs}$ and $\|K^{n-1}Q(u)h\|\le(1-\gs/2)^n\frac{16}{7\gs}\frac{|u|\sqrt V}2$. Since $\|\bar e\|=1$, (iii) follows.
\end{proof}

\paragraph{Remark on (5.12) of \citet{paulin_concentration_2015}.} Paulin's derivation of (5.12) contains two misprints: $\langle f,Zf\rangle$ equals $(\sigma^2+V)/2$, and each term with $p=n$ in his expansion (5.11) is bounded by $\langle f,Zf\rangle/\gamma^{n-2}$; moreover the terms with $p<n$ contribute at most $D'(n)V/(2\gamma^{n-2})$ with $D'(n)=\sum_{p<n}p^{-1}\binom{n-1}{p-1}\binom{2p-2}{p-1}$. Recounting gives $|\beta_n|\le\{\mathrm{Cat}_{n-1}\sigma^2+D(n)V\}/(2\gamma^{n-2})$, where $\mathrm{Cat}_{n-1}=n^{-1}\binom{2n-2}{n-1}$ and $D(n)=D'(n)+\mathrm{Cat}_{n-1}$. Since $\mathrm{Cat}_{n-1}\le0.2\cdot5^{n-2}$ and $D(n)\le0.6\cdot5^{n-2}$ for $n\ge4$, (5.12) holds as stated. 

\begin{lemma}[Berry--Esseen bound for an independent part plus a reversible-chain part]\label{lem:be-chain}
Assume \cref{asmp:mcmc}, and let $y,\eta,V,\sigma^2,\rho,a$ be as in \cref{lem:chf} and $n\ge2$. Let $x_1,\dots,x_m$ ($m\ge0$) be i.i.d.\ $[0,1]$-valued random variables with variance $v$, independent of the chain. Let $S:=\sum_{i\le m}(x_i-\EE x_i)+\sum_{j\le n}\eta(\Xj j)$ and $s^2:=mv+n\sigma^2$. Then
\[
\sup_x|\PP(S\le xs)-\Phi(x)|\le\frac{C_1}{s}+\frac1{7\pi}\Big(\frac2{\gs}+\frac12\Big)\Big(1-\frac{\gs}2\Big)^n,\qquad C_1:=\frac{3.05}{\kappa}+2.27a+\frac{2.59}{\gamma}+0.015,
\]
where $\kappa:=\min\{\gs/16,\,1/(4a)\}$.
\end{lemma}

\begin{proof}
Esseen's smoothing inequality \citep[XVI.3, Lemma~2]{feller_introduction_1971} with $T=\kappa s$ gives
\[
\sup_x|\PP(S\le xs)-\Phi(x)|\le\frac1\pi\int_{-\kappa}^{\kappa}\big|\varphi(u)-e^{-s^2u^2/2}\big|\frac{\dd u}{|u|}+\frac{24}{\pi\sqrt{2\pi}\,\kappa s},
\]
where $\varphi$ is the characteristic function of $S$, and $24/(\pi\sqrt{2\pi})\le3.05$. Let $\psi$ be the characteristic function of $x_1-\EE x_1$, so that $\varphi(u)=\psi(u)^m\{\lambda(u)^{n-1}A(u)+B_n(u)\}$ for $|u|\le\kappa$ by \cref{lem:chf}. Since $\EE|x_1-\EE x_1|^3\le v$, $|\psi(u)-1+vu^2/2|\le v|u|^3/6$ and $|\psi(u)-e^{-vu^2/2}|\le v|u|^3/6+v^2u^4/8$. Likewise \cref{lem:chf}(i) gives $|\lambda(u)-e^{-\sigma^2u^2/2}|\le a\sigma^2|u|^3+\sigma^4u^4/8$. Moreover $v\le1/4$ and $\sigma^2\le\cplus V\le\cplus/4<1/(2\gamma)$. For $|u|\le\kappa$ we have $a|u|\le1/4$ and $a\ge1/6$, hence $\max\{|\psi(u)|,e^{-vu^2/2}\}\le e^{-vu^2/4}$ and $\max\{|\lambda(u)|,e^{-\sigma^2u^2/2}\}\le e^{-\sigma^2u^2/4}$. Telescoping,
\[
\big|\psi^m\lambda^{n-1}-e^{-s^2u^2/2}\big|\le F e^{-s^2u^2/4}\Big\{\Big(\frac{mv}6+na\sigma^2\Big)|u|^3+\frac{(mv^2+n\sigma^4)u^4}8\Big\}+\frac{\sigma^2u^2}2e^{-(s^2-\sigma^2)u^2/2},
\]
where $F:=\exp\{(v+2\sigma^2)\kappa^2/4\}\le1.003$ accounts for the missing factors. We use $mv/6+na\sigma^2\le as^2$ and $mv^2+n\sigma^4\le(\cplus/4)s^2$. With $\int_\RR u^2e^{-s^2u^2/4}\dd u=4\sqrt\pi/s^3$, $\int_\RR|u|^3e^{-s^2u^2/4}\dd u=16/s^4$ and $\int_\RR|u|e^{-bu^2}\dd u=1/b$, these terms contribute at most $2.27a/s+F(\cplus/2\pi)/s^2+\sigma^2/\{\pi(s^2-\sigma^2)\}$, and the last term is at most $2\sigma^2/(\pi s^2)\le\cplus/(2\pi s^2)$ because $n\ge2$. By \cref{lem:chf}(ii), and since $|\psi|^m|\lambda|^{n-1}\le Fe^{-s^2u^2/4}$,
\[
\frac1\pi\int|\psi|^m|\lambda|^{n-1}|A-1|\frac{\dd u}{|u|}\le\frac F\pi\int_\RR e^{-s^2u^2/4}\Big(\frac{16}{7\gamma}+\frac{|u|V}4+\frac{|u|^3V^2}{64}\Big)\dd u\le\frac{2.59}{\gamma s}+\frac{0.08}{s^2}+\frac{0.005}{s^4}.
\]
By \cref{lem:chf}(iii), $\kappa\le\gs/16$ and $\sqrt V\le1/2$, the contribution of $B_n$ is at most $\frac{2\kappa}\pi\cdot\frac{16}{7\gs}(\frac2{\gs}+\frac12)\frac12(1-\frac{\gs}2)^n\le\frac1{7\pi}(\frac2{\gs}+\frac12)(1-\frac{\gs}2)^n$. Collecting terms, the bound holds with $3.05/\kappa+2.27a+2.59/\gamma$ in place of $C_1$ plus a remainder of at most $(0.32\cplus+0.085)/s^2$ for $s\ge1$. If $s<3.05/\kappa$, the claimed bound exceeds one. Otherwise $s\ge3.05\cdot16/\gs\ge24.4\cplus$ (as $\cplus<2/\gamma\le2/\gs$) and $s\ge48.8$, and the remainder is at most $(0.32/24.4+0.085/48.8)/s\le0.015/s$.
\end{proof}

\begin{proof}[Proof of \cref{thm:mcmc-BE}]
Let $F(t)=\PP(\Delta\le t)=\PP(W_t\le0)$ (\cref{thm:concentration}(a)), let $s(t)$ be as in \cref{lem:mcmc-basic}(b), and let $\tau:=1/(4+4\zeta_P)$. If $\Var_p(\pistar)=0$, the posterior draws contribute a constant to $W_t$, and the argument below applies with \cref{lem:be-chain} replaced by the Berry--Esseen bound for independent summands.

\emph{Central region $|t|\le\tau$.} $W_t-\mu(t)$ has the form of $S$ in \cref{lem:be-chain}, with $x_i=1-\pi_t(\Yi i)$ and $y=1-\pi_t$. Its parameter $\rho=\Var_p(\pi_t)/\sigma^2_P(\pi_t)$ is at most $1/\cminus$ by \cref{lem:mcmc-basic}(a), and $C_1$ is increasing in $\rho$. With $a(t):=-\mu(t)/s(t)$ and $s(t)\ge e^{-\zeta_P\tau}\sqrt{v_P}$,
\[
|F(t)-\Phi(a(t))|\le e^{\zeta_P\tau}\,\frac{C_1(1/\cminus)}{\sqrt{v_P}}+\frac{0.12}{\gs}e^{-\gs\Sone/2},
\]
where we used $(1-\gs/2)^{\Sone}\le e^{-\gs\Sone/2}$ and $\gs\le1$. Write $y=t/\sigNP$ and $a(t)=\varrho(t)y$. By \cref{lem:meanvar}(a) and \cref{lem:mcmc-basic}(b), $\varrho(t)=\{-\mu(t)/(N\Hov t)\}\{s(0)/s(t)\}\in[e^{-(1+\zeta_P)|t|},e^{(1+\zeta_P)|t|}]$. As in the proof of \cref{thm:BE}, the mean value theorem and $|t|=\sigNP|y|$ give
\[
|\Phi(a(t))-\Phi(y)|\le(1+\zeta_P)e^{3(1+\zeta_P)\tau}\sigNP\sup_ww^2\phi(w)=0.2935\,e^{3/4}(1+\zeta_P)\sigNP\le0.63(1+\zeta_P)\sigNP .
\]

\emph{Tails $|t|>\tau$.} For $t>\tau$, both $1-F(t)$ and $1-\Phi(t/\sigNP)$ lie in $[0,\max\{1-F(\tau),1-\Phi(\tau/\sigNP)\}]$. By \cref{thm:mcmc-conc}(a), $1-F(\tau)\le e^{-N\Hov\kappa_P(\tau)}$. By \cref{lem:mcmc-basic}(c), $\sigNP^2=v_P/(N\Hov)^2\le\cplus/(N\Hov)$, and $\kappa_P(\tau)\le\tau^2/(2\cplus)$, so $1-\Phi(\tau/\sigNP)\le e^{-\tau^2/(2\sigNP^2)}\le e^{-N\Hov\kappa_P(\tau)}$. The case $t<-\tau$ is symmetric.

Combining both regions and using $e^{\zeta_P\tau}\le e^{1/4}\le1.29$ proves the theorem with
\begin{equation}\label{eq:CP}
C_P:=1.29\Big\{\frac{3.05}{\kappa}+2.27a+\frac{2.59}{\gamma}+0.015\Big\},\qquad a:=\frac{1.64+2.58/\cminus}{\gamma},\qquad\kappa:=\min\Big\{\frac{\gs}{16},\frac1{4a}\Big\}.
\end{equation}
For a positive chain, $\cminus\ge1$ and $\gs=\gamma$, so $a\le4.22/\gamma$ and $3.05/\kappa=\max(48.8/\gamma,12.2a)\le51.5/\gamma$. Hence $C_P\le1.29(51.5+9.6+2.6+0.015)/\gamma\le83/\gamma$.
\end{proof}

\begin{proof}[Proof of \cref{thm:mcmc-sharp}]
The chain is reversible with respect to $(1-\varepsilon,\varepsilon)$, and its eigenvalues are $1$ and $1-\gamma\in[0,1)$. So it is positive, with right and absolute spectral gap $\gamma$ and $\gamma_-=2-\gamma$. Since $L^2_0(p)$ is one-dimensional and spanned by an eigenfunction, $\tau_P(u)=(2-\gamma)/\gamma$ for every non-constant $u$. With $\pistar=0,\tfrac12,1$ on $\{0,1,2\}$, $\Var_p(\pistar)=\Var_g(\pistar)=\varepsilon(1-\varepsilon)/4$. Hence $v_P=S\varepsilon(1-\varepsilon)\{1+(2-\gamma)/\gamma\}/4=S\varepsilon(1-\varepsilon)/(2\gamma)=\neff/\gamma$, using $\neff=S\varepsilon(1-\varepsilon)/2$ from \cref{thm:sharp}.
(a) If the chain never visits $1$, then $\Zhat\in\{0,\infty\}$, and this has probability $(1-\varepsilon)(1-\varepsilon\gamma)^{S-1}\ge(1-\varepsilon)\exp\{-\varepsilon\gamma(S-1)/(1-\varepsilon\gamma)\}\ge\frac12e^{-2\varepsilon\gamma S}=\frac12e^{-4\gamma N\Hov}$. Here we used $\log(1-x)\ge-x/(1-x)$, $\varepsilon\le1/2$ and $N\Hov=S\varepsilon/2$.
(b) If $n_{\mathrm{post}}=0<n_{\mathrm{prop}}$, then $\chat=+\infty$ (proof of \cref{thm:sharp}). So $\PP(\Delta\le x\sigNP)\le1-\PP(n_{\mathrm{post}}=0)\PP(n_{\mathrm{prop}}\ge1)$ for every $x$, while $\Phi(x)\to1$. With $x_0:=\varepsilon\gamma\le1/4$ and $S=\lceil1/x_0\rceil$, $\PP(n_{\mathrm{post}}=0)=(1-\varepsilon)(1-x_0)^{S-1}\ge\frac34(1-x_0)^{1/x_0}\ge(\frac34)^5$, because $x\mapsto(1-x)^{1/x}$ is decreasing. Also $\PP(n_{\mathrm{prop}}\ge1)=1-(1-\varepsilon)^S\ge1-e^{-\varepsilon S}\ge1-e^{-1/\gamma}\ge1-e^{-1}$. The Kolmogorov distance is therefore at least $(3/4)^5(1-e^{-1})\ge0.15$. Finally $v_P=S\varepsilon(1-\varepsilon)/(2\gamma)\ge(1-\varepsilon)/(2\gamma^2)\ge3/(8\gamma^2)$, so $0.15\le C/\sqrt{v_P}$ forces $C\ge0.15\sqrt{3/8}/\gamma\ge0.09/\gamma$.
\end{proof}

\begin{proof}[Proof of \cref{thm:mcmc-mcse}]
Let $\widehat n_P(c)=\Stwo\widehat V_{\mathrm{prop}}(c)+\Sone\widehat\omega^2$, so that, as in \cref{eq:ratiodecomp}, $\sighatP/\sigNP=\{\widehat n_P(\chat)/v_P\}^{1/2}\Hov/\Hzero(\chat)$. Consider four events.
\emph{E1}: $|\Delta|\le t_1:=2.9\sqrt{\ell_P}$. For $t\le t_1\le0.3625$ we have $(1-e^{-t})/t\ge0.8386$, $\cosh^2(t/2)\le1.0333$ and $b_P\le5\cplus$. Using $\sqrt{\ell_P}\le1/8$,
\[
N\Hov\kappa_P(t_1)\ge\frac{0.7033\cdot8.41\,\ell_PN\Hov}{2.0666\cplus+10\cplus\cdot2.9/8}=\frac{5.915}{5.692}\,L\ge L,
\]
so $\PP(E1^c)\le2e^{-L}=\delta/6$ by \cref{thm:mcmc-conc}(a).
\emph{E2}: $|\Hzero(\cstar)/\Hov-1|\le\eta:=2.2\sqrt{\ell_P}\le0.275$. Here $N\Hzero(\cstar)=\sum_j\pistar(\Xj j)$ has mean $N\Hov$, and $\Sone\Var_p(\pistar)\le\Sone\EE_p\pistar=N\Hov$. The moment generating function bound of the proof of \cref{thm:mcmc-conc}(a), applied to $\pm(\pistar-\EE_p\pistar)$, gives
\[
\PP(E2^c)\le2\exp\Big\{-\frac{\eta^2N\Hov}{2(\cplus+b_P\eta)}\Big\}\le2\exp\Big\{-\frac{4.84\,L}{2(1+5\cdot0.275)}\Big\}\le\delta/6.
\]
\emph{E3}: $|\{\Stwo\widehat V_{\mathrm{prop}}(\cstar)\}^{1/2}-\{\Stwo\Var_g(\pistar)\}^{1/2}|\le2\sqrt L$; by the argument for E3 in the proof of \cref{thm:mcse}, $\PP(E3^c)\le\delta/6$.
\emph{E4}: $|\widehat\omega^2/\sigma^2_P(\pistar)-1|\le\varepsilon_\omega$; $\PP(E4^c)\le\delta_\omega$ by assumption.
On $E1\cap E2\cap E3\cap E4$, whose probability is at least $1-\delta/2-\delta_\omega$: by \cref{eq:varlip}, $\widehat n_P(\chat)/\widehat n_P(\cstar)\in[e^{-2|\Delta|},e^{2|\Delta|}]$, because only the proposal term depends on $c$; and by \cref{eq:ratio}, $\Hzero(\chat)/\Hzero(\cstar)\in[e^{-|\Delta|},e^{|\Delta|}]$. By E3, E4, $|\sqrt{1+x}-1|\le|x|$ for $|x|\le1$, and the triangle inequality in $\RR^2$, $|\widehat n_P(\cstar)^{1/2}-v_P^{1/2}|\le2\sqrt L+\varepsilon_\omega\{\Sone\sigma^2_P(\pistar)\}^{1/2}\le r_Pv_P^{1/2}$. Hence, using $-\log(1-x)\le1.39x$ for $x\le1/2$ and $-\log(1-\eta)\le1.17\eta$ for $\eta\le0.275$,
\[
\Big|\log\frac{\sighatP}{\sigNP}\Big|\le2|\Delta|-\log(1-r_P)-\log(1-\eta)\le(5.8+2.58)\sqrt{\ell_P}+1.39\,r_P\le8.4\sqrt{\ell_P}+1.4\,r_P .
\]
For coverage, on this event $\sighatP\in\sigNP[e^{-r},e^{r}]$. By \cref{thm:mcmc-BE}, applied at $\pm x\sigNP$ (the bound also holds for left limits), $|\PP(|\Delta|\le x\sigNP)-(2\Phi(x)-1)|\le2B_P$ for every $x$. The claim follows as in the proof of \cref{thm:mcse}.
\end{proof}

\section{Proofs for Sections~\ref{sec:orth} and~\ref{sec:diagnostics}}\label{app:proofs-orth}

\begin{proof}[Proof of \cref{prop:orth}]
(a) Both terms equal $\int fg_\vartheta h_{\zeta,\vartheta}$, which is finite because $fg_\vartheta h_{\zeta,\vartheta}=g_\vartheta\,f/(\sone f+\stwo\zeta g_\vartheta)\le g_\vartheta/\sone$. (b) Divide (a) by $\EE_p[gh_\zeta]=\Z^{-1}\int fgh_\zeta>0$. (c) $\EE_g[fh_z]=K(z)$ and $z\EE_p[gh_z]=(z/\Z)K(z)$.
\end{proof}

\begin{proof}[Proof of \cref{thm:iteration}]
(a) By \cref{lem:esteq}(a),
\[
n(e^x)=(\sone\Stwo)^{-1}\textstyle\sum_i\{1-\pi_x(\Yi i)\},\qquad d(e^x)=(\stwo\Sone)^{-1}e^{-x}\sum_j\pi_x(\Xj j).
\] Differentiating the logarithms with $\partial_x\pi_x=\pi_x(1-\pi_x)$ gives $\partial_x\log n=-\mathsf B$ and $\partial_x\log d=-1+\sum_j\pi_x(1-\pi_x)/\sum_j\pi_x=-\mathsf A$. The population versions of the two weightings are proportional to $p\,gh_z$ and $g\,fh_z$, which coincide by \cref{prop:orth}.
(b) $\mathsf A$ and $\mathsf B$ are averages, with positive weights on draws in $A$, of numbers in $(0,1)$, so $|\varphi'|<1$ pointwise. As $x\to-\infty$ both tend to $0$, and as $x\to+\infty$ both tend to $1$ (\cref{lem:esteq}(c)). Hence $\varphi'(x)\to0$ at $\pm\infty$, and the continuous function $|\varphi'|$ has supremum $L<1$. Fixed points of $\varphi$ are roots of $W$ by \cref{lem:esteq}(a). The Banach fixed-point theorem gives global geometric convergence, and $|x_t-\chat|\le|x_t-x_{t+1}|+|\varphi(x_t)-\varphi(\chat)|\le|x_{t+1}-x_t|+L|x_t-\chat|$ gives the a posteriori bound.
(c) At the root, $\sum_j\widehat\pi_j=\sum_i(1-\widehat\pi_i)=N\Hzero$. Hence $\mathsf A-\mathsf B=\{\sum_j\widehat\pi_j^2-\sum_i\widehat\pi_i(1-\widehat\pi_i)\}/(N\Hzero)=\{N\Hzero-N\Hhat\}/(N\Hzero)$. The last expression in (c) is the same numerator written with labels.
(d) By (c) and \cref{eq:ratio},
\[
1-T'(\Zhat)=\frac{\Hhat(\chat)}{\Hzero(\chat)}\in\frac{\Hhat(\cstar)}{\Hzero(\cstar)}\,\big[e^{-2|\Delta|},e^{2|\Delta|}\big].
\] Let $\ell'=\log(6/\delta)/(N\Hov)\le1/64$. As in the proof of \cref{thm:mcse}, with probability at least $1-2\delta/3$ we have $|\Delta|\le1.642\sqrt{\ell'}$ and $|\Hzero(\cstar)/\Hov-1|\le1.4606\sqrt{\ell'}$. Moreover, with probability at least $1-\delta/3$, $|\Hhat(\cstar)/\Hov-1|\le0.7303\sqrt{\ell'}$, by Bernstein's inequality for $\sum_{\text{all}}\pistar(1-\pistar)$, whose summands lie in $[0,1/4]$ and whose variance is at most $N\Hov/4$. Combining, $|\log(1-T')|\le(3.284+1.787+0.804)\sqrt{\ell'}\le5.9\sqrt{\ell'}$.
\end{proof}

\begin{proof}[Proof of \cref{thm:proposal}]
Write $F_N(c,\vartheta)=N^{-1}W_N(c,\vartheta)=\stwo\mathbb P^\epsilon_{\Stwo}[1-\pi_{c,\vartheta}\circ\tau_\vartheta]-\sone\mathbb P_{\Sone}[\pi_{c,\vartheta}]$, where $\mathbb P^\epsilon_{\Stwo}$ and $\mathbb P_{\Sone}$ are the empirical measures of the base variates and of the posterior draws. Let $F$ be its expectation for fixed $(c,\vartheta)$. By \cref{prop:orth}, $F(\cstar,\vartheta)=0$ for all $\vartheta$, and by \cref{lem:meanvar}(b), $\partial_cF(\cstar,\vartheta)=-\Hov(\vartheta)$. Since the logistic function is $1/4$-Lipschitz and $\pi_{c,\vartheta}=\sigma(c+\log\{\stwo g_\vartheta/(\sone f)\})$, (P1) implies that $\{\pi_{c,\vartheta}\}$ and $\{\pi_{c,\vartheta}\circ\tau_\vartheta\}$, indexed by a compact neighbourhood of $(\cstar,\vartheta_0)$, are Lipschitz in the parameter with square-integrable Lipschitz constants. They are therefore Glivenko--Cantelli and Donsker \citep[Example~19.7]{vandervaart_asymptotic_1998}. Uniform convergence of $F_N$ and monotonicity in $c$ give $\chat(\widehat\vartheta)\xrightarrow{\,\Pr\,}\cstar$. Asymptotic equicontinuity \citep[Lemma~19.24]{vandervaart_asymptotic_1998} then gives
\[
0=F_N(\chat,\widehat\vartheta)=(F_N-F)(\cstar,\vartheta_0)+o_\PP(N^{-1/2})+F(\cstar,\widehat\vartheta)-\{\Hov(\vartheta_0)+o_\PP(1)\}(\chat-\cstar),
\]
where we used $F(\cstar,\widehat\vartheta)=0$, (P2), and continuity of $\partial_cF$. Hence $\chat(\widehat\vartheta)-\cstar=(F_N-F)(\cstar,\vartheta_0)/\Hov(\vartheta_0)+o_\PP(N^{-1/2})$. The same representation holds for $\chat(\vartheta_0)$, which proves the first claim. The remaining claims follow from the central limit theorem for the two independent averages in $(F_N-F)(\cstar,\vartheta_0)$, whose variance is $\sigN^2(\vartheta_0)\Hov(\vartheta_0)^2$, and from the continuous mapping theorem applied to $\Hzero$ and $\neffhat/N$.
\end{proof}

\begin{proof}[Proof of \cref{lem:block}]
$W_{(-g)}=W-W_g$ and $W(\chat)=0$, so $W_{(-g)}(\chat)=-W_g(\chat)$. The mean value theorem applied to $W_{(-g)}$, whose derivative is $-\widehat{\mathcal I}_{(-g)}$, between $\chat$ and its root $\chat_{(-g)}$ gives the identity. Finally, $\sum_gW_g=W$.
\end{proof}

\begin{proof}[Proof of \cref{thm:jackknife}]
Let $\bar W=G^{-1}\sum_gW_g$. By the ergodic theorem on each of the $G$ blocks, and equicontinuity in $c$ of $\pi_c(1-\pi_c)$, both $\widehat{\mathcal I}_{(-g)}(c)/N\to(1-G^{-1})\Hov_c$ and $\widehat{\mathcal I}_g(c)/N\to G^{-1}\Hov_c$ uniformly near $\cstar$, where $\Hov_{\cstar}=\Hov$. All $G+1$ estimators are consistent. From $W(\chat)=0$ we get $\chat-\cstar=W(\cstar)/\{N\Hov(1+o_\PP(1))\}$, and the mean value theorem then gives $W_g(\chat)=W_g(\cstar)-\bar W(\cstar)+o_\PP(\sqrt N)$. By \cref{lem:block}, jointly in $g$,
\[
\chat_{(-g)}-\chat=-\frac{G}{G-1}\,\frac{W_g(\cstar)-\bar W(\cstar)}{N\Hov}+o_\PP(N^{-1/2}).
\]
Since $\EE W_g(\cstar)=0$, the proposal part is i.i.d., and the posterior part satisfies the functional CLT independently of it, we have $(N^{-1/2}W_g(\cstar))_{g\le G}\Rightarrow(v\zeta_g/\sqrt G)_{g\le G}$ with $v^2=\stwo\Var_g\pistar+\sone\omega^2$ and $\zeta_g$ i.i.d.\ $\mathcal N(0,1)$. Note that $\sigma_{\mathrm{as}}^2=v^2/(N\Hov^2)$. Substituting, $\JackVar=\sigma_{\mathrm{as}}^2\sum_g(\zeta_g-\bar\zeta)^2/(G-1)+o_\PP(N^{-1})$ and $\max_g|\chat_{(-g)}-\chat|/\sigma_{\mathrm{as}}\Rightarrow\frac{\sqrt G}{G-1}\max_g|\zeta_g-\bar\zeta|$. Finally, $R_g-\bar R=-(G-1)(\chat_{(-g)}-\bar c_{(-)})$ is a common positive multiple of $\zeta_g-\bar\zeta$ up to $o_\PP(N^{-1/2})$, so the studentized maximum converges to the Grubbs statistic. Slutsky's theorem with $\sighat/\sigma_{\mathrm{as}}\to1$ completes the proof. The marginal claim in (b) follows from $\Var(\zeta_g-\bar\zeta)=(G-1)/G$.
\end{proof}

\begin{proof}[Proof of \cref{prop:onesided}]
(a) For numbers $x_1,\dots,x_S\in[0,1]$ with mean $\bar x$, $\sum_k(x_k-\bar x)^2=S\bar x(1-\bar x)-\sum_kx_k(1-x_k)\le S\bar x$. Applied to the posterior responsibilities this gives $\Sone\widehat V_{\mathrm{post}}\le\frac{\Sone}{\Sone-1}\sum_j\widehat\pi_j=\frac{\Sone}{\Sone-1}N\Hzero$, and applied to $\{1-\widehat\pi_i\}$ it gives $\Stwo\widehat V_{\mathrm{prop}}\le\frac{\Stwo}{\Stwo-1}N\Hzero$; adding proves the first inequality. If $\chat$ is finite, $N\Hzero=\mathsf a(\chat)=\mathsf b(\chat)$, where $\mathsf a(c)=\sum_j\pi_c(\Xj j)$ is non-decreasing and $\mathsf b(c)=\sum_i\{1-\pi_c(\Yi i)\}$ is non-increasing in $c$ (\cref{lem:esteq}). Hence $N\Hzero\le\mathsf a(\cstar)=M_{\mathrm{post}}$ if $\chat\le\cstar$ and $N\Hzero\le\mathsf b(\cstar)=M_{\mathrm{prop}}$ if $\chat\ge\cstar$. If $\chat=\pm\infty$, one of the samples misses $A$ (\cref{lem:esteq}(c)); with responsibilities at $\chat$ defined as limits, every posterior responsibility is then $0$ and every proposal responsibility is $1$, so $\neffhat=N\Hzero=0$.

(b) $M_{\mathrm{post}}$ and $M_{\mathrm{prop}}$ are sums of independent $[0,1]$-valued variables with means $\Sone\EE_p\pistar=N\Hov$ and $\Stwo\EE_g(1-\pistar)=N\Hov$ (\cref{lem:overlap}(a)). Let $S$ be such a sum with mean $\mu$. For $u>0$, $e^{ux}\le1+x(e^u-1)$ on $[0,1]$ gives $\EE e^{uS}\le\exp\{\mu(e^u-1)\}$, and Markov's inequality with $u=\log(x/\mu)$ gives $\PP(S\ge x)\le e^{-\mu}(e\mu/x)^x$ for $x\ge\mu$. The first claim follows from (a) and a union bound over the two sums. For the second, $\Var(S)\le\EE S=\mu$, and Bernstein's inequality gives $\PP(S\ge\mu+y)\le\exp\{-y^2/(2\mu+2y/3)\}\le\delta/2$ for $y=\sqrt{2\mu L}+2L/3$ with $L=\log(2/\delta)$. Since $\mu+\sqrt{2\mu L}+2L/3\le(\sqrt\mu+\sqrt{2L/3})^2$, the claim follows from (a) and a union bound.

(c) $N\Hov=\neff/\beta$ and $\beta=1-\Iov\ge1/2$. For the numerical claim, $\varsigma\le200/99$, so $x:=10/\varsigma\ge4.95$. The bound in (b) increases in $\mu=N\Hov$ on $(0,x]$ and decreases in $x$ on $[\mu,\infty)$, so it is at most $2e^{-1}(e/4.95)^{4.95}<0.04$.
\end{proof}

\begin{proof}[Proof of \cref{thm:calib-null}]
Write $\pi=\pistar$, $b:=\EE_{\mix_\star}[\pi]$, $\Hzero(c)=N^{-1}\sum_j\pi_c(\Xj j)$ and $\Hhat(c)=N^{-1}\sum_{\text{all}}\pi_c(1-\pi_c)$, so that $\Tp=1-\Hhat(\chat)/\Hzero(\chat)$. We use $\sone p=(1-\pi)\mix$ and $\stwo g=\pi\mix$ throughout. Let $A_1:=\sone\EE_p[\pi(1-\pi)]$ and $B_1:=\stwo\EE_g[\pi(1-\pi)]=\EE_\mix[\pi^2(1-\pi)]$, so that $A_1+B_1=\Hov$ and $b=B_1/\Hov$. Since $p$, $g$, $\sone$, $\stwo$ are fixed, $\chat-\cstar=W(\cstar)/(N\Hov)+O_\PP(N^{-1})$, and $\Hzero(\cstar)$, $\Hhat(\cstar)$ are $\Hov+O_\PP(N^{-1/2})$. The derivatives in $c$ of $\pi_c$ and $\pi_c(1-\pi_c)$ are $\pi_c(1-\pi_c)$ and $\pi_c(1-\pi_c)(1-2\pi_c)$, with population means $A_1$ and $A_1-B_1$ for the two averages, and all second derivatives are bounded. A Taylor expansion therefore gives
\[
\Tp=\frac{\Hzero(\cstar)-\Hhat(\cstar)+(\chat-\cstar)B_1}{\Hov}+O_\PP(N^{-1}).
\]
Substituting $N\{\Hzero(\cstar)-\Hhat(\cstar)\}=\sum_j\pi^2(\Xj j)-\sum_i\pi(1-\pi)(\Yi i)$ and the expansion of $\chat-\cstar$,
\begin{equation}\label{eq:T-linear}
N\Hov\,\Tp=\sum_{j=1}^{\Sone}\phi_{\mathrm{post}}(\Xj j)+\sum_{i=1}^{\Stwo}\phi_{\mathrm{prop}}(\Yi i)+O_\PP(1),\qquad \phi_{\mathrm{post}}=\pi(\pi-b),\quad\phi_{\mathrm{prop}}=-(1-\pi)(\pi-b).
\end{equation}
Both functions have mean zero: $\sone\EE_p\phi_{\mathrm{post}}=\EE_\mix[\pi(1-\pi)(\pi-b)]=0$ and $\stwo\EE_g\phi_{\mathrm{prop}}=-\EE_\mix[\pi(1-\pi)(\pi-b)]=0$. Their total variance is $\Sone\EE_p\phi_{\mathrm{post}}^2+\Stwo\EE_g\phi_{\mathrm{prop}}^2=N\EE_\mix[\pi(1-\pi)(\pi-b)^2]=N\Hov\Var_{\mix_\star}(\pi)$, and (a) follows from the central limit theorem.

(b) Deleting block $g$ leaves $\sone$ and $\stwo$ unchanged, so \cref{eq:T-linear} holds for $\Tp_{(-g)}$ with $N(1-G^{-1})$ in place of $N$ and the sums restricted to the other blocks. Let $U_g$ be the contribution of block $g$ to the right-hand side of \cref{eq:T-linear} and $\bar U=G^{-1}\sum_gU_g$. Then $\Tp=G\bar U/(N\Hov)+O_\PP(N^{-1})$ and $\Tp_{(-g)}-\bar\Tp_{(-)}=-(U_g-\bar U)/\{N(1-G^{-1})\Hov\}+O_\PP(N^{-1})$. Hence $\widehat{\mathrm{se}}_{\mathrm{jack}}^2(\Tp)=G\,s_U^2/(N\Hov)^2+O_\PP(N^{-3/2})$ with $s_U^2=\sum_g(U_g-\bar U)^2/(G-1)$, and
\[
t_\Tp=\frac{\sqrt G\,\bar U}{s_U}+o_\PP(1).
\]
The $U_g$ are i.i.d.\ with mean zero and variance $N\Hov\Var_{\mix_\star}(\pi)/G>0$, so $(U_g)_g$, suitably scaled, converges to i.i.d.\ standard normals, and the continuous mapping theorem gives $t_{G-1}$. For Markov chain output, the same expansion holds with the ergodic theorem in place of the law of large numbers, and the functional CLT for $(\pistar,{\pistar}^2)$ makes the scaled block sums asymptotically i.i.d.\ normal, which is all that the last step uses.
\end{proof}

\begin{proof}[Proof of \cref{prop:calib-alt}]
(a) $N^{-1}W(c)$ is an average of functions that are bounded and monotone in $c$, so it converges to $\widetilde F(c)$ almost surely, uniformly in $c$. $\widetilde F$ is continuous and strictly decreasing, because $\tilde p(A)>0$ and $g(A)>0$, with $\widetilde F(-\infty)>0>\widetilde F(\infty)$. Hence $\chat\to\tilde c$, and by uniform convergence $\Hzero(\chat)\to\widetilde H_0>0$ and $\Hhat(\chat)\to\sone\EE_{\tilde p}[\tilde\pi(1-\tilde\pi)]+\stwo\EE_g[\tilde\pi(1-\tilde\pi)]$, which gives the first expression for $\Tp_\infty$. Under the pooled law, $\EE[L-\tilde\pi]=\widetilde F(\tilde c)=0$ and $\EE[(L-\tilde\pi)\tilde\pi]=\stwo\EE_g[\tilde\pi(1-\tilde\pi)]-\sone\EE_{\tilde p}[\tilde\pi^2]$, which gives the second. The expansion in the proof of \cref{thm:calib-null}, carried out around $\tilde c$ and $\Tp_\infty$, shows that $\Tp$ and the $\Tp_{(-g)}$ are asymptotically linear with bounded influence functions. Hence $\widehat{\mathrm{se}}_{\mathrm{jack}}(\Tp)=O_\PP(N^{-1/2})$, and $|t_\Tp|\xrightarrow{\,\Pr\,}\infty$ when $\Tp_\infty\ne0$.

(b) $\widetilde F(c)=F(c)-\epsilon\sone\EE_p[h\pi_c]$, where $F$ is the population function under $p$, with $F(\cstar)=0$ and $F'(\cstar)=-\Hov$. Since $\sone\EE_p[h\pistar]=\EE_\mix[\pistar(1-\pistar)h]=\Hov\EE_{\mix_\star}[h]$, the implicit function theorem gives $\tilde c-\cstar=-\epsilon\EE_{\mix_\star}[h]+O(\epsilon^2)$. Let $\mathcal N(c,\epsilon)=\sone\EE_{\tilde p}[\pi_c^2]-\stwo\EE_g[\pi_c(1-\pi_c)]$ be the numerator of $\Tp_\infty$ at $c$. It vanishes at $(\cstar,0)$, its $\epsilon$-derivative there is $\sone\EE_p[h{\pistar}^2]=\Hov\EE_{\mix_\star}[\pistar h]$, and its $c$-derivative there is $2\EE_\mix[{\pistar}^2(1-\pistar)^2]-\EE_\mix[{\pistar}^2(1-\pistar)(1-2\pistar)]=\EE_\mix[{\pistar}^2(1-\pistar)]=\Hov\EE_{\mix_\star}[\pistar]$. Hence $\mathcal N(\tilde c,\epsilon)=\epsilon\Hov\{\EE_{\mix_\star}[\pistar h]-\EE_{\mix_\star}[\pistar]\EE_{\mix_\star}[h]\}+O(\epsilon^2)$, and dividing by $\widetilde H_0=\Hov+O(\epsilon)$ proves the claim.

(c) Responsibilities depend on a draw only through $f/g$, and they equal $0$ for posterior draws outside $A$ and $1$ for proposal draws outside $A$. Integrating the condition over $A$ gives $\tilde p(A)=\lambda p(A)$. Let $p''=\lambda p$ on $A$ and $p''=\tilde p$ on $A^c$, a probability density. On $A$ the laws of $f/g$ under $\tilde p$ and under $p''$ both equal $\lambda$ times its law under $p$, and both put mass $\tilde p(A^c)$ outside $A$, so the responsibilities of i.i.d.\ draws from $\tilde p$ and from $p''$ have the same joint law. Let $f''=(\Z/\lambda)p''$. Then $f''=f$ on $A$, and $f''=0=f$ where $g>0$ and $p=0$. Hence responsibilities computed from $f''$ coincide with those computed from $f$ at every point where a draw can fall, and the problem $(p'',f'',g)$ is correctly specified with evidence $\int f''=\Z/\lambda$. The bridge estimator is consistent in that problem, so $\Zhat\to\Z/\lambda$.
\end{proof}

\section{Further related work}\label{app:related}

This appendix collects connections omitted from \cref{sec:related} for space.

\paragraph{Noise-contrastive and density-ratio estimation.} The statistical theory of NCE covers consistency, asymptotic normality and comparisons of asymptotic variances \citep{gutmann_noise_2012,riou_noise_2018,uehara_analysis_2018,chehab_optimal_2022}. In particular, NCE is more efficient than importance sampling, and annealing turns an exponential dependence on the distance between proposal and target into a polynomial one \citep{chehab_provable_2023}. A second line of work explains why NCE fails when data and noise are far apart. The loss landscape becomes flat \citep{liu_analyzing_2022}, Gaussian noise can make the Hessian exponentially ill-conditioned in the dimension \citep{lee_pitfalls_2023}, and classifier-based density-ratio estimators suffer from a ``density chasm'' \citep{rhodes_telescoping_2020,srivastava_estimating_2023}. These results concern asymptotic variances, conditioning or optimization in multi-parameter models. Finite-sample theory for logistic-regression maximum likelihood with random design \citep[e.g.][]{chardon_finite_2024} addresses a different problem, the estimation of slopes rather than of an intercept with a known offset.

\paragraph{Free-energy computation.} With MBAR weights $W_{nk}=\widehat p_k(\theta_n)/\sum_lS_l\widehat p_l(\theta_n)$ over the pooled draws, the MBAR overlap matrix \citep{klimovich_guidelines_2015} is $O_{kl}=S_l\sum_nW_{nk}W_{nl}$. For two states this gives
\[
O=\begin{pmatrix}1-\Hhat/\sone&\Hhat/\sone\\\Hhat/\stwo&1-\Hhat/\stwo\end{pmatrix},
\]
whose eigenvalues are $1$ and $1-\Hhat/(\sone\stwo)$. The ``overlap scalar'' of the matrix (one minus its second eigenvalue, as implemented in \texttt{pymbar}) therefore equals $\Hhat/(\sone\stwo)$, our estimate of $\Iov$. We also checked the identity numerically. \citet{reinhardt_small_2021} studied BAR with few samples in one-dimensional test systems, using the overlap $\Omega=\int2pq/(p+q)$, which is our $\Iov$ for equal sample sizes. They found that BAR is then beaten by the simpler ``linear'' (simple overlap sampling) estimator, and that the sample size per state at which BAR overtakes it follows $n\approx b\,\Omega^{-a}$, with $a\approx0.8$ and $b\approx6$--$8.5$ in all four of their systems. BayesMBAR \citep{ding_bayesian_2024} provides Bayesian error bars in this regime, and \citet{li_understanding_2023} give an asymptotic error analysis of MBAR for correlated samples. For one-sided free-energy perturbation, \citet{wu_phase_2005} proposed overlap-based bias heuristics, and \citet{jarzynski_rare_2006} showed that the required number of samples grows exponentially in the dissipated work, the physics counterpart of \citet{chatterjee_sample_2018}.

\begin{table}
\centering\small
\caption{\Cref{thm:calib-null} under correct sampling ($2000$ replications per row, $G=20$). Ratio of the standard deviation of $\Tp$ to $\mathrm{sd}_{\mix_\star}(\pistar)/\sqrt{N\Hov}$, rejection rate of $|t_\Tp|>t_{19,0.975}$, and standard deviation of $t_\Tp$ ($1.057$ for $t_{19}$).}\label{tab:R2}
\begin{tabular}{lrrrcrr}
\toprule
Posterior and proposal & $\Sone$ & $\Stwo$ & $\neff$ & sd ratio & rejection & sd$(t_\Tp)$\\
\midrule
$\mathcal N(0,1)$, $\mathcal N(2,1)$ & 2000 & 2000 & 247.5 & 0.99 & 0.051 & 1.05\\
$\mathcal N(0,1)$, $\mathcal N(2,1)$ & 1000 & 3000 & 187.4 & 0.98 & 0.048 & 1.04\\
$\mathcal N(0,1)$, $\mathcal N(4,1)$ & 2000 & 2000 & 63.9 & 1.02 & 0.057 & 1.08\\
Cauchy, $\mathcal N(0,2^2)$ & 2000 & 2000 & 85.8 & 1.01 & 0.055 & 1.07\\
\bottomrule
\end{tabular}
\end{table}

\paragraph{Further non-asymptotic and Markov chain results.} Berry--Esseen bounds for general M-estimators are classical \citep{bentkus_berry_1997}, but they involve unspecified constants and moment conditions. Concentration inequalities for additive functionals of reversible chains with a spectral gap go back to \citet{lezaud_chernoff_1998}, and the rate $n^{-1/2}$ in the central limit theorem for Markov chains is due to \citet{bolthausen_berry_1982}. Among bounds with explicit constants, \citet{mann_berry_1996} treated countable chains by perturbation theory, \citet{lezaud_chernoff_2001} continuous-time reversible processes, and \citet{kloeckner_effective_2019} chains with a spectral gap on a general Banach algebra of functions, with a constant that is cubic in $\|f\|/\sigma$. From \citet{paulin_concentration_2015} we use the warm-start bounds and the bounds on the Taylor coefficients of the perturbed leading eigenvalue; see the remark following the proof of \cref{lem:chf}.

\section{Additional results}\label{app:repro}

\begin{table}
\centering\small
\caption{Universality study (\cref{sec:exp-universality}), $4000$ replications per row. Coverage of nominal $95\%$ delta-method intervals; KS distance of $\Delta/\sigN$ to $\mathcal N(0,1)$ (Monte Carlo noise level $\approx0.014$); ratio of the empirical standard deviation of $\Delta$ to $\sigN$; median of $|\log(\sighat/\sigN)|$; median of $\neffhat/\neff$.}\label{tab:E6}
\setlength{\tabcolsep}{4pt}
\resizebox{\textwidth}{!}{%
\begin{tabular}{llrrrccccc}
\toprule
Family & parameters & $\Sone=\Stwo$ & $\Hov$ & $\neff$ & coverage & KS & $\mathrm{sd}(\Delta)/\sigN$ & med $|\log\frac{\sighat}{\sigN}|$ & med $\neffhat/\neff$\\
\midrule
F2 & $\delta=2$ & 10 & 1.12e-01 & 1.24 & 0.943 & 0.014 & 0.994 & 0.105 & 0.949\\
F2 & $\delta=2$ & 30 & 1.12e-01 & 3.71 & 0.952 & 0.007 & 0.983 & 0.061 & 0.990\\
F2 & $\delta=2$ & 100 & 1.12e-01 & 12.37 & 0.950 & 0.016 & 0.999 & 0.033 & 0.999\\
F2 & $\delta=2$ & 1{,}000 & 1.12e-01 & 123.73 & 0.946 & 0.015 & 1.009 & 0.010 & 1.001\\
F2 & $\delta=2$ & 10{,}000 & 1.12e-01 & 1237.30 & 0.950 & 0.011 & 1.003 & 0.003 & 1.000\\
F2 & $\delta=4$ & 20 & 1.71e-02 & 0.64 & 0.893 & 0.019 & 0.971 & 0.218 & 0.499\\
F2 & $\delta=4$ & 60 & 1.71e-02 & 1.92 & 0.936 & 0.014 & 0.996 & 0.097 & 0.892\\
F2 & $\delta=4$ & 200 & 1.71e-02 & 6.39 & 0.954 & 0.017 & 0.979 & 0.050 & 0.972\\
F2 & $\delta=4$ & 2{,}000 & 1.71e-02 & 63.89 & 0.950 & 0.022 & 1.010 & 0.017 & 0.994\\
F2 & $\delta=6$ & 300 & 1.04e-03 & 0.62 & 0.888 & 0.024 & 0.912 & 0.292 & 0.366\\
F2 & $\delta=6$ & 1{,}000 & 1.04e-03 & 2.07 & 0.928 & 0.011 & 0.999 & 0.095 & 0.877\\
F2 & $\delta=6$ & 3{,}000 & 1.04e-03 & 6.20 & 0.944 & 0.010 & 0.996 & 0.049 & 0.955\\
F2 & $\delta=6$ & 30{,}000 & 1.04e-03 & 62.01 & 0.954 & 0.011 & 0.990 & 0.015 & 0.994\\
F1 & $d=10{,}\ \sigma=1.5$ & 10 & 1.31e-01 & 1.25 & 0.942 & 0.025 & 1.008 & 0.114 & 0.988\\
F1 & $d=10{,}\ \sigma=1.5$ & 30 & 1.31e-01 & 3.74 & 0.944 & 0.014 & 1.013 & 0.064 & 0.996\\
F1 & $d=10{,}\ \sigma=1.5$ & 100 & 1.31e-01 & 12.47 & 0.953 & 0.011 & 0.983 & 0.035 & 1.001\\
F1 & $d=10{,}\ \sigma=1.5$ & 1{,}000 & 1.31e-01 & 124.67 & 0.954 & 0.008 & 0.988 & 0.011 & 1.001\\
F1 & $d=40{,}\ \sigma=1.3$ & 100 & 8.80e-02 & 11.40 & 0.952 & 0.013 & 0.997 & 0.037 & 0.992\\
F1 & $d=40{,}\ \sigma=1.3$ & 300 & 8.80e-02 & 34.21 & 0.952 & 0.010 & 0.995 & 0.021 & 0.998\\
F1 & $d=40{,}\ \sigma=1.3$ & 1{,}000 & 8.80e-02 & 114.03 & 0.952 & 0.015 & 0.994 & 0.012 & 1.000\\
F1 & $d=40{,}\ \sigma=1.3$ & 10{,}000 & 8.80e-02 & 1140.27 & 0.955 & 0.021 & 0.977 & 0.004 & 1.000\\
F3 & $d=20{,}\ \nu=3$ & 10 & 1.28e-01 & 1.25 & 0.926 & 0.047 & 1.087 & 0.155 & 0.969\\
F3 & $d=20{,}\ \nu=3$ & 30 & 1.28e-01 & 3.75 & 0.944 & 0.033 & 1.047 & 0.090 & 0.988\\
F3 & $d=20{,}\ \nu=3$ & 100 & 1.28e-01 & 12.49 & 0.953 & 0.021 & 0.990 & 0.049 & 0.997\\
F3 & $d=20{,}\ \nu=3$ & 1{,}000 & 1.28e-01 & 124.93 & 0.949 & 0.012 & 1.009 & 0.016 & 1.000\\
F3 & $d=100{,}\ \nu=3$ & 100 & 7.13e-02 & 10.20 & 0.951 & 0.032 & 1.008 & 0.070 & 0.994\\
F3 & $d=100{,}\ \nu=3$ & 300 & 7.13e-02 & 30.59 & 0.948 & 0.016 & 1.013 & 0.041 & 0.998\\
F3 & $d=100{,}\ \nu=3$ & 1{,}000 & 7.13e-02 & 101.95 & 0.946 & 0.014 & 1.009 & 0.022 & 1.000\\
F3 & $d=100{,}\ \nu=3$ & 3{,}000 & 7.13e-02 & 305.85 & 0.950 & 0.012 & 0.998 & 0.013 & 1.000\\
\bottomrule
\end{tabular}}
\end{table}

\begin{table}
\centering\small
\caption{MCMC universality with AR(1) posterior chains ($\gamma=1-\phi$) and independent proposals ($2000$ replications). Sample sizes target the $\neff^P$ header values (scaled by $1.5$ for F1). Top: Coverage of nominal $95\%$ intervals using the spectral MCSE (Geyer's estimator), the oracle variance $\sigma_{N,P}^2$, and a naive variance ignoring autocorrelation. Bottom: Accuracy of the long-run variance estimate (median $\widehat n_P/v_P$) and the median overlap certificate $\neffhat^P$ for the three smallest $\neff^P$ values.}\label{tab:E9}
\setlength{\tabcolsep}{3.5pt}
\resizebox{\textwidth}{!}{%
\begin{tabular}{llrccccccccc}
\toprule
& & & \multicolumn{6}{c}{coverage, spectral MCSE, at increasing $\neff^P$} & & \\
\cmidrule(lr){4-9}
Family & $\phi$ & $\tau_W$ & $1$ & $2$ & $4$ & $8$ & $20$ & $80$ & oracle & ignoring autocorr.\\
\midrule
F2, $\delta=2$ & 0 & 1.00 & 0.943 & 0.952 & 0.952 & 0.956 & 0.955 & 0.951 & 0.945--0.953 & 0.948\\
F2, $\delta=2$ & 0.5 & 1.87 & 0.875 & 0.917 & 0.930 & 0.942 & 0.943 & 0.946 & 0.944--0.956 & 0.846\\
F2, $\delta=2$ & 0.8 & 4.55 & 0.833 & 0.877 & 0.914 & 0.931 & 0.937 & 0.955 & 0.943--0.957 & 0.649\\
F2, $\delta=2$ & 0.95 & 18.02 & 0.768 & 0.849 & 0.882 & 0.918 & 0.938 & 0.942 & 0.946--0.956 & 0.362\\
\midrule
F2, $\delta=4$ & 0 & 1.00 & 0.913 & 0.939 & 0.942 & 0.947 & 0.949 & 0.956 & 0.945--0.955 & 0.940\\
F2, $\delta=4$ & 0.5 & 1.46 & 0.900 & 0.931 & 0.935 & 0.943 & 0.946 & 0.954 & 0.950--0.955 & 0.891\\
F2, $\delta=4$ & 0.8 & 3.13 & 0.844 & 0.883 & 0.906 & 0.936 & 0.928 & 0.945 & 0.946--0.961 & 0.728\\
F2, $\delta=4$ & 0.95 & 11.76 & 0.770 & 0.833 & 0.893 & 0.913 & 0.949 & 0.938 & 0.945--0.962 & 0.429\\
\midrule
F1, $d=10$, $\sigma=1.5$ & 0 & 1.00 & 0.948 & 0.945 & 0.947 & 0.956 & 0.949 & 0.956 & 0.945--0.954 & 0.947\\
F1, $d=10$, $\sigma=1.5$ & 0.5 & 1.22 & 0.923 & 0.954 & 0.951 & 0.952 & 0.944 & 0.948 & 0.941--0.960 & 0.924\\
F1, $d=10$, $\sigma=1.5$ & 0.8 & 2.20 & 0.881 & 0.917 & 0.926 & 0.940 & 0.945 & 0.955 & 0.944--0.960 & 0.808\\
F1, $d=10$, $\sigma=1.5$ & 0.95 & 7.33 & 0.815 & 0.882 & 0.894 & 0.912 & 0.949 & 0.944 & 0.937--0.962 & 0.526\\
\bottomrule
\end{tabular}}

\vspace{6pt}
\begin{tabular}{llrccccccccc}
\toprule
& & & \multicolumn{6}{c}{median $\widehat n_P/v_P$ at increasing $\neff^P$} & \multicolumn{3}{c}{median $\neffhat^P$}\\
\cmidrule(lr){4-9}\cmidrule(lr){10-12}
Family & $\phi$ & $\tau_W$ & $1$ & $2$ & $4$ & $8$ & $20$ & $80$ & $1$ & $2$ & $4$\\
\midrule
F2, $\delta=2$ & 0 & 1.00 & 0.95 & 0.97 & 1.03 & 1.02 & 1.02 & 1.02 & 1.0 & 1.9 & 3.9\\
F2, $\delta=2$ & 0.5 & 1.87 & 0.61 & 0.75 & 0.86 & 0.93 & 0.98 & 1.00 & 1.3 & 2.3 & 4.4\\
F2, $\delta=2$ & 0.8 & 4.55 & 0.43 & 0.62 & 0.78 & 0.87 & 0.95 & 0.98 & 1.6 & 2.8 & 4.8\\
F2, $\delta=2$ & 0.95 & 18.02 & 0.34 & 0.54 & 0.71 & 0.81 & 0.91 & 0.98 & 2.0 & 3.1 & 5.2\\
\midrule
F2, $\delta=4$ & 0 & 1.00 & 0.72 & 0.89 & 0.98 & 1.00 & 1.02 & 1.01 & 0.7 & 1.7 & 3.5\\
F2, $\delta=4$ & 0.5 & 1.46 & 0.62 & 0.75 & 0.85 & 0.91 & 0.98 & 1.00 & 1.0 & 1.9 & 3.9\\
F2, $\delta=4$ & 0.8 & 3.13 & 0.40 & 0.59 & 0.71 & 0.83 & 0.91 & 0.98 & 1.3 & 2.4 & 4.5\\
F2, $\delta=4$ & 0.95 & 11.76 & 0.29 & 0.48 & 0.64 & 0.78 & 0.90 & 0.96 & 1.7 & 2.9 & 5.0\\
\midrule
F1, $d=10$, $\sigma=1.5$ & 0 & 1.00 & 0.98 & 1.00 & 1.01 & 1.02 & 1.02 & 1.01 & 1.4 & 3.0 & 5.8\\
F1, $d=10$, $\sigma=1.5$ & 0.5 & 1.22 & 0.86 & 0.93 & 0.97 & 1.01 & 1.02 & 1.02 & 1.6 & 3.0 & 5.9\\
F1, $d=10$, $\sigma=1.5$ & 0.8 & 2.20 & 0.61 & 0.76 & 0.88 & 0.95 & 1.00 & 1.03 & 2.0 & 3.6 & 6.6\\
F1, $d=10$, $\sigma=1.5$ & 0.95 & 7.33 & 0.41 & 0.62 & 0.75 & 0.87 & 0.98 & 1.03 & 2.9 & 4.4 & 7.6\\
\bottomrule
\end{tabular}
\end{table}

\paragraph{Checks of \cref{prop:onesided,thm:calib-null,prop:calib-alt}.} 

\begin{table}[!htbp]
\centering\small
\caption{\Cref{prop:onesided} under poor overlap ($10^4$ replications per row). Largest ratio $\neffhat/(\varsigma\max\{M_{\mathrm{post}},M_{\mathrm{prop}}\})$ (at most $1$ by part (a)); frequency of $\neffhat\ge10$ and its bound from part (b); $99\%$ quantile of $\neffhat$ and the bound of part (b) with $\delta=0.01$.}\label{tab:R1}
\setlength{\tabcolsep}{4pt}
\resizebox{\textwidth}{!}{%
\begin{tabular}{lrrrrcrrrr}
\toprule
Posterior and proposal & $\Sone$ & $\Stwo$ & $\neff$ & $N\Hov$ & max ratio & $\PP(\neffhat\ge10)$ & bound & $q_{0.99}(\neffhat)$ & bound\\
\midrule
$\mathcal N(0,1)$, $\mathcal N(5,1)$ & 1000 & 1000 & 9.30 & 9.48 & 0.63 & 0.347 & 1 & 14.9 & 49.2\\
$\mathcal N(0,1)$, $\mathcal N(6,1)$ & 1000 & 1000 & 2.07 & 2.08 & 0.74 & 0 & 0.46 & 5.1 & 22.1\\
$\mathcal N(0,1)$, $\mathcal N(7,1)$ & 1000 & 1000 & 0.36 & 0.36 & 0.97 & 0 & 0.0004 & 1.8 & 12.3\\
$\mathcal N(0,1)$, $\mathcal N(6,1)$ & 400 & 1600 & 1.62 & 1.62 & 0.81 & 0 & 0.21 & 4.2 & 19.9\\
Cauchy, $\mathcal N(8,0.5^2)$ & 150 & 150 & 2.28 & 2.35 & 0.66 & 0 & 0.67 & 5.6 & 23.5\\
$\mathcal N(0,1)$, $\mathcal N(0,0.05^2)$ & 40 & 40 & 3.01 & 3.70 & 0.52 & 0 & 1 & 5.4 & 29.7\\
$\mathcal N(0,1)$, $\mathcal N(0,10^2)$ & 30 & 30 & 3.24 & 4.74 & 0.44 & 0 & 1 & 5.0 & 34.0\\
\bottomrule
\end{tabular}}
\end{table}

\Cref{tab:R1} checks \cref{prop:onesided} in poor-overlap configurations, two of them with importance ratios of infinite variance (Cauchy posterior with a narrow off-centre proposal, and a proposal twenty times narrower than the posterior). The deterministic bound (a) held in every replication, with largest ratio $0.97$, and $\neffhat\ge10$ never occurred when $\neff\le3.2$. \Cref{tab:R2} checks \cref{thm:calib-null}: the standard deviation of $\Tp$ matches $\mathrm{sd}_{\mix_\star}(\pistar)/\sqrt{N\Hov}$ to within $3\%$, and $t_\Tp$ has the size and spread of $t_{19}$, including for a Cauchy posterior. For \cref{prop:calib-alt}(b) we take $p=\frac12\mathcal N(-1.5,0.4^2)+\frac12\mathcal N(1.5,0.8^2)$, $g=\mathcal N(0,1.6^2)$ and a sampler that puts weight $\frac12+\epsilon$ on the narrow component. By quadrature, $(\tilde c-\cstar)/(-\EE_{\mix_\star}h)$ is $1.000$, $1.001$, $1.003$ and $1.005$, and $\Tp_\infty/\Cov_{\mix_\star}(\pistar,h)$ is $1.00$, $1.01$, $1.02$ and $1.04$, for $\epsilon=0.005$, $0.02$, $0.05$ and $0.1$. With $\Sone=\Stwo=2000$ and $\epsilon\in\{0.01,0.02,0.04\}$, the biases are $0.21$, $0.43$ and $0.86$ MCSEs, and the rejection rates of $t_\Tp$ are $4.5\%$, $6.2\%$ and $8.6\%$, against $5.3\%$, $6.3\%$ and $10.4\%$ predicted by the noncentral $t_{19}$ law of \cref{rem:calib-power} ($1000$ replications; $3.8\%$ under the null). For the blind spot, $p=0.7\,\mathcal N(0,1)+0.3\,\mathcal N(10,1)$, $g=\mathcal N(0,1.5^2)$, and the sampler draws from $\mathcal N(0,1)$. By quadrature $\tilde c-\cstar=-0.35660$, against $\log0.7=-0.35667$, and $\Tp_\infty=5.5\times10^{-6}$. With $\Sone=\Stwo=2000$ the bias is $21$ MCSEs and the coverage of nominal $95\%$ intervals is $0.000$, while $t_\Tp$ rejects in $4.5\%$ of $1000$ replications.

\end{document}